\documentclass[journal,transmag]{IEEEtran}

\usepackage{amssymb}
\usepackage{latexsym}
\usepackage{amsmath}
\usepackage{amsthm}
\usepackage{fullpage}
\usepackage{url}
\usepackage{graphicx}
\usepackage{nopageno}
\usepackage{algorithm}
\usepackage{url}
\usepackage[noend]{algpseudocode}
\usepackage{dsfont}

\usepackage{tikz}
\usetikzlibrary{arrows,positioning}

\usepackage{textpos}
\usepackage{booktabs}

\usepackage{paralist}
\usepackage{multirow}
\usepackage{subfigure}
\usepackage{caption}
\graphicspath{{./}{./}}

\usepackage{color}

\newcommand{\G}{\mathcal{G}}
\newcommand{\V}{\mathcal{V}}
\newcommand{\E}{\mathcal{E}}

\DeclareMathOperator*{\vol}{vol}
\DeclareMathOperator*{\cut}{cut}

\newcommand{\spcut}{\tilde{\phi}}
\newcommand{\cond}{\phi}
\newcommand{\lcond}{\phi'}

\begin{document}
	
\title{An optimization approach to locally-biased graph algorithms}

\author{Kimon~Fountoulakis
\thanks{K.~Fountoulakis is with the International Computer Science Institute and the Department of Statistics, University of California Berkeley. E-mail: kfount@berkeley.edu.}
\and 
David F.~Gleich
\thanks{D.~F.~Gleich is with the Department of Computer Science, Purdue University. E-mail: dgleich@purdue.edu.} 	
\and 
Michael~W.~Mahoney
\thanks{M.~W.~Mahoney is with the International Computer Science Institute and the Department of Statistics, University of California Berkeley. E-mail: mmahoney@stat.berkeley.edu.}}

\date{}
\maketitle

% MWM: I'M NOT SURE WHETHER THE FOLLOWING SHOULD GO ABOVE OR BELOW MAKETITLE.
\begin{abstract}
Locally-biased graph algorithms are algorithms that attempt to find local or small-scale structure in a large data graph.
In some cases, this can be accomplished by adding some sort of locality constraint and calling a traditional graph algorithm; but more interesting are locally-biased graph algorithms that compute answers by running a procedure that does not even look at most of the input graph.
This corresponds more closely to what practitioners from various data science domains do, but it does not correspond well with the way that algorithmic and statistical theory is typically formulated.
Recent work from several research communities has focused on developing locally-biased graph algorithms that come with strong complementary algorithmic and statistical theory and that are useful in practice in downstream data science applications.
We provide a review and overview of this work, highlighting commonalities between seemingly-different approaches, and highlighting promising directions for future work.
\end{abstract}

%!TEX root = locally-biased-ieee.tex
\section{Introduction}
\label{sxn:intro}

Graphs, long popular in computer science and discrete mathematics, have received renewed interest recently in statistics, machine learning, data analysis, and related areas because they provide a useful way to model many types of relational data. 
In this way, graphs can be used to extract insight from data arising in many application domains. In biology, e.g., graphs are routinely used to generate hypotheses for experimental validation~\cite{Tuncbag-2016-glioblastoma}; and in neurscience, they are used to study the networks and circuits in the brain~\cite{Bassett-2006-small-world,Sporns-2002-network-analysis}.

Given their ubiquity, graphs and graph-based data have been approached from several different perspectives.
In computer science, it is common to develop algorithms, e.g., for connected component, minimum spanning tree, and maximum flow problems, to run on data that are modeled as a precisely-specified input graph. 
These algorithms are then characterized by the number of operations they use for the worst-possible input at any size. 
In statistics and machine learning, on the other hand, it is common to use graphs as models to perform inference about unseen data. In this case, one often hypothesizes an unseen graph with a particular structure, such as block structure, hierarchical structure, low-rank structure in the adjacency matrix, etc. Then one runs algorithms on the observed data in order to impute entries in the unobserved hypothesized graph. 
These methods may be characterized in terms of running time, but they are also characterized in terms of the amount of data needed to recover the hidden hypothesized structure.

%Many application areas, however, adopt a sort of hybrid approach that does not fit cleanly into the above graphs-as-data and graphs-as-models methodological taxonomy.
% For example, 
In many application areas where the end goal is to obtain some sort of domain-specific insight, e.g., such as in social network analysis, neuroscience, medical imaging, etc., one constructs graphs from primary data, and then one runs a computational procedure that does \emph{not} come with either of these traditional types of theoretical guarantees. 
As an example, consider the GeneRank method~\cite{morrison2005-generank}, where we have a set of genes related to an experimental condition in a microarray study. 
This set of genes is ``refined'' via a locally-biased graph algorithm closely related to those we will discuss.  
Importantly, this operational refinement procedure does \emph{not} come with the sort of theory traditional in statistics, machine learning, or computer science.
As anther example, e.g., in social network applications, one might run a random walk process for a few steps from a starting node of interest, and if the walk ``gets stuck'' then one might interpret this as evidence that that region of the graph is meaningful in the application domain~\cite{EH09_TR}. 
These are examples of the types of heuristics commonly-used in applications. 
By heuristic, we mean an algorithm in the sense that it performs a sequence of well-defined steps, but one where precise theory is lacking (although usually heuristics come with strong intuitive motivation and are justified in terms of some downstream application). 
In particular, typically heuristics do not explicitly optimize a well-defined objective function and typically they do not come with well-defined inferential guarantees. 
%one might have a small set of nodes that correspond to a candidate seed set of ground truth nodes -- such as a set of genes related to a microarray experiment -- and one might want to ``refine'' it to obtain a prediction that is more useful in an application domain -- such as 

%\footnote{By heuristic, we mean an algorithm, in the sense that it performs a sequence of well-defined steps, but for which theory is lacking. For example, typically it does not explicitly optimize some well-defined objective exactly or approximately, and typically it does not explicitly come with well-defined inferential guarantees.} and interprets the output of that procedure.  
%For example, one might run a random walk process for a few steps from a starting node of interest, and if one gets ``stuck,'' then one might interpret this as evidence for a cluster or community that is meaningful in a particular application domain. Alternatively, one might have a small set of nodes that correspond to a candidate seed set of ground truth nodes, and one might want to ``refine'' it to obtain a prediction that is more useful in a particular application domain. 

Note that, in both of these examples, the goal is to find ``local'' or ``small scale'' structure in the data graph. 
Both examples also correspond to what practitioners interested in downstream applications actually do. 
Existing algorithmic and statistical theory, however, has challenges with these local or small-scale structures. 
For instance, a very ``good'' algorithmic runtime on a graph is traditionally one that is \emph{linear} in the number of vertices and edges.
If the output of interest is only a vanishingly small fraction of a large graph, however, then this theory may not provide strong qualitative guidance on how these locally-biased methods behave in practice. 
Likewise, inferential methods often assume that the structures inferred constitute a substantial fraction of the graph, and many statistical techniques have challenges differentiating very small structure from random noise. 
%(\textcolor{red}{Michael one of the reviewers asked clarification about the last sentence. I think a bunch of citation of non-working methods will do. However, I am not sure what you had in mind.})

In this overview, we describe a class of graph algorithms that has proven to be very useful for identifying and interpreting small-scale local structure in large-scale data. 
For this class of algorithms, however, strong algorithmic and statistical theory has been developed. 
In particular, these graph algorithms are \emph{locally-biased} in one of several precisely-quantified senses. 
We will describe what we mean by this in more detail below, but, informally, this means that the algorithms are most interested in only a small part of a large graph. 
As opposed to heuristic operational procedures, however, many of these algorithms do come with strong worst-case algorithmic guarantees, and many of these algorithms also do come with statistical guarantees that prove they have \emph{implicit} regularization properties. 
This complementary algorithmic-statistical theory helps explain their improved performance in many practical applications.

While the approach of locally-biased graph algorithms is very general, it has been developed most extensively for the fundamental problem of finding locally-biased graph partitions, i.e., clusters or communities, and so we will focus on locally-biased approaches for the graph clustering problem.
Of course, this partitioning question is of interest in many more application-driven areas, where one is interested in finding useful or meaningful clusters as part of a data analysis pipeline.
%\cite{XXX-Citation-XXX}

\subsection{The rationale for local analysis in real-world data}

%To understand the motivation, recall that a fundamental and ubiquitous question---that has been studied in different ways by different research communities---has to do with partitioning an input graph into two or more pieces.
%his partitioning question is of interest in computer science for divide and conquer algorithms, in statistics and machine learning for controlling inference, and in many more application-driven areas where one is interested in finding useful or meaningful clusters as part of a data analysis pipeline.
%A common intuition is that the idealized case is essentially a graph bisection where a very large graph is repeated split into roughly two equally sized pieces. 
%Doing so involves creating an objective to optimize the tradeoff between partition size and partition quality that results in a well-balanced partition since strict bisection is rarely required or even possible. 
%Then this involves reading the entire graph into memory and running a procedure that examines all of the nodes and edges to output the partitions. 

\begin{figure}[t]
	\subfigure[A random geometric graph\label{fig:small-cluster-a}]{\includegraphics[width=0.5\linewidth]{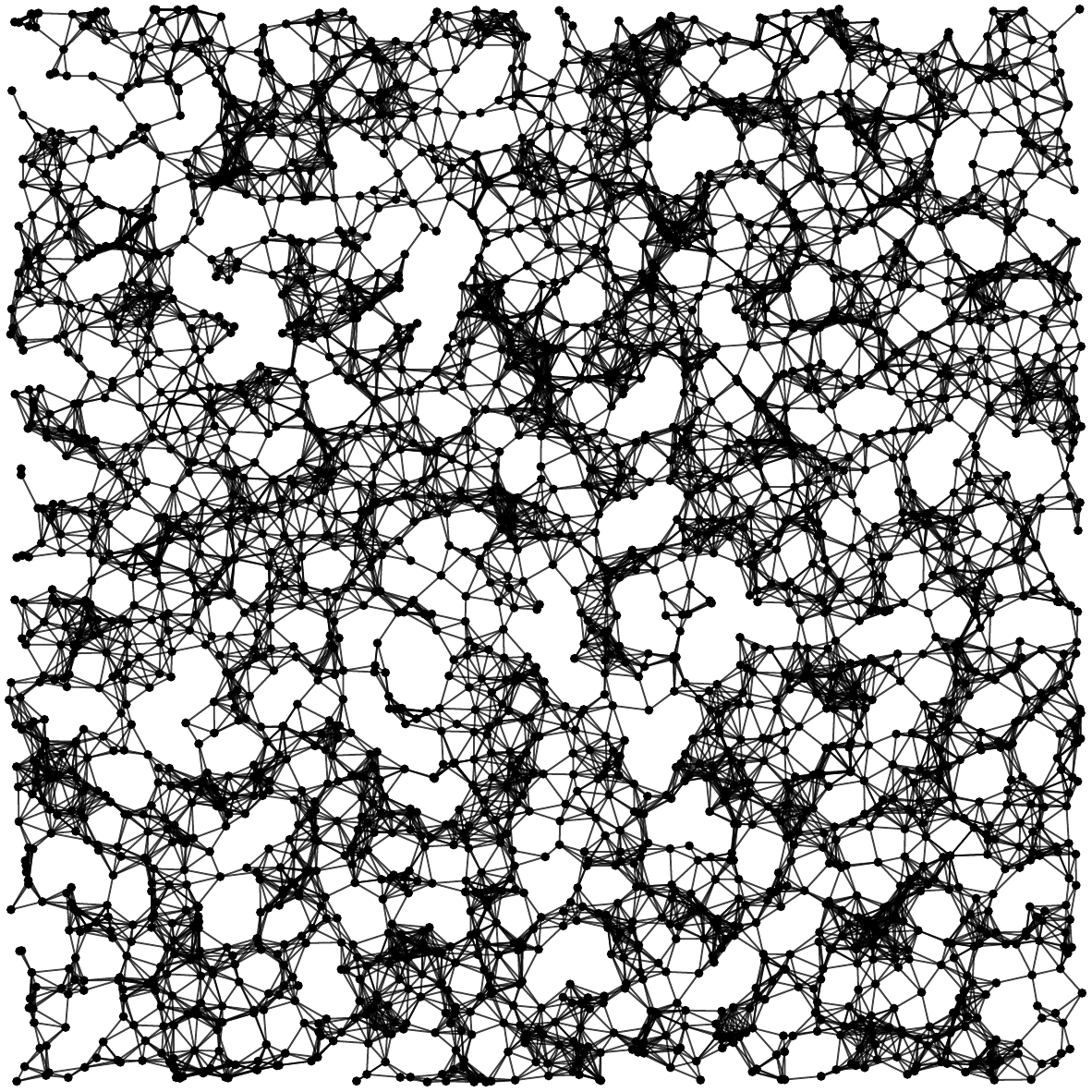}}%
	\subfigure[A typical data graph\label{fig:small-cluster-b}]{\includegraphics[width=0.5\linewidth]{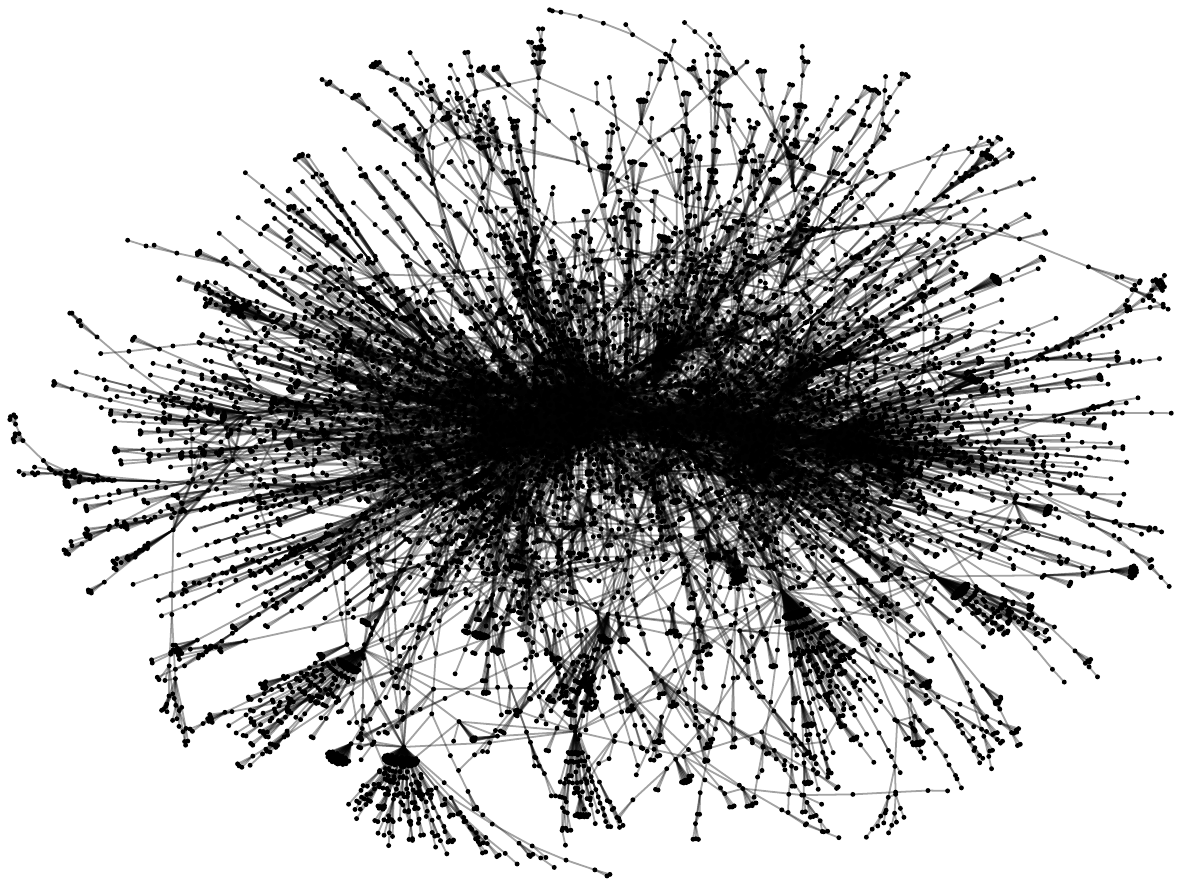}}
	\subfigure[The optimal conductance solution for the geometric graph bisects the graph into two large well-balanced pieces\label{fig:small-cluster-c}]{\includegraphics[scale=0.32]{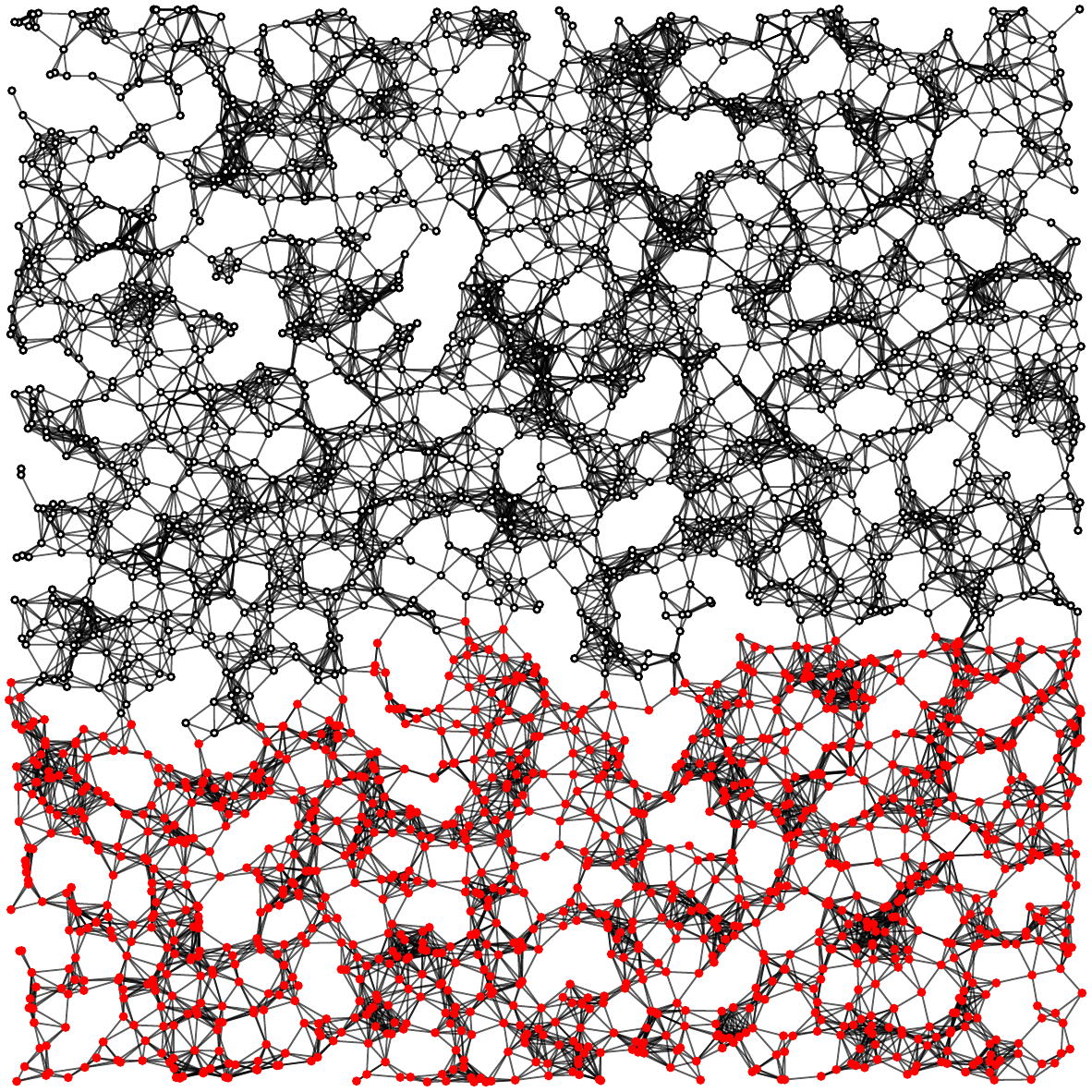}}
		\subfigure[The optimal conductance solution for a typical data graph. (Inset. A zoomed view of the subgraph where the two unfilled notes are the border with the rest of the graph.)\label{fig:small-cluster-d}]{\raisebox{0cm}{\includegraphics[width=0.5\linewidth]{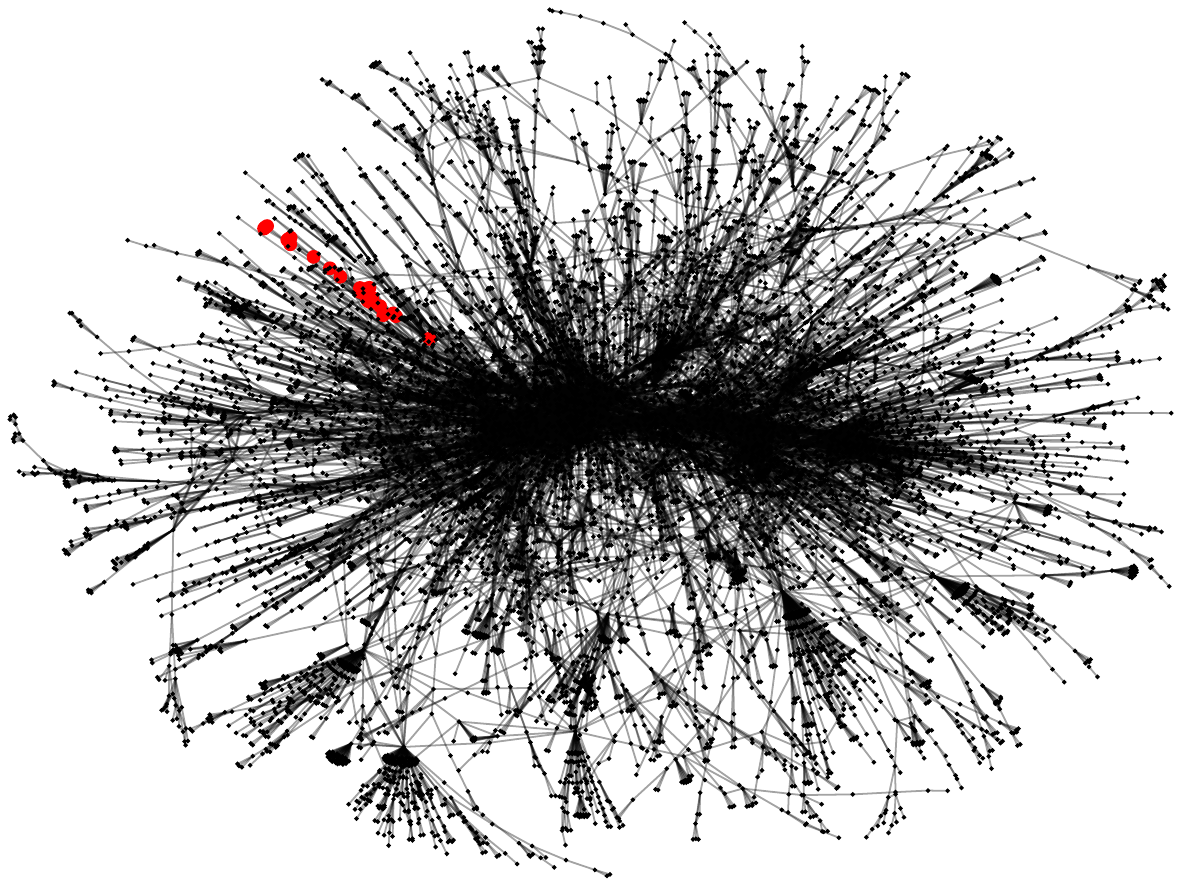}}\llap{\colorbox{white}{\includegraphics[scale=0.32]{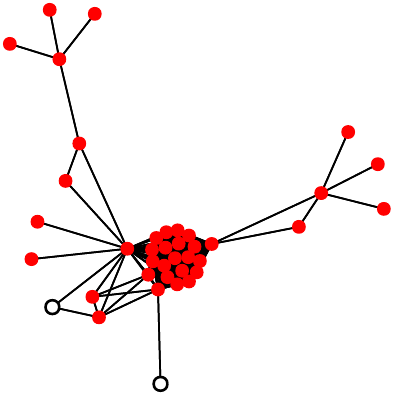}}}}
	\caption{At left, a geometric graph has a pleasing and intuitive layout in the two-dimensional plane. At right, a more typical data graph has a classic \emph{hairball} layout that shows little high level structure. Due to the evident lack of global structure in the data graph, locally-biased graph algorithms are often used in these contexts. 
	The solution of the Minimum-Conductance problem in the geometric graph is a large set of nodes, and it has conductance value $0.00464$.  The solution of the Minimum-Conductance problem in the more typical data graph is a small set of nodes, and it has conductance value $0.00589$. 
	The inset figure shows that this small graph is very dense and has only 3 edges leaving the set.}
	\label{fig:small-cluster}
\end{figure}

As a quick example of why local graph analysis is frequently used in data and data science applications, we present in Figure~\ref{fig:small-cluster} the results of finding the best partition of both a random geometric graph and a more typical data graph. 
Standard graph partitioning algorithms must operate on, or ``touch'', each vertex and edge of the graph to identify these partitions. 
The best partition of the geometric graph is around half the data, where it is reasonable to run an algorithm that touches all the data.  
On the other hand, the best partition of the data graph is very small, and in this case touching the entire graph to find it can be too costly in terms of computation time. 
%The best partition of the data graph is small. In this case a standard graph partitioning algorithm 
%will have to operate on all edges and nodes of the graph in order to find this small cluster. 
The local graph clustering techniques discussed in this paper can find this cluster touching only edges and nodes in the output cluster, greatly reducing the computation~time.  
%This is the motivation behind local graph clustering.

Far from being a pathology or a peculiarity, the finding that optimal partitions of real-world networks are often extremely imbalanced, thus leading to very small optimal clusters, is endemic to many of the graphs arising in large-scale data analysis~\cite{LLDM08_communities_CONF,LLDM09_communities_IM,LLM10_communities_CONF,Jeub15}.\footnote{An important applied question has to do with the meaningfulness, usefulness, etc., of such small clusters.  We do not consider those questions here, and instead we refer the interested reader to prior work~\cite{LLDM08_communities_CONF,LLDM09_communities_IM,LLM10_communities_CONF,Jeub15}.  Here, we instead focus on the algorithmic and statistical properties of these locally-biased algorithms.}  %Thus, the information contained in the optimal results from these partitioning problems on data-graphs is small compared with the size of the graph.

Let us now explain in more detail Figure~\ref{fig:small-cluster}.
Figure~\ref{fig:small-cluster-a} shows a graph with 3000 nodes that is typical of many graphs that possess a strong underlying \emph{geometry}, e.g., those used in computer graphics, computer vision, logistics planning, road network analysis, and so on. 
This particular graph is produced by generating 3000 random points in the plane and connecting all points within a small radius, such that the final graph is connected. 
The geometric graph can be nearly bisected by optimizing a measure known as conductance (we shall define this shortly), which is designed to balance partition size and quality. 
In Figure~\ref{fig:small-cluster-b}, we show a more typical data graph of around 10680 nodes~\cite{Moguna-2004-models}, where this particular data graph is based on the trust relationships in a PGP (Pretty Good Privacy) key-chain. 
Optimizing the same conductance objective function results in a tiny set, Figure \ref{fig:small-cluster-d}, and not the near-bisection as in the geometric case, Figure \ref{fig:small-cluster-c}. 
(We were able to use integer optimization techniques to directly solve the NP-hard problems at the cost of months of computation.)  
Many other examples of this general phenomenon can be found in prior work~\cite{LLDM08_communities_CONF,LLDM09_communities_IM,LLM10_communities_CONF,Jeub15}.

\subsection{Partitioning as a model problem}
The problem of finding good partitions or clusters is ubiquitous.  Representative examples 
include biomedical applications \cite{ICOYHS01,GN02}, internet and world-wide-web \cite{FFF99,AJB99,KKRRT99}, social graphs \cite{LLDM08_communities_CONF,scott,Traud11,UKBM11},
human communication graphs \cite{OSHSLKKB07}, human mobility graphs \cite{EEBL10},
voting graphs in political science \cite{PMNW05,multiplex_Mucha,MMP12,TKMP11,GHKV07}, protein interaction graphs \cite{JPD10}, material science \cite{BODP12,RCHSSKMN11}, neuroscience 
\cite{BWPMCG11,WBMPG12,BWRPMG13}, collaboration graphs \cite{ELP11}.

All of the features of locally-biased computations are present in this model partitioning problem. 
For example, while some of these algorithms read the entire graph as input but are engineered to return answers that are biased toward and meaningful for a smaller part of the input graph~\cite{MOV12_JMLR,andersen08soda,ZBLWS04,pan2004-cross-modal-discovery}, other algorithms can take as input a small ``seed set'' of nodes as well as an oracle with which to access neighbors of a node, and they return meaningful answers without even touching the entire graph~\cite{FCSRM16_TR,ACL06,KG14,ST13}.
Similarly, while these algorithms are often formulated in the language of theoretical computer science as approximation algorithms, i.e., they come with running time guarantees and can be shown to \emph{approximate} to within some quality-of-approximation factor some objective function of interest, e.g., conductance, in other cases one can prove statistical results such as that they \emph{exactly} solve a regularized version of that objective~\cite{GM14_ICML,Veldt-preprint-simple-local-flow,MO11-implementing,FCSRM16_TR}. 

Importantly, this statistical regularization is implicit rather than explicit. Typically, regularization is explicitly used in statistics, when fitting models with a large number of parameters, in order to avoid overfitting to the given data. It is also used to select answers biased towards specific sets---for example, sparse solution sets by using the Lasso~\cite{T96}. In the case of locally-biased graph algorithms, one simply runs a \emph{faster} algorithm for a problem. In some cases, the nature of the approximation that this fast algorithm makes can be related back to a regularized variant of the objective function. 
%Thus, using these locally-biased algorithms can often yield improved robustness on prediction and learning problems based on network data. 
%\begin{figure}[t]
%	\subfigure[The Isoperimetry solution for the geometric graph]{\includegraphics[scale=0.32]{U3A-bestset}}
%		\subfigure[The Isoperimetry solution for the data graph. (Inset. A zoomed view of the subgraph where the two unfilled notes are the border with the rest of the graph.)]{\raisebox{0cm}{\includegraphics[width=0.5\linewidth]{pgp-bestset}}\llap{\colorbox{white}{\includegraphics[scale=0.32]{pgp-bestset-zoom}}}}
%\caption{The solution of the Isoperimetry problem in the geometric graph shows a large set with isoperimetry ratio 0.00464. The solution of the Isoperimetry problem in the data graph shows a small set with an isoperimetry ratio of 0.00589. The inset figure shows that this small graph is very dense and has only 3 edges leaving the set. }
%\end{figure}

%It almost goes without saying that one is typically interested in the entire graph---e.g., one constructs an objective to optimize that involves all of the nodes and edges, one is interested in very large well-balanced clusters, etc.---and one typically must at least read in the entire graph as input to the algorithm.

%\item

\subsection{Overview}
In this paper, we will consider partitioning from the perspective of conductance, and we will survey a recent class of results about a common localizing construction. 
These methods will let us find the optimal sets in Figure~\ref{fig:small-cluster} without resorting to integer optimization (but also losing the proof of optimality). 
Depending on the exact construction, they will also come with a variety of helpful statistical properties. 
We'll conclude with a variety of different perspectives on these problems and some open questions. 

In Section \ref{sec:prelim} we describe assumptions, notation and preliminary results which we will use in this paper. In Section \ref{sec:global_cond_spcut} we discuss two global graph partitioning problems and their spectral relaxation. In Section \ref{sec:localgraphpart} we describe the local graph clustering application. In Section \ref{sec:empirical} we provide empirical evaluations for the global and local graph clustering algorithms which are described in this paper. Finally, in Section \ref{sec:disc_conclusion} we give our conclusions.

\section{Preliminaries and Notation}
\label{sec:prelim}

\textbf{Graph assumptions:}
We use the letter $\G$ to denote a given connected graph. We assume that $\G$ is undirected with no self-loops. 
Many of the constructions we will use operate on \emph{weighted graphs} and so we assume that each edge may have a positive capacity. Graphs that are unweighted should have all of their capacities set to $1$. 
 
\textbf{Nodes, edges and cuts:}
Let $\V = \{v_1,v_2\dots,v_n\}$ be a given set of $|\V|$ nodes of graph $\G$. We denote with $e_{ij}$ an edge in the graph between nodes $v_i$ and $v_j$. 
Let $\E$ be a given set of $|\E|$ edges of graph $\G$. A subset $S\subset\V$ of nodes can be used to define a partitioning 
of $\V$ into $S$ and $S^c:=\V \backslash S$. We define a cut as subset $E \subset \E$ which partitions the graph $\G$ into
two sets.
Given a partition $S\subset\V$ and $S^c$, then $E(S,S^c)=\{e_{ij}\in \E \ | \ v_i\in S \mbox{ and } v_j\in S^c\}$ is the set of edges with one side in $S$ and the other side 
in $S^c$. If the partition is clear from the context we write the cut set as $E$ instead of $E(S,S^c)$.\ Let $c_{ij}$ be a weight of the edge $e_{ij}$, then we define the cut $S$ as 
\begin{equation}
 \cut(S) := \cut(E(S,S^c)) := \sum_{e_{ij}\in E(S,S^c)} c_{ij}.
\end{equation}
The volume of a set $S$ is
\begin{equation}
 \vol(S) := \sum_{v_i \in S} \sum_{e_{ij} \in \E} c_{ij}.
\end{equation}
For simplicity of notation, we will drop the input $\G$ in $\V$ and $\E$ if it is clear from the context that we are referring to a single graph $\G$.

\textbf{Matrix notation:}
We denote with $A\in\mathbb{R}^{|\V|\times |\V|}$ the adjacency matrix for a given graph, where $A_{ij} = c_{ij}$ $\forall e_{ij}\in \E$ 
and zero elsewhere.
Let $d_i$ be the degree of node $v_i\in\V$, 
$D\in\mathbb{R}^{|\V|\times |\V|}$ be the degree diagonal matrix $D_{ii} = d_i$, 
$L=D-A$ be the graph Laplacian, and 
$\mathcal{L} = D^{-1/2}LD^{-1/2}$ be the 
symmetric
normalized graph Laplacian. 
Note that the volume of a subset $S$ is $\vol(S) = \sum_{v_i\in S} d_{i}$.
We denote with $B\in\mathbb{R}^{|\E|\times |\V|}$ the incidence matrix of the given graph $\G$.  Every row of the incidence matrix corresponds to an edge $e_{ij}\in \E$ in $\G$.  Assuming arbitrary ordering of the edges of the graph, in this paper we define the rows of the incidence matrix as $B_{e_{ij}} = e_i - e_j$ $\forall e_{ij}\in \E$, where $e_i\in\mathbb{R}^{|\V|}$ is equal to one at the $i^{th}$ position and zero elsewhere.
Finally, $C\in\mathbb{R}^{|\E|\times|\E|}$
is a diagonal matrix of the weights of each edge, i.e., $C_{ij} = c_{ij}$ $\forall i,j$. In this notation, the Laplacian matrix $L = B^T C B$. 

\textbf{Norms:}
For all $x \in \mathbb{R}^{|\E|}$, we define the 
weighted $\ell_1$ and $\ell_2$ norms $\|x\|_{1,C}: = \sum_{e_{ij}\in E} c_{ij} |x_{ij}|$ 
and $\|x\|_{2,C}^2: = \sum_{e_{ij}\in E} c_{ij} |x_{ij}|^2$, respectively. Given a partition $S,S^c$ and a vector
$x\in\{0,1\}^{|\V|}$ such that $x_i=1$ if $v_i\in S$ and $x_i = 0$ if $v_i\in S^c$, then $\cut(S) = \|Bx\|_{1,C}$. Moreover, notice that 
$\cut(S) = \|Bx\|_{2,C}^2 = x^T B^T C B x = x^T L x$.

\textbf{Miscellaneous:}
We use $[x;y;z]$ to denote a single column vector where the individual vectors $x,y,z$ are stacked in this order.
Finally, the vectors $0_{|\V|}$ and $1_{|\V|}$ are the all zeros and ones vectors of length $|\V|$, respectively.

%\section{Conductance partitioning}

%In this section, we will present ubiquitous optimization problems and algorithms for graph partitioning problems. 
%The optimization problems that we will discuss are combinatorial, which generally are NP-hard, but in special
%cases can be solved in polynomial time, for instance the Min-Cut problem. 
%The algorithms that we will present are global which means that they use all nodes and edges of a given graph. Moreover, the algorithms can be exact, i.e., they solve 
%the optimization problem to some predefined accuracy, or the algorithms can be inexact, they provide solutions which have a constant approximation error.

\section{ Sparsest-Cut, Minimum-Conductance, and Spectral~Relaxations}\label{sec:global_cond_spcut}
In this section, we present two ubiquitous combinatorial optimization problems:
Sparsest-Cut and Minimum-Conductance.
These problems are NP-hard \cite{SM90,LR99}, but they can be relaxed to tractable convex optimization problems~\cite{ARV09}. 
We discuss one of the commonly used relaxation techniques which will motivate part of our discussion for local graph clustering algorithms.
Both Sparsest-Cut and Minimum-Conductance give different ways of balancing the size of a partition with its quality.

Sparsest-Cut finds the partition that minimizes the ratio of the fraction of edges that are removed divided by the scaled product of volumes of the two disjoint sets of nodes defined by 
removing those edges. In particular, if we have a partition $(S, S^c)$, where $S^c$ is defined in Section \ref{sec:prelim}, then $\cut(S)$ is the number of edges
that are removed, and the scaled product of volumes $\vol(S)\vol(S^c)/\vol(\V)$ is the volume of the disjoint sets of nodes. 
%In this case, \textcolor{blue}{$\cut(S)/(\textcolor{red}{\vol(edges\_cut)})$} and $2vol(S)vol(S^c)/\vol(\V)^2$
%are fractions of these, respectively. 
Putting all together in an optimization problem, we obtain
%\begin{align*}
%\mbox{minimize}             & \ \tfrac{\cut(S)}{\frac{1}{\vol(\V)}\vol(S)\vol(S^c)} \\
%	  		\mbox{subject to} &\ S\subset \V.
%\end{align*}
%For convenience, we will remove constants and work~with 
\begin{align}\label{spcut}
\spcut(\G):= \mbox{minimize}             & \ \spcut(S):= \frac{\cut(S)}{\frac{1}{\vol(\V)}\vol(S)\vol(S^c)} \\\nonumber
	  		\mbox{subject to} &\ S\subset \V.
\end{align}
%for which $\spcut(\G)\in[0,2|\E|/|\V|^2]$ (\textcolor{red}{is this correct?}).\ %This problem is known as the Sparsest-Cut problem. 
We shall use the term \emph{expansion of a set} to refer to the ratio $\spcut(S)$, and the term \emph{expansion of the graph} to refer to $\spcut(\G)$, in which case this problem is known as the Sparsest-Cut problem. 

Another way of balancing the partition is
 \begin{align}\label{isop}
\cond(\G):= \mbox{minimize}             & \ \cond(S):= \frac{\cut(S)}{\min(\vol(S),\vol(S^c))} \\\nonumber
	  		\mbox{subject to} &\ S\subset \V.
\end{align} 
In this case, we divide by the minimum of $\vol(S)$ and $\vol(S^c)$, rather than their product.
We shall use the term \emph{conductance of a set} to refer to the ratio of cut to volume $\cond(S)$, and the term \emph{conductance of the graph} to refer to $\cond(\G)$, in which case this problem is known as the Minimum-Conductance problem. 

The difference between the two problems \eqref{spcut} and \eqref{isop} is that the former regularizes based on the number of connections lost among pairs of nodes,
while the latter regularizes based on the size of the small side of the partition. 
%Moreover, the two problems differ by a factor since 
%$$
%\frac{vol(S)vol(S^c)}{n} \le \min(vol(S),vol(S^c))\le \frac{2vol(S)vol(S^c)}{n}.
%$$
Optimal solutions to these problems differ by a factor of $2$:
\begin{equation}\label{eq:equiv_cond_sparsecut} 
\frac{1}{2}\tilde{\phi}(S) \le \phi(S) \le \tilde{\phi}(S),
\end{equation}
leading to the two objectives $\tilde{\phi}(S)$ and $\phi(S)$ being almost substitutable from a theoretical computer science perspective~\cite{ARV09}.
However, this does not mean that the actual obtained solutions by solving \eqref{spcut} and \eqref{isop}
are similar; in general, they are not. 
%We computed solutions to the Minimum-Isoperimetry problem using Gurobi for the two graphs from Figure~\ref{fig:graphs}. 
%Solutions to the Minimum-Isoperimetry problem (computed by exactly solving the integer problem)

%We now review common relaxations of \eqref{spcut} and \eqref{isop}. 

There are three major relaxation techniques for the NP-hard problems \eqref{spcut} and \eqref{isop}: spectral relaxation; all pairs multi-commodity flow or linear programming (LP) relaxation; and
semidefinite programming (SDP) relaxation. For detailed descriptions about the LP and SDP relaxations we refer the reader to \cite{LR99} and \cite{ARV09}, respectively. 
We focus here on spectral relaxation since similar relaxations are widely used for the development 
of local clustering methods, which we discuss in subsequent sections. 
%To the best of our knowledge there are no local graph clustering 
%algorithms which are based on LP or the SDP relaxations.
%However, it is important to mention that in terms of worst-case guarantees for approximating the optimal value of the problems \eqref{spcut} and \eqref{isop}
%the best among these three is the SDP relaxation with an approximation guarantee $\mathcal{O}(\sqrt{\log |\V|})$. The LP relaxation has approximation error $\mathcal{O}({\log |\V|})$.

\subsection{Spectral Relaxation}\label{subsec:spcut}
%A high-level interpretation of state-of-the-art relaxations of the latter two combinatorial problems is given in Figure \ref{fig:relaxDiagram}.
%In particular, the main characteristics of the relaxations are given in Figure \ref{fig:relaxDiagram} and how the relaxation are related to each other. 
%More details are provided in the text below.

%%\notered{Do we cite Luca for this derivation somehwere?}

%\input{ConductanceRelaxGraph}
%For every relaxation we discuss several specialized algorithms that solve the relaxed problems efficiently. 
Spectral graph partitioning is one of the best known relaxations of the Sparsest-Cut \eqref{spcut} and Minimum-Conductance \eqref{isop}. The relaxation is the same for both problems, although the diversity of derivations of spectral partitioning does not always make this connection clear. 
For a partition $(S, S^c)$, let's associate a vector $x\in\{c_1,c_2\}^{|\V|}$ such that $x_i=c_1$ if $v_i\in S$ and $x_i=c_2$ if $v_i\in S^c$. (For simplicity, think of
$c_1 = 1$ and $c_2 = 0$, so $x$ is the set indicator vector.) 
The spectral clustering relaxation uses a continuous relaxation of the set indicator vector in problems \eqref{spcut} and \eqref{isop} to produce an eigenvector.
The relaxed problem is
%\footnote{Strictly speaking, this generalized eigenvector problem is a non-convex optimization but all local solutions are equally important and have the same properties.} 
\begin{align}\label{spcutLD}
\lambda_2  := \mbox{minimize}   & \ \frac{\|Bx\|_{2,C}^2}{2\|x\|_{2,D}^2} \\\nonumber
\mbox{subject to} &\  1_{|\V|}^T D x = 0\\\nonumber
			   &\ x\in\mathbb{R}^{|\V|} - \{0_{|\V|}\}.
\end{align}
(The denominator $\|x\|_{2,D}^2 = \sum_i |x_i|^2 d_i$.)
To see why \eqref{spcutLD} is a continuous relaxation of \eqref{spcut} we make two observations. 

First, notice that for all $x$ in $\mathbb{R}^{|\V|} - \{0_{|\V|}\}$ such that $1_{|\V|}^T D x = 0$ the denominator in \eqref{spcutLD} satisfies $2\vol(\V)\|x\|_{2,D}^2 = \sum_{i=1}^{|\V|}\sum_{j=1}^{|\V|} d_id_j|x_i-x_j|^2$.
Therefore, $\lambda_2$ in \eqref{spcutLD} is equivalent to 
\begin{align}\label{eq:1434}
\lambda_2  = \mbox{minimize}   & \ \frac{\|Bx\|_{2,C}^2}{\frac{1}{\vol(\V)}\sum_{i=1}^{|\V|}\sum_{j=1}^{|\V|} d_id_j|x_i-x_j|^2} \\\nonumber
\mbox{subject to} &\  1_{|\V|}^T D x = 0\\\nonumber
			   &\ x\in\mathbb{R}^{|\V|} - \{0_{|\V|}\}.
\end{align}
The optimal value of the right hand side in \eqref{eq:1434} is equivalent to the optimal value of right hand side in the following expression
\begin{align}\label{eq:1435}
\lambda_2  = \mbox{minimize}   & \ \frac{\|Bx\|_{2,C}^2}{\frac{1}{\vol(\V)}\sum_{i=1}^{|\V|}\sum_{j=1}^{|\V|} d_id_j|x_i-x_j|^2} \\\nonumber
\mbox{subject to} & \ x\in\mathbb{R}^{|\V|} - \{0_{|\V|},1_{|\V|}\}.
\end{align}
To prove this notice that the objective function of the right hand side in \eqref{eq:1435} is invariant to constant shifts of $x$, i.e., $x$ and $x + c 1_{|\V|}$ have the same 
objective function, where $c$ is a constant. Therefore, if $\tilde{x}$ is an optimal solution of the right hand side in \eqref{eq:1435} then $\hat{x} = \tilde{x} - \frac{1^T_{|\V|}D\tilde{x}}{\vol(\V)} 1_{|\V|}$
has the same optimal objective value and also $1_{|\V|}^TD \hat{x} = 0$.

Second, by restricting the solution in \eqref{eq:1435}
in $\{0,1\}^{|\V|}$ instead of $\mathbb{R}^{|\V|}$ we get that $\cut(S) = \|Bx\|_{2,C}^2$ and $\sum_{i=1}^{|\V|}\sum_{j=1}^{|\V|} d_id_j|x_i-x_j|^2 = \vol(S)\vol(S^c)$.

Using these two observations, it is easy to see that \eqref{spcutLD} is a continuous relaxation of \eqref{spcut}. 
Using \eqref{eq:equiv_cond_sparsecut}, it is easy to see that \eqref{spcutLD} is a relaxation for \eqref{isop} as well.

%Let $\lambda_2$ be the Rayleigh quotient at the optimal solution.\footnote{Here, $\lambda_2$ is also the second smallest eigenvalue of the normalized Laplacian matrix $\mathcal{L}$. }
The quality of approximation of relaxation \eqref{spcutLD} to Sparsest-Cut \eqref{spcut} is given by Cheeger's inequality \cite{LS88,SJ89} 
$${\lambda_2}/{\vol(\V)}\le \spcut(\G) \le {(8 \lambda_2)^{1/2}}/{\vol(\V)} $$
while the approximation guarantee for the Minimum-Conductance problem \eqref{isop} is 
$$
{\lambda_2}/{2}\le \cond(\G) \le (2 \lambda_2)^{1/2},
$$
which can be found in \cite{bM89}. (A generalization of these bounds holds for arbitrary vectors~\cite{Mihail}.) Both of these approximation ratios can be realized by rounding procedures described below. 

Another form of relaxation is the combinatorial model relaxation, which is formulation as problem \eqref{spcutLD} by ignoring the orthogonality constraint and restricting $x\in\{0,1\}^{|\V|}$
instead of $x\in\mathbb{R}^{|\V|}$. An extensive study of the spectral and the combinatorial mode relaxation can be found in \cite{Hoc13}, while empirical comparisons between these relaxations are discussed in \cite{HLB13}.

\subsection{Rounding}
\label{subsec:rounding}
In practice, the solution obtained by the spectral relaxation is unlikely to lie in $\{0,1\}^{|\V|}$, i.e., it is unlikely to be the indicator vector of a set. Therefore, it is necessary to have an efficient post-processing procedure 
where the solution is rounded to a set. 
%Similarly, one has to convert the semi-metric obtained by the LP relaxation to some partition $S$, $S^c$. 
At the same time 
it is important to guarantee that the rounded solution has good worst-case guarantees in terms of the conductance or sparsest cut objective.
%In this subsection we describe briefly rounding procedures for the spectral, LP and SDP which are theoretically guaranteed
%to produce partitions with low Sparsest-Cut and/or Minimum-Isoperimetry values. 
%In this subsection we emphasize on the rounding procedure for spectral relaxation since it is used
%for spectral local clustering algorithms. To the best of our knowledge there are no local graph clustering 
%algorithms which are based on LP or the SDP relaxations. Therefore, their rounding procedures are outside of the scope
%of this paper.
%We discuss briefly the rounding procedures for LP and SDP but
%to the best of our knowledge these procedures are not used currently by any local graph clustering algorithm.

%\paragraph{Rounding for spectral relaxation}
One of the most efficient and theoretically justified rounding procedures for spectral relaxation is the \emph{sweep cut}.
The sweep cut procedure is summarized in the following steps.
\begin{compactenum}\footnotesize
\item Input: the solution $x\in\mathbb{R}^{|\V|}$ of \eqref{spcutLD}.
\item Sort the indices of $x$ in decreasing order with respect to the values of the components of $x$. 
Let $i_1,i_2,\dots,i_{|\V|}$ be the sorted indices.
\item Using the sorted indices generate a collection of sets $\mathcal{S}_{j}:=\{i_1,i_2,\dots,i_j\}$ for each $j\in \{1,2,\dots,|\V|\}$.
\item Compute the conductance or sparsest-cut objective for each set $S_j$ 
and return the minimum. 
\end{compactenum} 
Notice that sweep cut can be used to obtain approximate solutions for both Sparsest-Cut and Minimum-Conductance. 
In fact, the proof for the upper inequalities of the approximation guarantees of spectral relaxation to
Sparsest-Cut and Minimum-Conductance are obtained by using the sweep cut procedure \cite{LS88,SJ89}. 

\section{Locally-biased graph partitioning methods}\label{sec:localgraphpart}

All of the algorithms described in Section~\ref{sec:global_cond_spcut} are ``global,'' in that they touch all of the nodes of the input graph at least once, and thus they have a running time that is at least linear in the size of the input graph. 
Informally, locally-biased graph clustering algorithms find clusters near a specified seed set of vertices, in many cases without even touching the entire input graph.
In this section, we will describe several local graph clustering algorithms, each of which has somewhat different properties.  

To understand the seemingly-quite-different algorithms we will discuss, we will distinguish local graph clustering algorithms based on three major features.
\begin{compactenum}
\item \textit{Weakly or strongly local algorithms}. Weakly local algorithms are those that are biased toward a local part of the graph but may ``touch'' the entire input graph during the computation---i.e., they formally have running times that scale with the size of the graph. Strongly-local graph algorithms are those that only access a small portion of the entire input graph in order to perform their computation---i.e., they formally have running times that are linear in the size of the output or input set of nodes but independent of the size of the input graph.
We show that the difference between weakly and strongly local algorithms often translates to whether we penalize the solution by adding an $\ell_1$ norm penalty implicitly
to the objective function and/or by restricting the feasible region of the problem by adding implicitly a locality constraint. Both ways result in strongly local algorithms. 
%The $\ell_1$ norm penalty guarantees that the output solution is sparse.
%Algorithms like Local FlowImprove and $\ell_1$-regularized Page-Rank exploit this property by working only within the support of the optimal solution.
\item \textit{Localizing bias.} Current local graph clustering algorithms are supervised, i.e., one has to give a reference set of seed nodes. 
We discuss two major ways that this information is incorporated in the problem. First, the bias to the input is incorporated in the objective function of the problem through a localizing construction we will call the \emph{reference cut graph}.
Second, the bias to the input is incorporated to the problem as a constraint.
\item \textit{$\ell_1$ and $\ell_2$ metrics.} The final major feature that distinguishes locally-biased algorithms is how the cut measure is treated. Via the cut metric~\cite{DezaLaurent97}, this can be viewed as embedding the vertices of $\G$ and evaluating their distance in the embedding. Two distances are explicitly or implicitly used in locally-biased algorithms:  the $\ell_1$ and the $\ell_2$ metric spaces. 
The former results in local flow algorithms, and the latter results in local spectral algorithms. The distinction between these two is very important 
in practice, as we show by empirical evaluation in subsequent sections.  
\end{compactenum}

\noindent
The local graph clustering algorithms that we consider in the following sections and their basic properties with respect to the above three features are given in Table \ref{table1}.
\begin{table}  
\centering
\begin{tabular}{lcccc}
\toprule
 \multicolumn{1}{c}{Methods} & Locality  & Bias & Metric \\ 
\midrule
 Flow-Improve \cite{AL08_SODA}                 & Weak & Objective   &  $\ell_1$ \\ 
 MOV \cite{MOV12_JMLR}                               & Weak & Constraint &  $\ell_2$ \\ 
 Local Flow-Improve \cite{OZ14}        & Strong & Objective  &  $\ell_1$ \\ 
 MQI \cite{kevin04mqi}                                & Strong & Constraint  & $\ell_1$  \\ 
 spectral MQI \cite{Chung07_localcutsLAA}                   & Strong & Constraint &  $\ell_2$ \\ 
 $\ell_1$-reg. Page-Rank \cite{ACL06,GM14_ICML} & Strong & Objective  & $\ell_2$ \\ 
\bottomrule
 \end{tabular}
 \caption{State-of-the-art local graph clustering methods and their properties with respect to the three features that are discussed in Section \ref{sec:localgraphpart}.}
\label{table1}
\end{table}

\subsection{A general localizing construction}

We describe these locally-biased graph methods in terms of an augmented graph we call the \emph{reference cut graph}. 
We should emphasize that this is a conceptual construction to highlight the similarities and differences between locally-biased algorithms in Table~\ref{table1}---in particular, these algorithms do \emph{not} explicitly construct the reference cut graph.

Let $h,g\in\mathbb{R}^{|\V|}$, $h,g\ge 0$, and $g-h \ge 0$, and let $\alpha$, $\beta$, and $\gamma$ be parameters specified below..  
Then the reference cut graph is constructed from a simple, undirected graph $\G$ as follows:
\begin{compactenum}\footnotesize
	\item Add a source and sink node $s$ and $t$.
	\item Add edges from $s$ to each node in $\V$.
	\item Add edges from each node in $\V$ to $t$.
	\item Weight each original edge by $\gamma$.
	\item Weight the edges from $s$ to $\V$ by $\alpha h$, where $\alpha \ge 0$.
	\item Weight the edges from $t$ to $\V$ by $\beta (g - h)$, $\beta \ge 0$.
\end{compactenum}
%(The change to the methods for finding expansion cuts often involves removing the degree weighting of the end-points.)
Let $H := diag(h)$, $G := diag(g)$ and $Z = G-H$.  Then we can also view the augmented graph through its incidence matrix and weight matrix:
\[ \tilde{B} = \begin{bmatrix} 1_{|\V|} & -I_{|\V|}               & 0  \\
			   		  0 & B            		  & 0 \\
					  0 & I_{|\V|}               & -1_{|\V|}
\end{bmatrix},
\
\tilde{C}= \begin{bmatrix} \alpha H & 0                  & 0  \\
			   		           0 & \gamma C   & 0 \\
					          0 & 0                   & \beta Z
\end{bmatrix}, 
\]
respectively.
%Based on the previous we can construct the Laplacian matrix $\tilde{L} = \tilde{B}^T \tilde{C} \tilde{B}$ of the augmented graph
%as follows
%\[
%\tilde{L} = \begin{bmatrix}    \alpha e^T h & -\alpha h                    & 0  \\
%			   		 -\alpha h       & \beta G + (\alpha - \beta)H+ \gamma L & -\beta(g - h) \\
%					  0                  & -\beta(g - h)             & \beta e^T(g -h)
%\end{bmatrix}.
%\]
The above construction might look overly complicated. However, we will see in the following subsections 
that it simplifies for local spectral and flow graph clustering algorithms with specific settings for $h,g$ and $\gamma$.

\subsection{Weakly-Local and Strongly-Local flow methods}
%(\textcolor{red}{Kimon: one problem that I can see here is that the reader might not be familiar with flow methods at all. Therefore,
%the stuff below would not make sense to them. How can we explain them better? Why is flow related to conductance? This stuff seem to be
%explained for each method separately, but somehow we need to mention them earlier or at least mention where we explain them})

Although finding the set of minimum conductance is NP-hard in general, there are a number of cases and variations that admit polynomial time algorithms and can be solved via Max-Flow/Min-Cut or a parametric Max-Flow method.  
These algorithms begin with a reference set $R$ of nodes, and they return a smaller (or not much larger) set of nodes that is a better partition in terms of the conductance ratio $\cond$. Typically, the returned value is \emph{optimal} for a variation on the Minimum-Conductance
and/or the Sparsest-Cut objective. 
The methods themselves are highly flexible and apply to other variations of Minimum-Conductance and Sparsest-Cut. 
For the sake of simplicity, we will describe them for conductance. 

\begin{figure}
 \subfigure[A simple graph]{\includegraphics[width=0.3\linewidth]{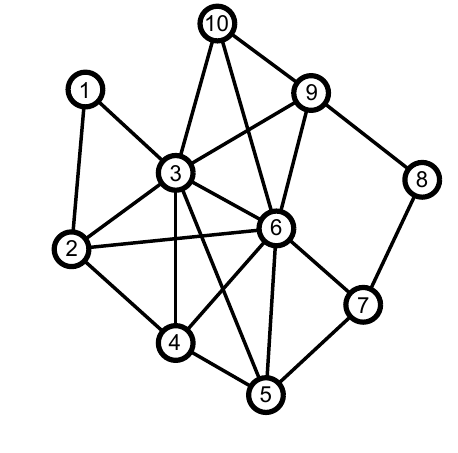}}
 \hfill
 \subfigure[Adding the source $s$ and sink $t$]{\includegraphics[width=0.3\linewidth]{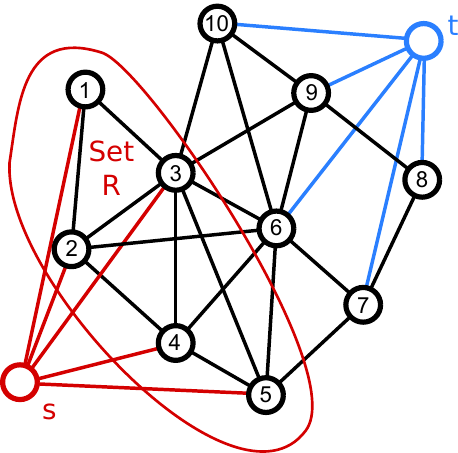}}
 \hfill
 \subfigure[The reference cut graph, with weights indicated]{\includegraphics[width=0.3\linewidth]{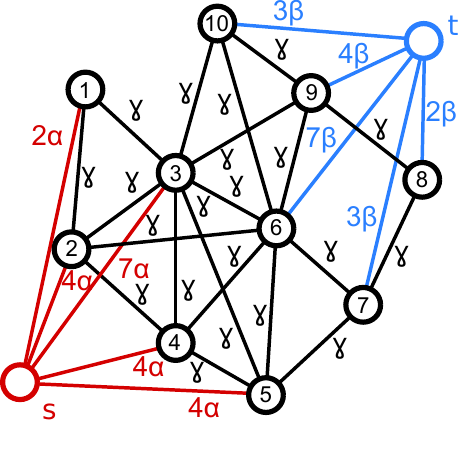}}
 \caption{The construction of the reference cut graph begins by adding a source node $s$ connected to the reference set $R$ and a sink node $t$ connected to the rest of the graph. Then we add weights to the network based on the degrees and three parameters $\alpha, \beta, $ and $\gamma$. Each edge to the source node is weighted by $\alpha \cdot \text{degree}$, each edge to the sink node is weighted by $\beta \cdot \text{degree}$, and each internal edge is weighted by $\gamma$. Note that one of the choices of $\alpha, \beta, $ or $\gamma$ will be 1, but various papers adopt different choices, and so we leave it general.}
 \label{fig:cut-graph}
\end{figure}

All of the following procedures adopt the following meta-algorithm starting with working set $W$ initialized to an input reference set of nodes $R$, values $\alpha_1, \beta_1, \gamma_1$
and vectors $h = d_R$, $g=d$,
where $d_R$ is a vector of length $|\V|$ with components equal to $d_i$'s for nodes $v_i\in R$ and zeros for nodes $v_i\in R^c$. 
Figure~\ref{fig:cut-graph} illustrates the construction of an augmented graph based on the previous setting of $\alpha,\beta,\gamma$
and $h,g$.

\noindent \textbf{\footnotesize The Local-Flow Meta-algorithm}
\begin{compactenum} \footnotesize
 \item Initialize $W_1 = R$, $k = 1$, $h=d_R$, $g=d$ and $\alpha_1, \beta_1, \gamma_1$ based on $R$.
 \item Create the reference cut graph $\tilde{B}_k, \tilde{C}_k$ based on $R$ and $\alpha_k, \beta_k, \gamma_k$.
 \item Solve the $s,t$-Min-Cut problem associated with $\tilde{B}_k,\tilde{C}_k$
 \item Set $W_{k+1}$ to be the $s$-side of the cut
 \item\label{step:cond-or-variant} Check if $W_{k+1}$ has smaller conductance (or some variant of it, as we will make specific in the text below) than before and stop if not and return $W_k$
 \item Update $\alpha_{k+1}, \beta_{k+1}, \gamma_{k+1}$ based on $W_{k+1}$ and $R$
 \item Set $k \to k+1$
 \item Repeat starting from state 2
 
\end{compactenum}

Next, we describe several procedures that are instantiations of this basic Local-Flow Meta-algorithm.

\paragraph{MQI} The first algorithm we consider is the MQI procedure due to Lang and Rao~\cite{kevin04mqi}. This method is designed to take the reference set $R$ with $\vol(R) \le \vol(G)/2$ and identify a subset of it $S \subseteq R$. The method instantiates the Local-Flow Meta-algorithm using $\alpha_k = \cut(W_k), \gamma_k = \vol(W_k), \beta_k = \infty$ and $h=d_R$, $g=d$.\ The idea with this method is that the reference cut graph will have an $s,t$-Min-Cut value strictly less than $\alpha_k \gamma_k$ if and only if there is a strict subset $S \subset W_k$ that has conductances less than $\alpha_k / \gamma_k$. (See~\cite{kevin04mqi} for the proof.) If there is such a set, then the result $W_{k+1}$ will be a set with conductance less than $\alpha_k / \gamma_k$. Since $\alpha_k$ and $\gamma_k$ are picked based on the current working set $W_k$, at each step the algorithm monotonically improves the conductance. Also, each step minimizes the objective
\begin{align*}
	\mbox{minimize}   & \;   \| \tilde{B}x \|_{1,\tilde{C}_k}  \label{eq:mqi-l1} \\
	\mbox{subject to} & \; x_i = 0 \ \forall v_i\in R^c, x_s = 1, x_t = 0.
\end{align*}

In fact, when the algorithm terminates, that means that there is no subset of $R$ with conductance less than $\alpha_k / \gamma_k$. Hence, we have solved the following variation on the conductance problem
\begin{align*}
\mbox{minimize}   & \;   \cond(S)= \frac{\cut(S)}{\vol(S)} \\
\mbox{subject to} & \;  S \subseteq R. 
\end{align*}
The key difference from the standard conductance problem is that we have restricted ourselves to a \emph{subset} of the reference set $R$. Procedurally, this is guaranteed because the edges connecting $R^c$ to $t$ have weight infinity, so they will never be cut. 
Thus, operationally, MQI is always a strongly local algorithm since it only operates within the input seed set $R$. Nodes connected to $t$ with weight infinity can be agglomerated or merged into a mega-sink node $T$. The resulting graph has the same size as $R$ along with the source and sink. (This is how the MQI construction is described in the original paper.) The MQI problem can be solved using Max-Flow method on the resulting graph a logarithmic number of times \cite{kevin04mqi}. Therefore, the running time 
for solving MQI depends on the Max-Flow algorithm that is used. Details about running times of Max-Flow algorithms can be found in~\cite{GT14}.

\begin{figure}
	\centering
\subfigure[MQI]{\includegraphics[width=0.30\linewidth,trim=80mm 33mm 00mm 57mm,clip]{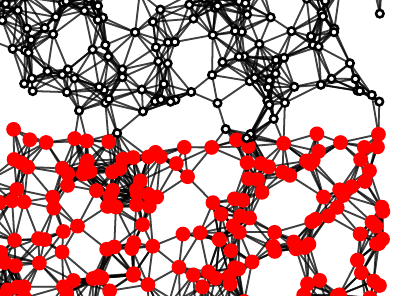}}
\subfigure[Flow-Improve]{\includegraphics[width=0.30\linewidth,trim=80mm 33mm 00mm 57mm,clip]{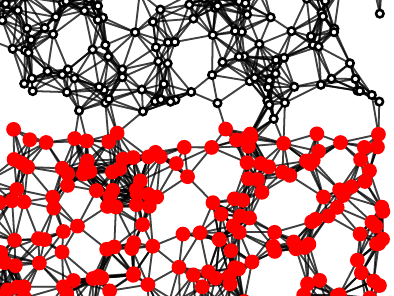}}
\subfigure[The difference]{\includegraphics[width=0.30\linewidth,trim=80mm 33mm 00mm 57mm,clip]{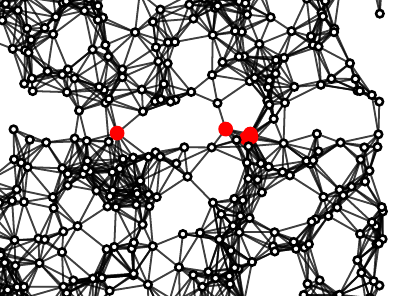}}
\caption{The results of running MQI and Flow-Improve on the reference set produced by a spectral partitioning method on the geometric graph (i.e., run global spectral \S\ref{subsec:spcut}; then round with a sweep-cut \S\ref{subsec:rounding}; and then refine using MQI and Flow-Improve). The Flow-Improve method identifies the optimal set from Figure~\ref{fig:small-cluster} in this case, whereas MQI cannot because it searches only within the given set $R$.}
\label{fig:mqi-vs-flow-improve}
\end{figure}

\paragraph{Flow-Improve} The Flow-Improve method due to Andersen and Lang~\cite{AL08_SODA} was inspired by MQI and designed to address the weakness that the algorithm will always find an output set within the reference set $R$, i.e., that is a subset of $R$. (As an illustration, see Figure~\ref{fig:mqi-vs-flow-improve}.) Again, Flow-Improve takes as input a reference set $R$ with volume less than half the graph. The idea behind Flow-Improve is that we want to find a set with conductance at least as good as $R$ and that also is highly correlated with $R$. To do this, consider the following variant of conductance
\[ \cond_R(S) = \frac{\cut(S)}{\vol(S \cap R) - \theta \vol(S \cap R^c)} \] 
where $\theta = \vol(R)/\vol(R^c)$,
and where the value is $\infty$ if the denominator is negative.
For any set $S$, $\cond_R(S) \ge \cond(S)$. Thus, this modified conductance score is an upper-bound on the true conductance. Again, we are able to show that the Local-Flow Meta-algorithm can solve for the exact value of $\cond_R(S)$ in polynomial time. To do so, instantiate that algorithm with $\alpha_k = \cond_R(W_k)$, $\beta_k = \theta \cond_R(W_k)$, $\gamma_k = 1$ and $h=d_R$, $g=d$.\ The value of a cut on set $S$ in the resulting graph is 
\[ \cut(S) + \alpha_k \vol(R) - \alpha_k [\vol(S \cap R) - \theta \vol(S \cap R^c)]. \]
(See~\cite{AL08} for the justification.) Andersen and Lang show that the algorithm monotonically reduces $\cond_R(W_{k})$ at each iteration as well. Each iteration now solves the $s,t$-Min-Cut problem
\begin{align}
	\mbox{minimize}   & \;   \| \tilde{B}x \|_{1,\tilde{C}_k}  \label{eq:FlowImprove} \\
	\mbox{subject to} & \;  x_s = 1, x_t = 0. \nonumber
\end{align}
In order to match precisely their Flow-Improve procedure, we would need to modify our Meta-algorithm to check the value of $\cond_R(W_k)$, instead of conductance (at Step~\ref{step:cond-or-variant} of the Local-Flow Meta-algorithm above), for monotonic decrease. 
%(In our experience, checking conductance will give the same output.) 
The authors also show that this procedure terminates in a finite number of iterations. 

At termination, the Flow-Improve algorithm has exactly solved $\mbox{minimize} \;   \cond_R(S), S \subseteq V$.
This can be considered a locally-biased variation of the conductance objective, where we penalize departure from the reference set $R$. Consequently, the solutions will tend to identify small conductance sets nearby $R$. 

Flow-Improve is a very useful algorithm, but it has two small weaknesses. The first is that it is a weakly local algorithm. At each step, we have to solve a Min-Cut problem that is the size of the original graph. The second is that the Min-Cut problems do not have integer weights. (Note that $\theta$ will almost never be an integer.) Most fast Max-Flow/Min-Cut procedures and implementations assume integer weights.  For instance, many implementations of the push-relabel method (hipr~\cite{goldberg95_push}) only allows integer weights. Boykov and Kolmogorov's solver is a notable exception~\cite{Boykov-2004-maxflow}. Similarly to MQI, the running time of solving the Max-Flow/Min-Cut problem depends on the particular solver that is used.
A summary of Max-Flow/Min-Cut methods can be found in \cite{GT14}.

\begin{figure}
	\subfigure[Reference set $R$.]{\includegraphics[trim=0 60mm 90mm 10mm, clip,width=0.24\linewidth]{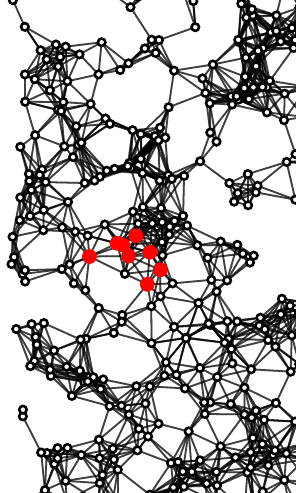}}
	\subfigure[The Flow-Improve result.]{\includegraphics[trim=0 60mm 90mm 10mm, clip,width=0.24\linewidth]{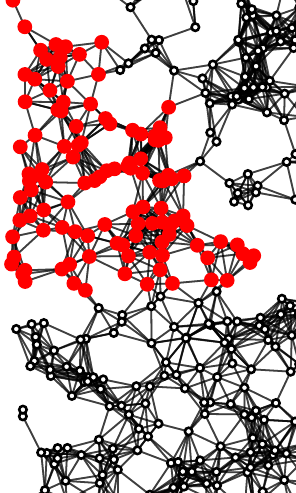}}
	\subfigure[Local-Flow-Improve $\kappa  \!= \! e^3$]{\includegraphics[trim=0 60mm 90mm 10mm, clip,width=0.24\linewidth]{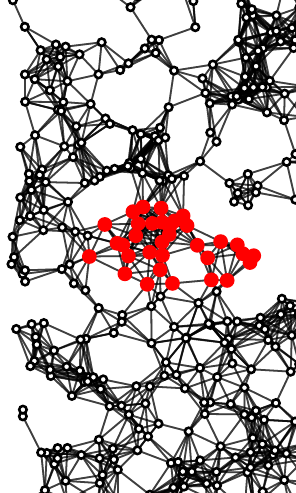}}
	\subfigure[Local-Flow-Improve $\kappa  \!= \! e^5$]{\includegraphics[trim=0 60mm 90mm 10mm, clip,width=0.24\linewidth]{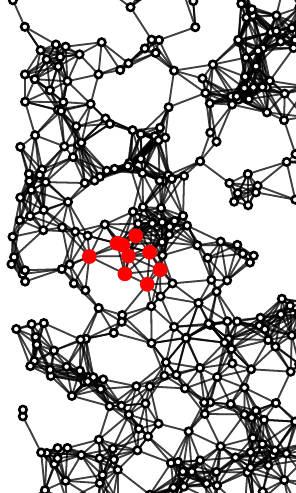}}
\caption{The results of running Flow-Improve compared with Local-Flow-Improve with a reference set $R$. The Flow-Improve result returns a fairly large set whereas the Local-Flow-Improve results produce successively smaller sets as the penalty $\kappa$ increases. When $\kappa = e^5$ then the result simply fills in the hole in the reference~set.}
\label{fig:local-flow-improve}	
\end{figure}

\paragraph{Local-Flow-Improve} The Local-Flow-Improve algorithm due to Orecchia and Zhu~\cite{OZ14} sought to address the weak-locality of the Flow-Improve method and create a strongly local flow based method. This involved two key innovations: a modification to the construction and objective that enables strong locality; and an algorithm to realize that strong locality. This Local-Flow-Improve method essentially interpolates between MQI and Flow-Improve. In one limit, it is strictly local to the reference graph and exactly reproduces the MQI output. In the other limit, it is exactly Flow-Improve.  To do this, Orecchia and Zhu alter the objective function used for Flow-Improve to place an additional penalty on deviating from the set $R$. They describe this as increasing the weight of connections $\beta_k$ in the reference cut graph by scaling these by a value $\kappa \ge 1$. If $\kappa = 1$, then their construction is exactly that of Flow-Improve. If $\kappa = \infty$, then this construction is equivalent to that of MQI. The effect of $\kappa$ is illustrated in Figure~\ref{fig:local-flow-improve}.

In terms of the optimization framework, their modification corresponds to using 
\[ \cond_R'(S;\kappa) = \frac{\cut(S)}{\vol(S \cap R) - \theta \kappa \vol(S \cap R^c)} \ \]
where $\kappa \ge 1$ and $\theta = \vol(R)/\vol(R^c)$ as in Flow-Improve, and again the value is $\infty$ if the denominator is negative. This result corresponds to instantiating the Local-Flow Meta-algorithm using $\alpha_k = \cond_R'(W;\kappa)$, $\beta_k = \cond_R'(W;\kappa) \theta \kappa$ and $h=d_R$, $g=d$. 

The second innovation is that they describe an algorithm to solve the Min-Cut problem on the reference cut graph that does not need to explore the entire graph. This second piece used a novel modification of Dinic's procedure~\cite{D70} to compute a Max-Flow/Min-Cut that exploited the locality. We refer interested readers back to Orecchia and Zhu for details of this second somewhat complicated construction. In our recent work~\cite{Veldt-preprint-simple-local-flow}, however, we describe a simplified framework for the Local-Flow-Improve method that shows that the strong locality in their modification results from implicitly regularizing the Flow-Improve objective with a $\ell_1$ norm regularizer.
(This will mirror strongly local spectral results in the forthcoming spectral section.) In fact, our recent work~\cite{Veldt-preprint-simple-local-flow} shows that each iteration exactly solves 
\begin{align}
	\mbox{minimize}   & \;   \| \tilde{B} x\|_{1,\tilde{C}_k'} + \varepsilon \| D x \|_1   \label{eq:FlowImprove-l1} \\
	\mbox{subject to} & \;  x_s = 1, x_t = 0, \nonumber
\end{align}
where $\tilde{C}_k'$ is a small perturbation on the above definition and $\varepsilon$ is a locality parameter.
The volume of the output cluster $S$ of the method in \cite{Veldt-preprint-simple-local-flow} is bounded $\vol(S) \le \vol(R)(1+2/\epsilon) + E(R,R^c)$, where 
$\epsilon:= \vol(R)/\vol(R^c) + \delta$ and $\delta\ge 0$ is a constant.

That work also describes a simple procedure to realize the strong locality that leverages any Max-Flow/Min-Cut solver on a sequence of sub-problems whose size is bounded independent of the graph. 

\subsection{Weakly-and-Strongly local spectral methods}

There are spectral analogues for each of the three flow-based constructions on the augmented graph. 
The development of these ideas occurred in parallel, largely independently, and it was not obvious until recently that the ideas were very related. 
Here, we make these connections explicit. 
Of the three flow constructions, the simplest is the MQI objective. We begin with it. 

\paragraph{SpectralMQI} The method we call SpectralMQI was proposed as the Dirichlet partitioning problem in~\cite{Chung07_localcutsLAA}. Given a graph $\G$ and a subset of vertices, consider finding the set $S$ of minimal local conductance 
$\lcond(S) = \cut(S)/\vol(S)$
such that $S \subseteq R$, where, again, $R$ is a reference set specified in the input. 
Note that the only difference from conductance is that we don't have the minimum in the denominator. 
A spectral algorithm to find an approximate minimizer of this is to solve the generalized eigenvector problem
\begin{align*}
\lambda_R  = \mbox{minimize}   & \quad \frac{\|Bx\|_{2,C}^2}{\|x\|_{2,D}^2} \\
\mbox{subject to} &\quad  x_{i} = 0 \ \forall v_i \in R^c.
\end{align*}
The solution vector $x$ and value $\lambda_R$ are related to the smallest eigenvalue of the sub-matrix of the normalized Laplacian corresponding to the nodes in $R$. (Note that we take the sub-matrix of the normalized Laplacian, rather than the normalized Laplacian on the sub-graph induced by $R$.) A sweep-cut over the eigenvector $x$ produces a set $S$ that satisfies a Cheeger inequality with respect to the best possible solution~\cite{Chung07_localcutsLAA}. 
This definition of local conductance is also called $\texttt{NCut'}$ by Hochbaum~\cite{Hochbaum-2010-ncutp}, who gave a polynomial time algorithm to compute it that is closely related to the MQI procedure. In this case, if $R$ has volume less than half the graph, then this is exactly the spectral analogue of the MQI procedure and the result is a Cheeger-like bound on the overall conductance. 

\paragraph{MOV} 
The Mahoney-Orecchia-Vishnoi (MOV) objective~\cite{MOV12_JMLR} is a spectral analogue of the FlowImprove method, with a few subtle differences. The goal is to find a small Rayleigh quotient, as in \eqref{spcutLD}, that is highly correlated with an input vector $z\in\mathbb{R}^{|V|}$, where $z$ represents the seed or local-bias. Given this, the MOV objective is
\begin{align*}
\mbox{minimize}   & \ \frac{\|Bx\|_{2,C}^2}{\|x\|_{2,D}^2} \\\nonumber
\mbox{subject to} &\  1_{|\V|}^T D x = 0 \\\nonumber
			   &\  (z^T D x)^2 \ge \kappa
			   &\ x\in\mathbb{R}^{\V}.
\end{align*}
The solution of this problem represents an embedding of the nodes in $\V$ which is locally biased, i.e., large values for components/nodes that are considered important and small or zero values for the rest. 
%The MOV solution is based on the result of taking the spectral minorant of this $s,t$-MinCut problem~\cite{GM14_ICML,GM15_KDD}. 
%It is proved in \cite{GM14_ICML,GM15_KDD} 

According to \cite{MOV12_JMLR}, there is a constant $\rho$, i.e., the optimal dual variable for the locally-biased constraint, such that the solution to the MOV problem satisfies
$ (L+\rho D) x = \rho D z$.\ The null space of $L$ is the vector $1_{|\V|}$, and assuming that $1_{|\V|}Dz = 0$, then the solution to the previous system is unique.
%In order to map to Rayleigh quotients in \eqref{spcutLD}, the MOV translates this system into an equivalent problem 
%$ (L + \rho D) y = D s$ where $1_{|\V|}^T D s = 0$.\footnote{(\textcolor{red}{I need to correct this}) In this case, $s =z- \delta 1_{|\V|}$ with $\delta = \vol(\V)/\vol(R)$.}  (This is possible, and equivalent up to a shift $x = y + \delta 1_{\V}$, because the null-space of $L$ is known.) 
One final detail is that the MOV construction fixes $\| x \|_2 = 1$. Consequently, the MOV solution is
\begin{equation} \label{eq:mov}
 x = c (L + \rho D)^\dagger D z \quad c = (\| (L + \rho D)^\dagger D z \|_2)^{-1}.
\end{equation}
In the MOV paper~\cite{MOV12_JMLR}, they show that $\rho$ can be chosen such that $x^T D z = \kappa$, the desired correlation strength with the input vector $z$, through a simple bisection procedure. 
Solving the linear system \eqref{eq:mov} results in a weakly-local method that satisfies another Cheeger-like inequality.  Recent extensions show that it is possible to get multiple locally-biased vectors that are akin to eigenvectors from this setup~\cite{HM14_JRNL,LBM16_TR}.
The methodology is able to leverage the large number of Laplacian system solvers \cite{V12} that can find an approximate solution to \eqref{eq:mov} in nearly linear time.

The pseudo-inverse allows us to ``pass through'' $\rho = 0$ and approach $\rho = -\lambda_2$. (This system is singular at $\rho = 0$ and $\rho = \lambda_2$.) What is interesting is that taking the limit $\rho \to -\lambda_2$ directly maps to the spectral relaxation~\eqref{spcutLD}. Thus, the $\rho$ parameter interpolates between the global spectral relaxation~\eqref{spcutLD} and a spectral-version of the Min-Cut problem in each step of FlowImprove. 

Based on the reference cut graph, the MOV objective
is $\mbox{minimize} \ \|\tilde{B}\tilde{x}\|_{2,\tilde{C}}^2$, where $\tilde{x} : = [1; x; 0]$.  
The reference graph cut setting is $\gamma=1$, $g=d$, $h=Dz$ and $\alpha=\beta=\rho\ge0$ controls the desired strength of the correlation to the input vector $z$.\ 
Notice that the MOV problem is a spectral, i.e., $\ell_2$, version of the $s,t$-Min-Cut problem. This observation was made first in \cite{GM14_ICML,GM15_KDD}. 
If $\rho$ is extremely large, the solution to the above problem would have perfect correlation with the input vector $h$. 
As $\rho \to 0$, we decrease the effective correlation with the input vector $h$. (These arguments are formal in~\cite{MOV12_JMLR}.)

%\subsection{Strongly-local flow and spectral methods}
%
%The weakly-local algorithms just described have the advantage that they correspond to a well-defined objective and they provide a method to ``engineer in'' local information.
%(Although this has been in the form of seed sets of nodes for local clusters, one can easily imagine extensions of this to tread that as a prior, etc.)
%It nevertheless remains a challenge to work with very large graphs, e.g., those consisting of millions up to billions of nodes, basically since even touching the entire graph in these cases is prohibitive.
%(Actually, touching it isn't, since graphs of that size can fit in memory, but if one runs the usual data analysis and machine learning analytics pipelines, where one wants to explore many parts of the graph, look at cross validation experiments, etc., then it quickly becomes prohibitive.)
%
%Let's start with strongly local-spectral methods.  
%These were introduced by \cite{ST08b_TR}, then \cite{ACL06}, then other \cite{} (\textcolor{red}{give more citations}).
%Here, we will focus in former two citations. 

\paragraph{$\ell_1$-Regularized Page-Rank}
The $\ell_1$-regularized Page-Rank problem was initially studied in \cite{GM14_ICML} and then further refined in \cite{FCSRM16_TR}.
In the latter work, the problem is defined as
\begin{align}\label{eq:l1pr}
\mbox{minimize} & \quad \frac{1}{2}\|\tilde{B}\tilde{x}\|_{\tilde{C},2}^2 + \epsilon \|Dx\|_1,
\end{align}
where $\tilde{x} : = [1; x; 0]$.\ The reference cut graph setting for \eqref{eq:l1pr} is $g=d$ and $h\ge 0$ is a vector that satisfies $\|h\|_1=1$ and $\|h\|_\infty \ge \epsilon$.\ 
The latter condition is to guarantee that the solution to \eqref{eq:l1pr} is not the zero vector. Moreover, $\alpha = \beta $ and $\gamma = (1-\alpha)/2$. 
Similarly to $z$ for MOV, the vector $h$ controls the input seed set and the weights of nodes in that set. The larger the weights the more the solution will be correlated 
with the corresponding nodes in the input seed set.
The solution vector to problem \eqref{eq:l1pr} is component-wise non-negative and the parameter $\alpha$ controls how much energy is concentrated close to
the input seed set. Formally, based on theoretical guarantees in \cite{ACL06} the vector $h$ should be an indicator vector for a single seed node, around which there is a 
target cluster of nodes $C$. The algorithm is not guaranteed to find the exact target cluster $C$, but if $C$ has conductance less than $\alpha/10$ then it is guaranteed to return a cluster 
with conductance of $\mathcal{O}(\sqrt{\alpha \log(\vol(C))})$. We refer the reader to \cite{ACL06} for a detailed description of the theoretical graph clustering guarantees.

The idea of $\ell_1$-regularized Page-Rank graph clustering initially appeared in \cite{ACL06} in the form of implicit regularization. In particular, the authors in \cite{ACL06}
suggest solving a personalized Page-Rank linear system approximately. In \cite{FCSRM16_TR,GM14_ICML}, the authors noticed that the termination criteria 
in \cite{ACL06} are related to the first-order optimality conditions of the above $\ell_1$-regularized Page-Rank problem, and they draw the connection 
to explicit $\ell_1$ regularization.
It is shown in \cite{FCSRM16_TR} that solving the $\ell_1$-regularized Page-Rank problem has the same Cheeger-like worst-case approximation guarantees to the Minimum-Conductance 
problem as the original procedure in \cite{ACL06}. 
However, there is an important technical difference: one advantage of solving the $\ell_1$-regularized problem is that the locality of the solution is a property of the optimization
problem as opposed to a property of an algorithm. In particular, by solving the $\ell_1$-regularized problem it is guaranteed to obtain the same solution 
regardless of the algorithm used.  In comparison, applying the procedure in \cite{ACL06}, where the output depends on the setting of the procedure, i.e., the strategy for choosing 
nodes to be updated at every iteration, leads to somewhat different solutions, depending on the specific settings chosen.

Let $x^*$ be the optimal solution of \eqref{eq:l1pr} and $S^*$ be the set of nodes where $x^*$ is non-zero.
In \cite{FCSRM16_TR}, it is shown that many standard optimization algorithms such as iterative soft-thresholding or block iterative soft-thresholding solve 
\eqref{eq:l1pr} with running time $\mathcal{O}(\vol(S^*)/\alpha)$, which can be independent of the volume of the whole graph $\vol(\V)$. This opens up the possibility of the use of these algorithms more generally. For details about the 
algorithms, we refer the reader to~\cite{FCSRM16_TR}.

\usetikzlibrary{arrows,shapes,backgrounds}

\tikzstyle{every picture}+=[remember picture]
\tikzstyle{na} = [baseline=-.5ex]

\section{Empirical Evaluation}
\label{sec:empirical}
%We mentioned in Section \ref{sec:localgraphpart} that all local graph clustering algorithms have three major 
%factors: bias towards the input seed set, weak vs.~strong locality, and spectral vs.~flow computations (which is equivalently  $\ell_2$ vs. $\ell_1$
%metrics for node distances). 
%The first factor is related to how local-bias is formulated as an the optimization problem. 
%The second factor is fundamental to locally-biased algorithms and either would come as the result of some apriori domain knowledge or the result of another algorithmic process. (For instance, the bias in Figure~\ref{fig:mqi-vs-flow-improve} was the result of the global spectral partition, which we sought to improve.) Or by imposing a sparsity inducing regularization term, i.e., $\ell_1$-norm, in the optimization problem, for instance see algorithms Local-Flow-Improve and $\ell_1$-regularized Page-Rank.

In this section, we illustrate differences among global, weakly local, and strongly local solutions to the problems discussed in Section~\ref{sec:localgraphpart}. Additionally, we discuss differences between spectral and flow methods (which is equivalently  $\ell_2$ vs. $\ell_1$
metrics for node distances). 
%The first factor is fundamental to locally-biased algorithms and either would come as the result of some apriori domain knowledge or the result of another algorithmic process.

To do so, we make use of the following real-world undirected and unweighted networks.
\begin{compactitem}
\item \textbf{US-Senate}. Each node in this network is a Senator that served in a single term (two years) of Congress. Our data cover the period from year $1789$ to $2008$. 
Senators appear as multiple nodes if they served in multiple terms. Edges are based on the similarity of voting records between Senators and thresholded at the maximum similarity such that the graph remains connected. Edge-weights are discarded. For a detailed discussion of this data-set we refer the reader to \cite{multiplex_Mucha}.
This graph has $8974$ nodes and $153804$ edges. This graph has two large clusters with small conductance ratio, i.e., downward-slopping network community profile; see Figure $6$ in \cite{Jeub15} for details.
The first cluster consists of all the nodes before the year $1913$ and the second cluster consists of nodes after that year. 
%%% MM: LETS AVOID HISTORY.
%%%The two main clusters represent in history the period before and after the $17th$ Amendment to the United States Constitution.
\item \textbf{CA-GrQc}. The data for this graph is a general relativity and quantum cosmology collaboration network. Details can be found in the Stanford Network Analysis Project.\footnote{\url{http://snap.stanford.edu/data}}
This graph has $4158$ nodes and $13422$ edges. This graph has many clusters of small size with small conductance ratio, while large clusters have large conductance ratio, 
i.e., upward-slopping network community profile; see Figure $6$ in \cite{Jeub15} for details.
\item \textbf{FB-Johns55}. 
This graph is a Facebook anonymized data-set on a particular day in September $2005$ for a student social network at John Hopkins university. The graph is unweighted and it represents ``friendship" ties.
The data form a subset of the Facebook100 data-set from \cite{TKMP11,TMP12}.
This graph has $5157$ nodes and $186572$ edges. This is an expander-like graph, all small and large clusters have about the same conductance ratio, i.e., flat network community profile; 
see Figure $6$ in \cite{Jeub15} for details.
\item \textbf{US-Roads}. The data for this graph is from the National Highway Planning Network \cite{LLDM08_communities_CONF}. Each node in this network is an intersection between two highways and the edges represent segments of the highways themselves.
\end{compactitem}
%The results from running these locally-biased algorithmsare shown in Figures \ref{fig:roads}, \ref{fig:senate} and \ref{fig:cagqrc}, for the US-ROADS, US-SENATE
%and CA-GRQC, respectively. The figures are separated in three layers, which are distinguished with horizontal black lines. 
%Algorithms which are in the same layer have the same locality property, either weakly- or strongly-local, and they 
%are spectral and flow analogues as were discussed in Section \ref{sec:localgraphpart}. 
%For all figures the first layer is about weakly-local algorithms, while the second and third layers are about strongly-local algorithms.
%In particular, the first layer shows the 
%solutions of Flow-Improve and MOV, the second layer shows the performance of Local-Flow-Improve and $\ell_1$-regularized Page-Rank
%and the third layer shows the performance of MQI and SpectralMQI.
Note that the small-scale vs. large-scale clustering properties of the first three networks have been characterized previously~\cite{Jeub15}.
In addition, it is known that US-Roads has a downward-sloping network community profile.

\paragraph{Global, weakly local, and strongly local solutions}
We first demonstrate differences among global, weakly local, and strongly local algorithms.
%In terms of optimization we saw in Section \ref{sec:localgraphpart} that weak locality manifests itself 
%by penalizing in some manner the differences between the solution and the input seed set in the objective function or by
%introducing a feasibility constraint which restricts the feasible region in some sense close to the input seed set.
%Strong locality manifests itself by taking a weakly local algorithm and adding an $\ell_1$ penalization term on the solution in the objective
%or by explicitly restricting the feasible region within the input set of nodes.
%This penalty or constraint guarantees that the solution is sparse, which is a property that the algorithms exploit in order
%to obtain running time which does not depend on the whole graph.
Let us start with a comparison among spectral algorithms. By comparing algorithms that 
use that same metric, i.e., $\ell_2$, to measure distances among nodes we minimize factors 
that can affect the solution, and we focus on weak vs. strong locality. In all figures we show the solution obtained by an algorithm 
without applying any rounding procedure. We illustrate the importance of the nodes by colouring and size; details 
are explained in the captions of the figures and in the text.
The layout for all graphs has been obtained using the force-directed algorithm \cite{H05}, which is available from the graph-tool project.\footnote{\url{https://graph-tool.skewed.de}}

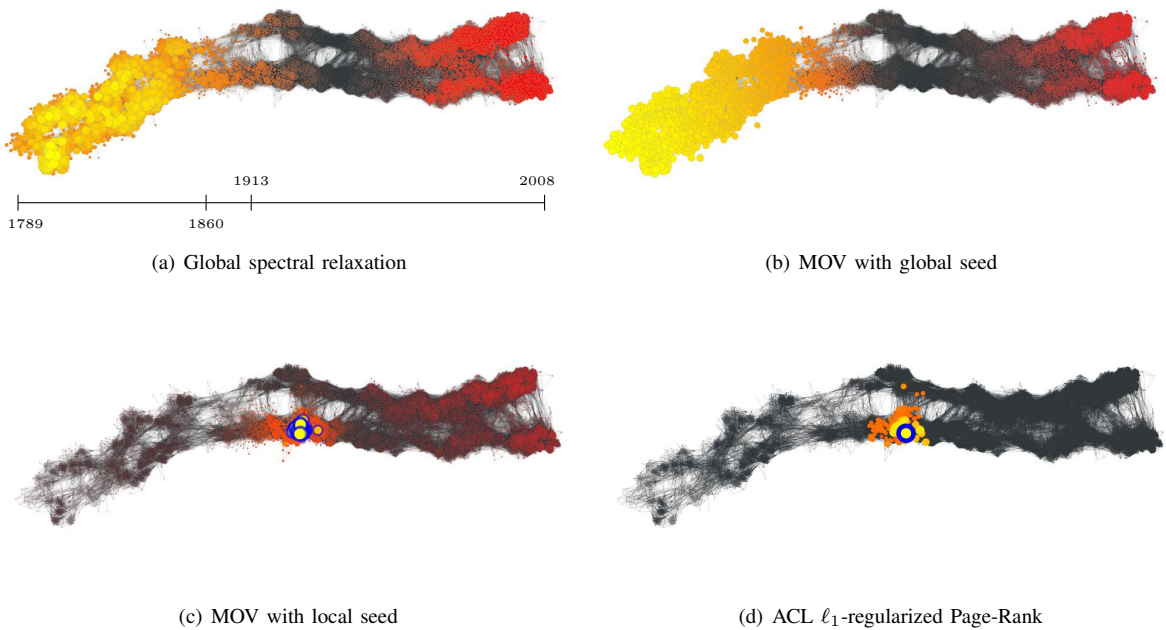
\begin{figure*}
\centering
	\subfigure[Global spectral relaxation]{\label{Fig_senate_spectral_1}\includegraphics[scale=0.22]{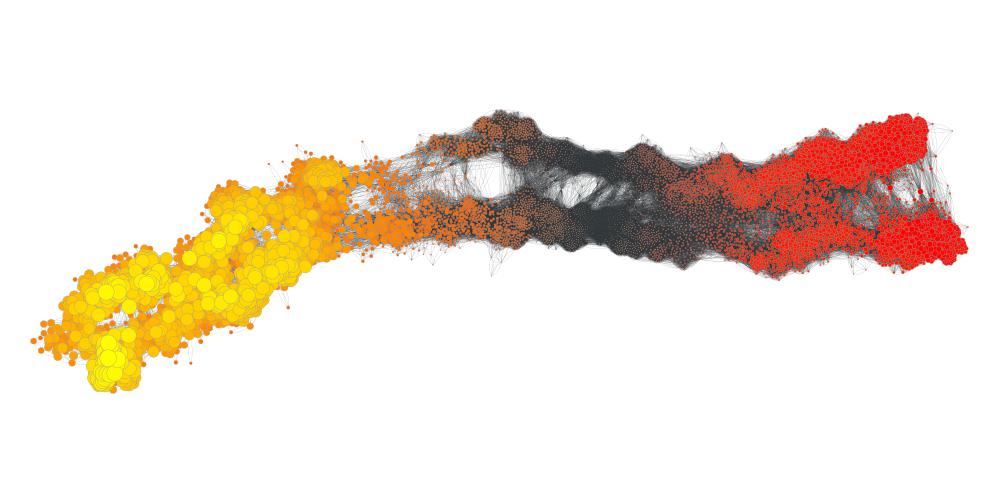} 
	\put(-220,0){ 
	\begin{tikzpicture}
	\draw (0,0) -- (7.0,0);
	\foreach \x in {0,2.5,3.1,7.0}
	\draw (\x cm,3pt) -- (\x cm,-3pt);
	\draw (0.1,0) node[below=3pt] (a) {\tiny $1789$} node[above=3pt] {};
	\draw (2.5,0)  node[below=3pt]  {\tiny $1860$} node[above=3pt] (c) {};
	\draw (3.1,0) node[below=3pt](d) {} node[above=3pt] {\tiny $1913$};
	\draw (6.9,0) node[above=3pt] (f) {\tiny $2008$} node[below=3pt] {};
	\end{tikzpicture}
	}
	}
	\subfigure[MOV with global seed]{\label{Fig_senate_spectral_2}
%	\begin{tikzpicture}
%	\put(-1,0){\draw[thick] (2.1,1.85) ellipse (0.42cm and 0.12cm);}
%	 \put(55,45){\tikz[na] \coordinate (s-cluster);}
%	\node [inner sep=0pt,above right] {
	\includegraphics[scale=0.22]{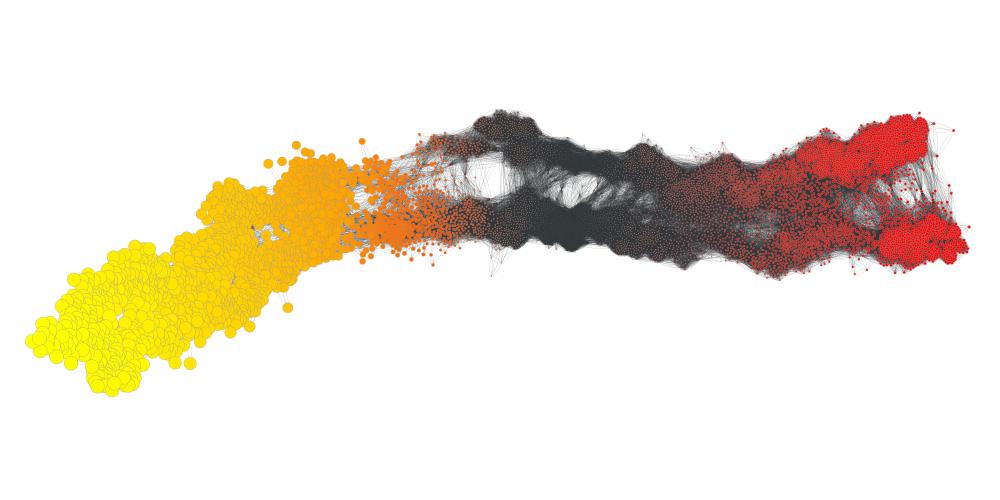}
%	};
%	\end{tikzpicture}
	}\\
	\subfigure[MOV with local seed]{\label{Fig_senate_spectral_3}
%	\begin{tikzpicture}
%	\put(60,60){ \tikz[na] \coordinate (s-clusterE);}
%	\node [inner sep=0pt,above right] {
	\includegraphics[scale=0.22]{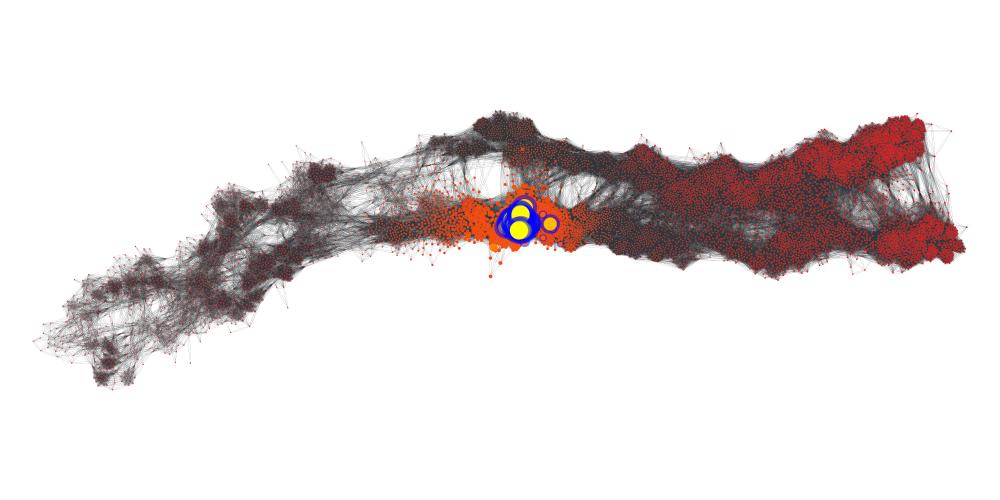}
%	};
%	\end{tikzpicture}
	}
	\subfigure[ACL $\ell_1$-regularized Page-Rank]{\label{Fig_senate_spectral_4}\includegraphics[scale=0.22]{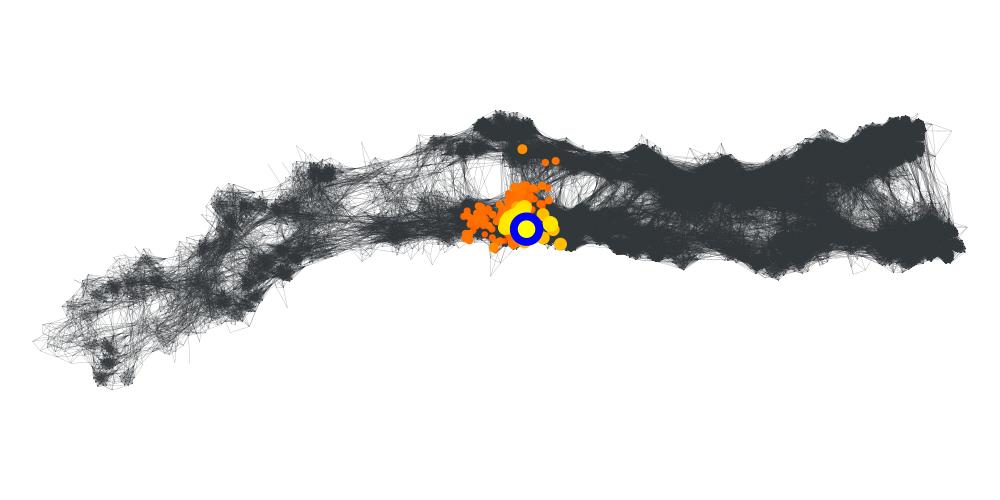}}
	\caption{US-Senate.\ This figure shows the solutions of (a) Spectral relaxation, (b) MOV with global bias, (c) MOV with local bias
	and (d) strongly-local $\ell_1$-regularized Page-Rank.
	We use a heat map to represent the weights of the nodes.
	For Spectral relaxation and MOV bright yellow means large positive and bright red means large negative. For $\ell_1$-regularized Page-Rank bright yellow means large positive and bright red means small positive.
	The blue halo around a node in Figures \ref{Fig_senate_spectral_3} and \ref{Fig_senate_spectral_4} means that this node is included in the seed set.
	The size of the nodes shows the weights of the solution in absolute value. 
	If a weight is nearly zero then the corresponding node is barely~visible. 
	%The black circle in Figure \ref{Fig_senate_spectral_2} shows a small cluster found by the MOV global solution. The arrow from Figure \ref{Fig_senate_spectral_2} to Figure \ref{Fig_senate_spectral_3}
	%shows that this small cluster can be found by MOV local as well as with $\ell_1$-regularized Page-Rank in Figure \ref{Fig_senate_spectral_4}.
	}	
	\label{Fig_senate_spectral}	
%\begin{tikzpicture}[overlay]
%        \path[->,black] (s-cluster) edge [bend right=10] (s-clusterE);
%\end{tikzpicture}
\end{figure*}

For US-Senate, the comparison is shown in Figure \ref{Fig_senate_spectral}.
Figures \ref{Fig_senate_spectral_1} and \ref{Fig_senate_spectral_2} show the solutions of global algorithms, Spectral relaxation 
and MOV global ($z=1_{|\V|}$ and then we orthogonalize $z$ with respect to $D1_{|\V|}$), respectively. 
As expected, the US-Senate graph has two large clusters, i.e., before the year $1913$ and after that year, that partition along the one-dimensional time axis. 
This global structure is nicely captured by Spectral relaxation and MOV global in Figures \ref{Fig_senate_spectral_1} and \ref{Fig_senate_spectral_2}, respectively.
%Moreover, notice that MOV global finds a small cluster which in Figure \ref{Fig_senate_spectral_2} is shown with a black ellipse around it. 

Given an input seed set, Figures \ref{Fig_senate_spectral_3} and \ref{Fig_senate_spectral_4} illustrate the weakly 
and strongly local solutions by MOV and $\ell_1$-regularized Page-Rank, respectively. 
For MOV in Figures \ref{Fig_senate_spectral_3} we set 
$z_i = 1$ for all $i$ in the input seed set and $z_i=0$ for all $i$ outside the input seed set. Then we orthogonalize $z$ with respect to $D\cdot1_{|\V|}$.
For $\ell_1$-regularized Page-Rank, we only give a single node as an input seed set, i.e., $h_i=1$ where $i$ is the input node and $h_i=0$ for all other nodes.
Moreover, we set the locality parameter $\epsilon$ large enough such that the solution is very sparse, i.e., strongly local.
In Figures \ref{Fig_senate_spectral_3} and \ref{Fig_senate_spectral_4}, we demonstrate the input seed sets by nodes with a blue halo around them. 
In Figure \ref{Fig_senate_spectral_3}, the cluster which is found by MOV consists of the nodes which have large mass concentration around the input seed set, i.e.,
the nodes around the input seed set that have large size and are coloured with a bright red shade.
MOV recovers this cluster by examining the whole graph; each node has a weight assigned to it in Figure \ref{Fig_senate_spectral_3}. 
On the other hand, a similar cluster is found in Figure \ref{Fig_senate_spectral_4} by using $\ell_1$-regularized Page-Rank without examining 
the whole graph. This is possible because nodes of the graph have zero weight assigned and need not be considered. \emph{This speed and data advantage, along with the sparsity-based implicit regularization~\cite{GM14_ICML}, are some of the reasons that strongly-local algorithms, such as $\ell_1$-PageRank, are used so often in practice~\cite{LLDM09_communities_IM,Jeub15}.}

\begin{figure}
\centering
	\subfigure[Spectral relaxation]{\label{Fig_grqc_spectral_1}\includegraphics[scale=0.111]{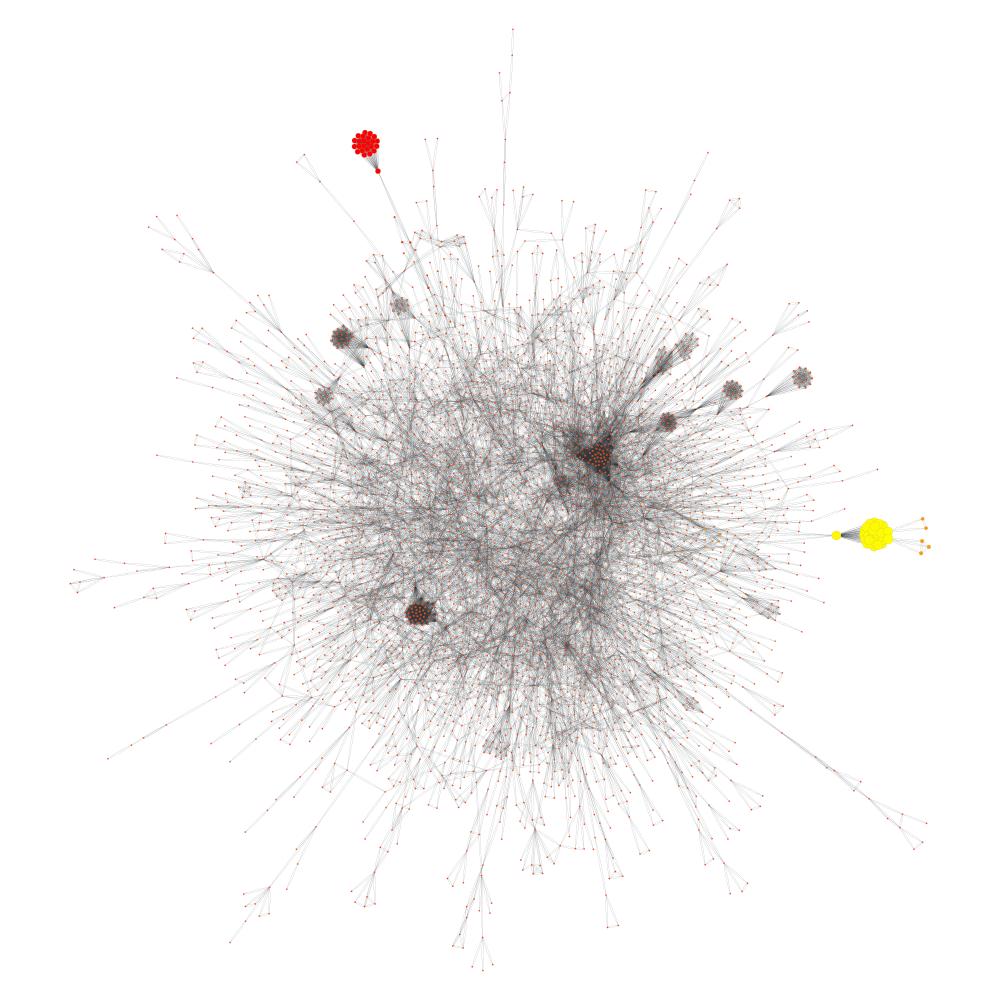}}
	\subfigure[MOV global]{\label{Fig_grqc_spectral_2}\includegraphics[scale=0.111]{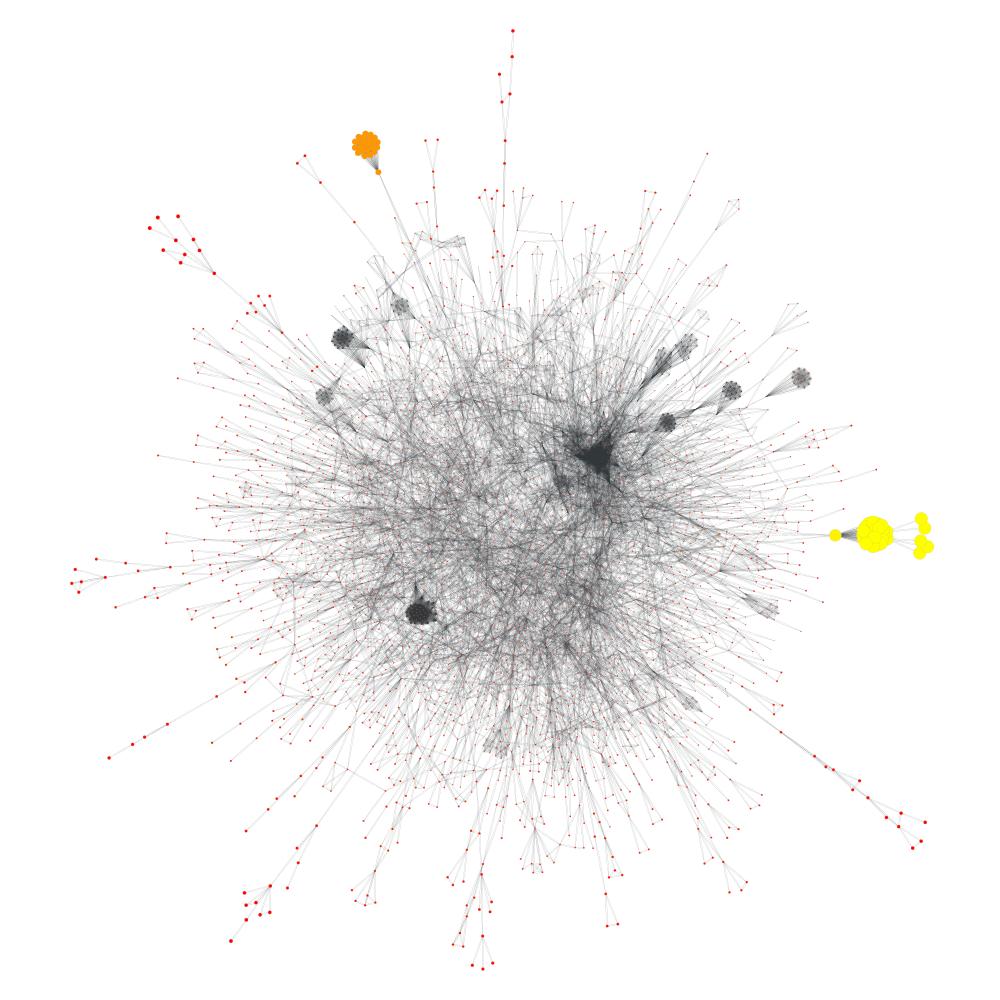}} \\
	\subfigure[MOV local, seed $1$]{
  	\begin{tikzpicture}[every node/.style={anchor=center}]
   	 \node(a) at (8,4){\label{Fig_grqc_spectral_3}\includegraphics[scale=0.095]{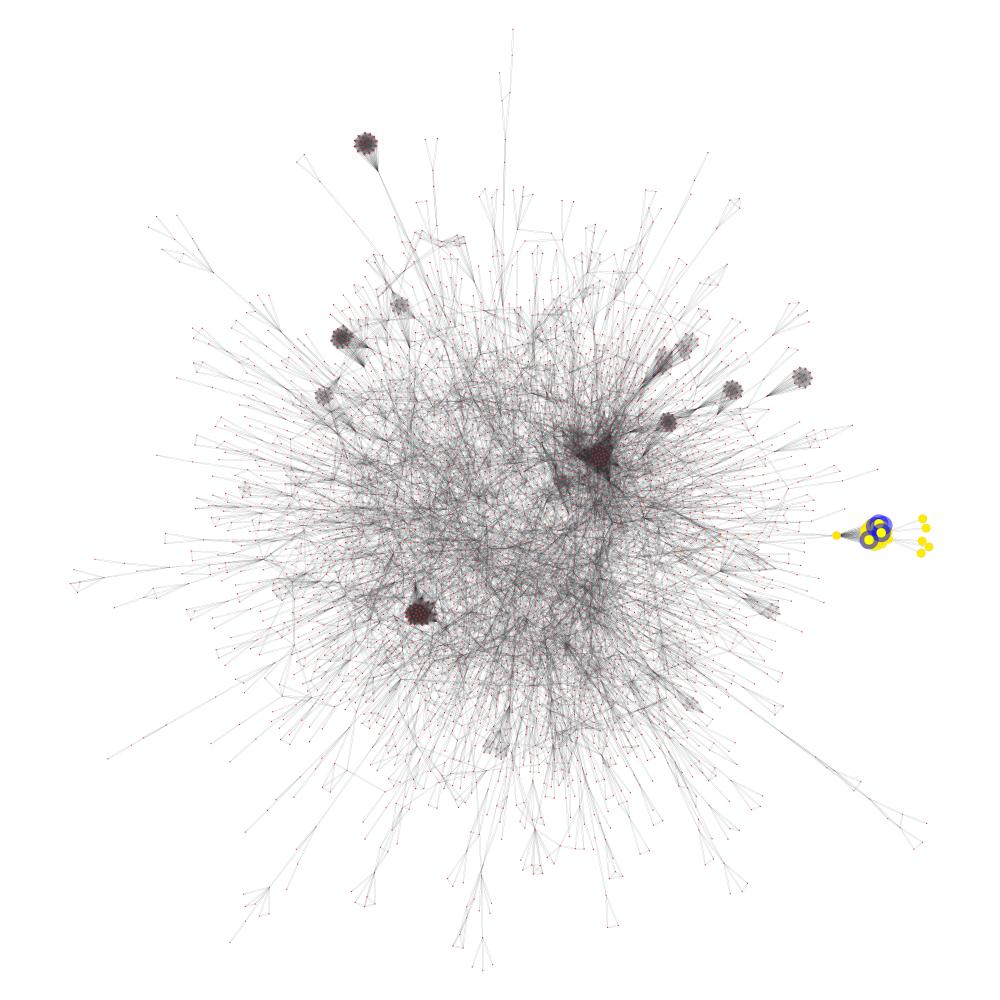}};
    	\node(b) at (7.5,3.3){\frame{\colorbox{white}{\includegraphics[scale=0.04]{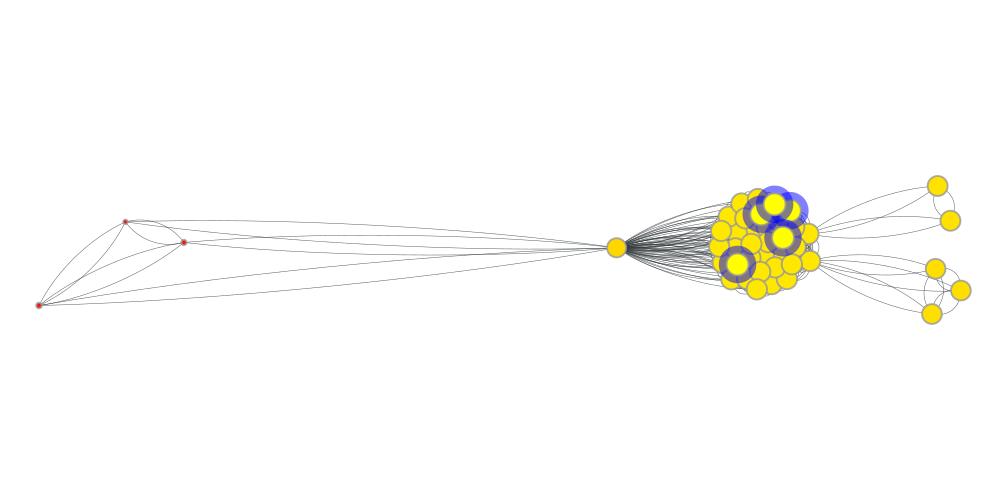}}}};
    	\draw[black] (9.0,4.05)rectangle (9.6,3.76);
    	\draw[black,->](8.33,3.3)--(9,3.75);
  	\end{tikzpicture}
	}
	\subfigure[MOV local, seed $2$]{
  	\begin{tikzpicture}[every node/.style={anchor=center}]
   	 \node(a) at (8,4){\label{Fig_grqc_spectral_4}\includegraphics[scale=0.095]{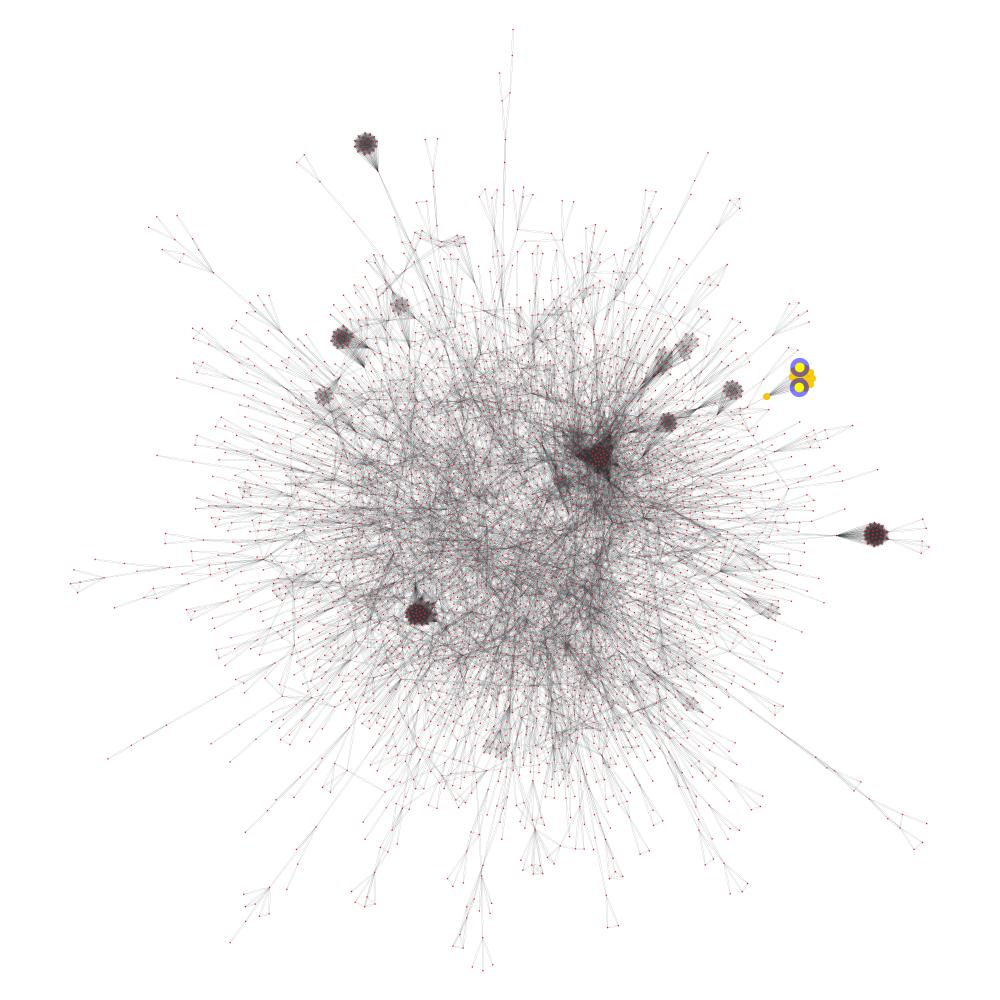}};
    	\node(b) at (7.5,3.3){\frame{\colorbox{white}{\includegraphics[scale=0.04]{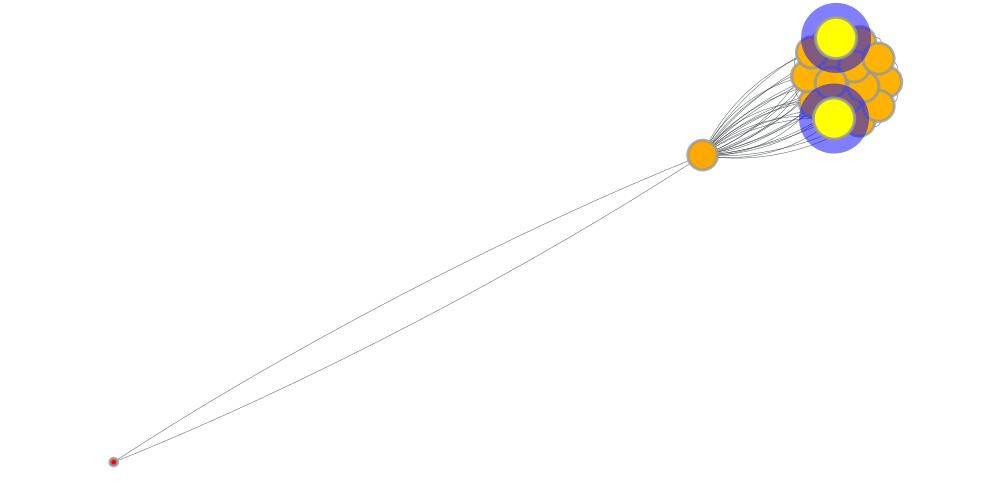}}}};
    	\draw[black] (8.8,4.53)rectangle (9.2,4.28);
    	\draw[black,->](8.33,3.3)--(8.8,4.28);
  	\end{tikzpicture}
	}
	\\
	\subfigure[$\ell_1$-regularized Page-Rank, seed $1$]{
  	\begin{tikzpicture}[every node/.style={anchor=center}]
   	 \node(a) at (8,4){\label{Fig_grqc_spectral_5}\includegraphics[scale=0.095]{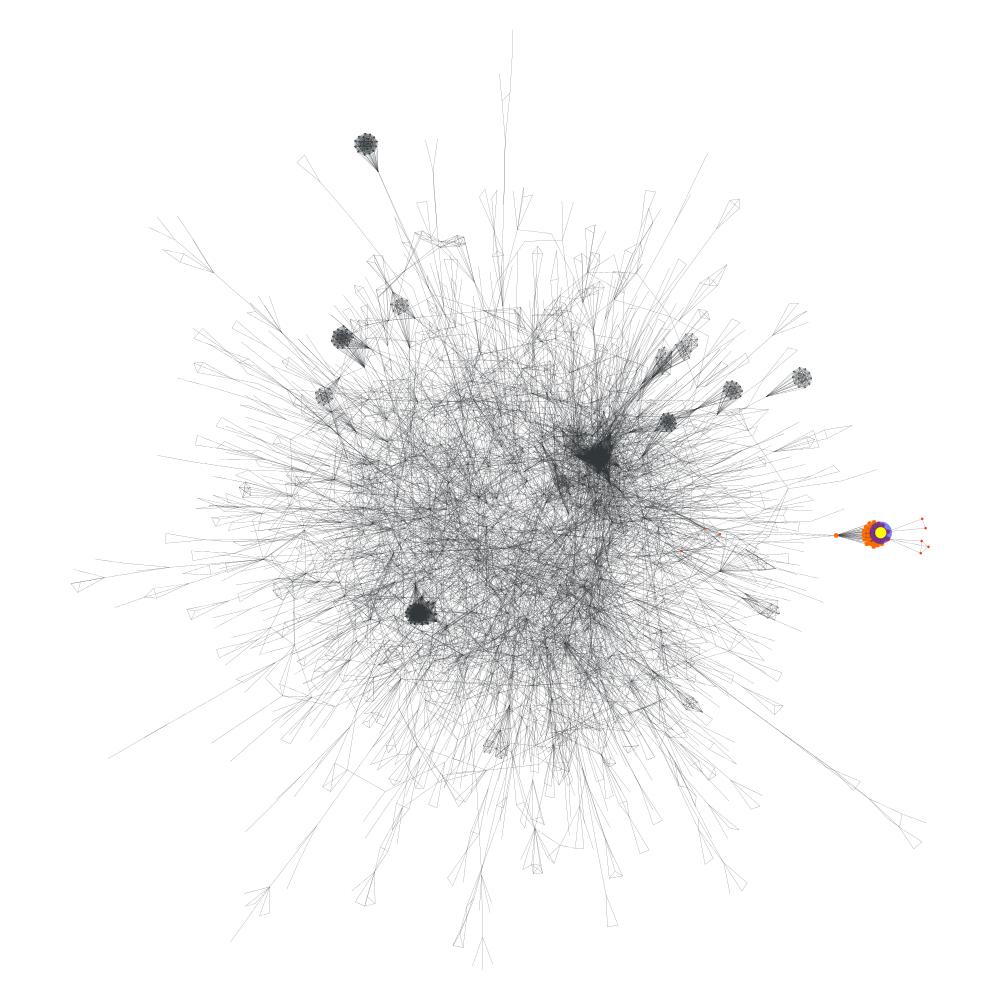}};
    	\node(b) at (7.5,3.3){\frame{\colorbox{white}{\includegraphics[scale=0.04]{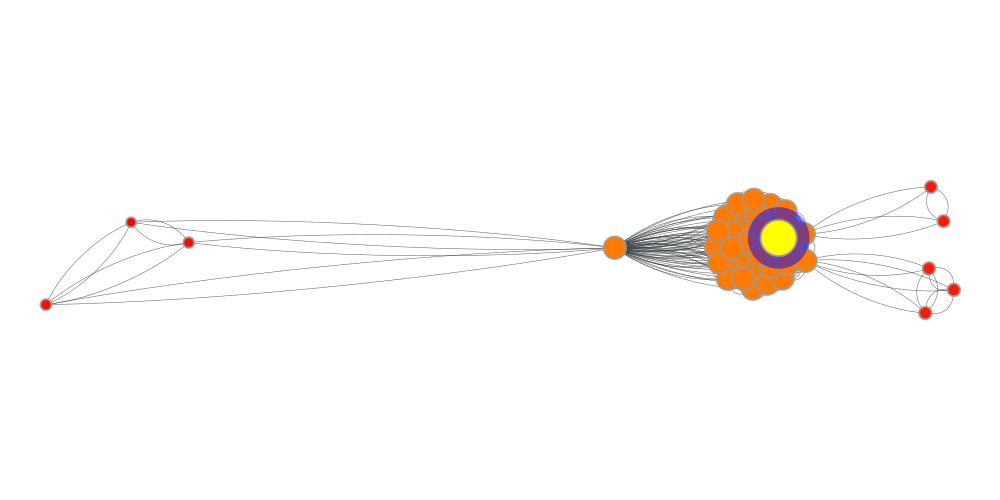}}}};
    	\draw[black] (9.0,4.05)rectangle (9.6,3.76);
    	\draw[black,->](8.33,3.3)--(9,3.75);
  	\end{tikzpicture}
	}
	\subfigure[$\ell_1$-regularized Page-Rank, seed $2$]{
  	\begin{tikzpicture}[every node/.style={anchor=center}]
   	 \node(a) at (8,4){\label{Fig_grqc_spectral_6}\includegraphics[scale=0.095]{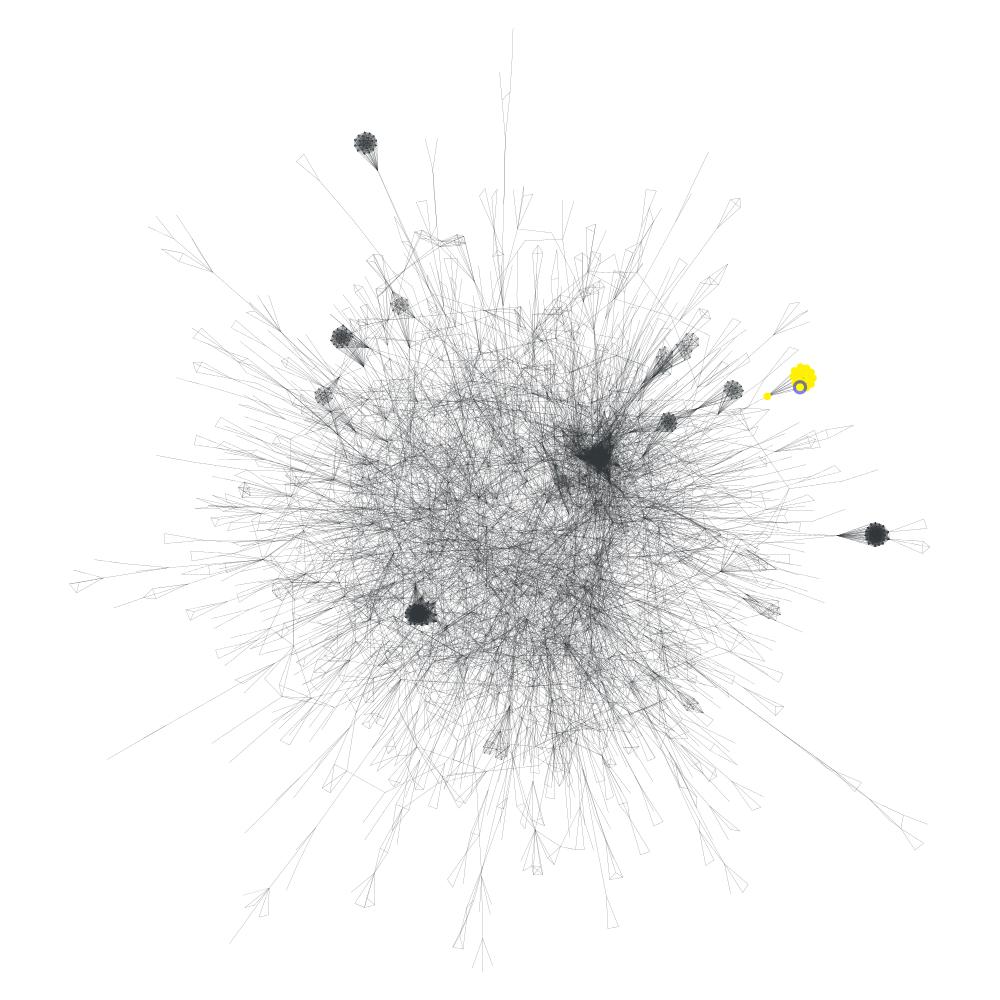}};
    	\node(b) at (7.5,3.3){\frame{\colorbox{white}{\includegraphics[scale=0.04]{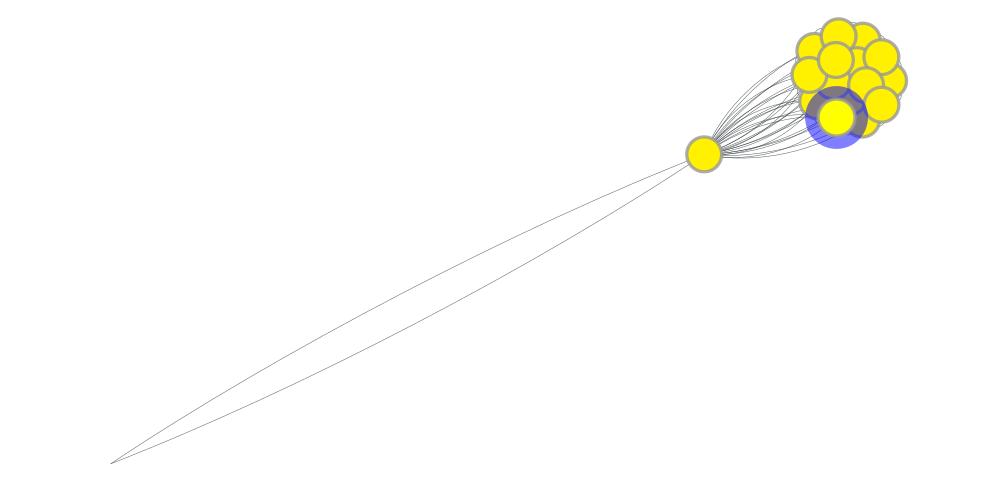}}}};
    	\draw[black] (8.8,4.53)rectangle (9.2,4.28);
    	\draw[black,->](8.33,3.3)--(8.8,4.28);
  	\end{tikzpicture}
	}
	\caption{CA-GrQc.\ This figure shows the solutions of Spectral relaxation, MOV with global input, MOV with local input
	and $\ell_1$-regularized Page-Rank.
	The meaning of the colours of the nodes and its sizes is the same as in Figure \ref{Fig_senate_spectral}.}
	\label{Fig_grqc_spectral}		
\end{figure}

In Figure \ref{Fig_grqc_spectral}, we present global, weakly local, and strongly local solutions for the less well-partitionable and thus less easily-visualizable CA-GrQc graph. 
As already mentioned in the description of this data-set, this graph has many small clusters with small conductance ratio and large clusters have large ratio.
This is also justified in our experiment by the fact that global methods, such as the Spectral relaxation and MOV global in Figures \ref{Fig_grqc_spectral_1} and \ref{Fig_grqc_spectral_2}, respectively, 
recover small clusters.
The two global procedures find small clusters which are presented in Figures \ref{Fig_grqc_spectral_1} and \ref{Fig_grqc_spectral_2} with red, orange and yellow colours. However, since there are many small clusters
of small conductance ratio, 
one might want to find different clusters than the ones obtained by Spectral relaxation and MOV global. This is possible using localized procedures such as MOV and $\ell_1$-regularized Page-Rank. Given two different seed sets we demonstrate  
in Figures \ref{Fig_grqc_spectral_3} and \ref{Fig_grqc_spectral_4} that MOV successfully finds other clusters than the ones obtained by the global methods. The same is shown in Figures \ref{Fig_grqc_spectral_5}
and \ref{Fig_grqc_spectral_6} for $\ell_1$-regularized Page-Rank. Notice that MOV assigns weights (perhaps small) to all the nodes of the graph; on the other hand, $\ell_1$-regularized Page-Rank, as a strongly local procedure,
assigns weights only to a small number of nodes, without examining all of the graph. 

\begin{figure}
\centering
	\subfigure[Spectral relaxation]{\label{Fig_fb_spectral_1}\includegraphics[scale=0.11]{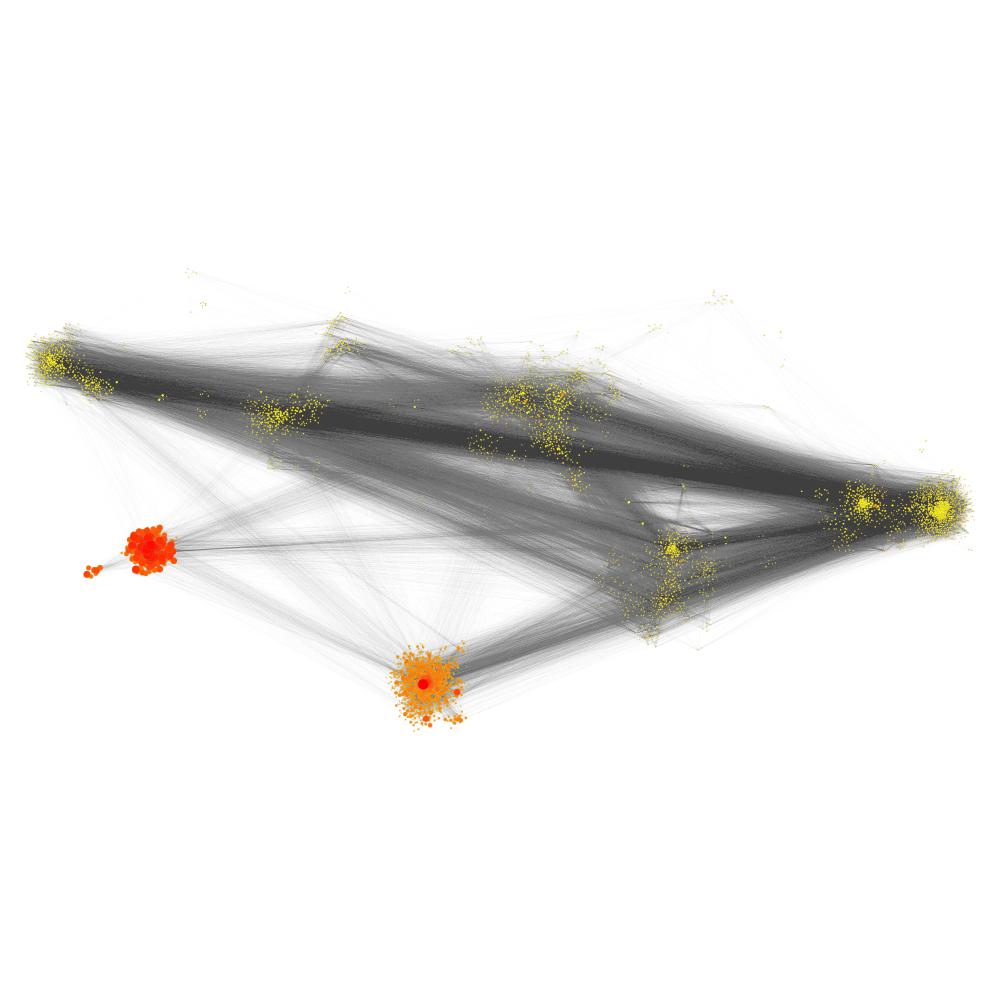}}
	\subfigure[MOV global]{\label{Fig_fb_spectral_2}\includegraphics[scale=0.11]{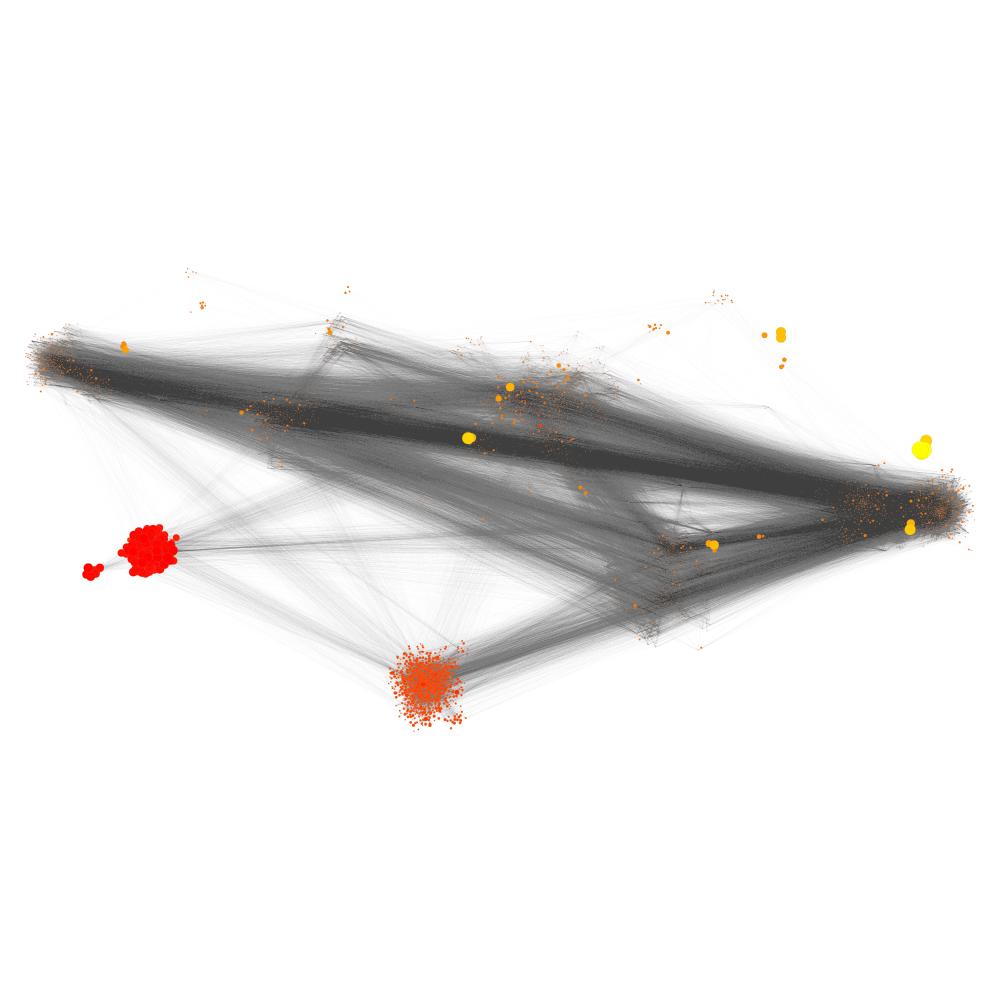}} \\
	\subfigure[MOV local]{\label{Fig_fb_spectral_3} \includegraphics[scale=0.11]{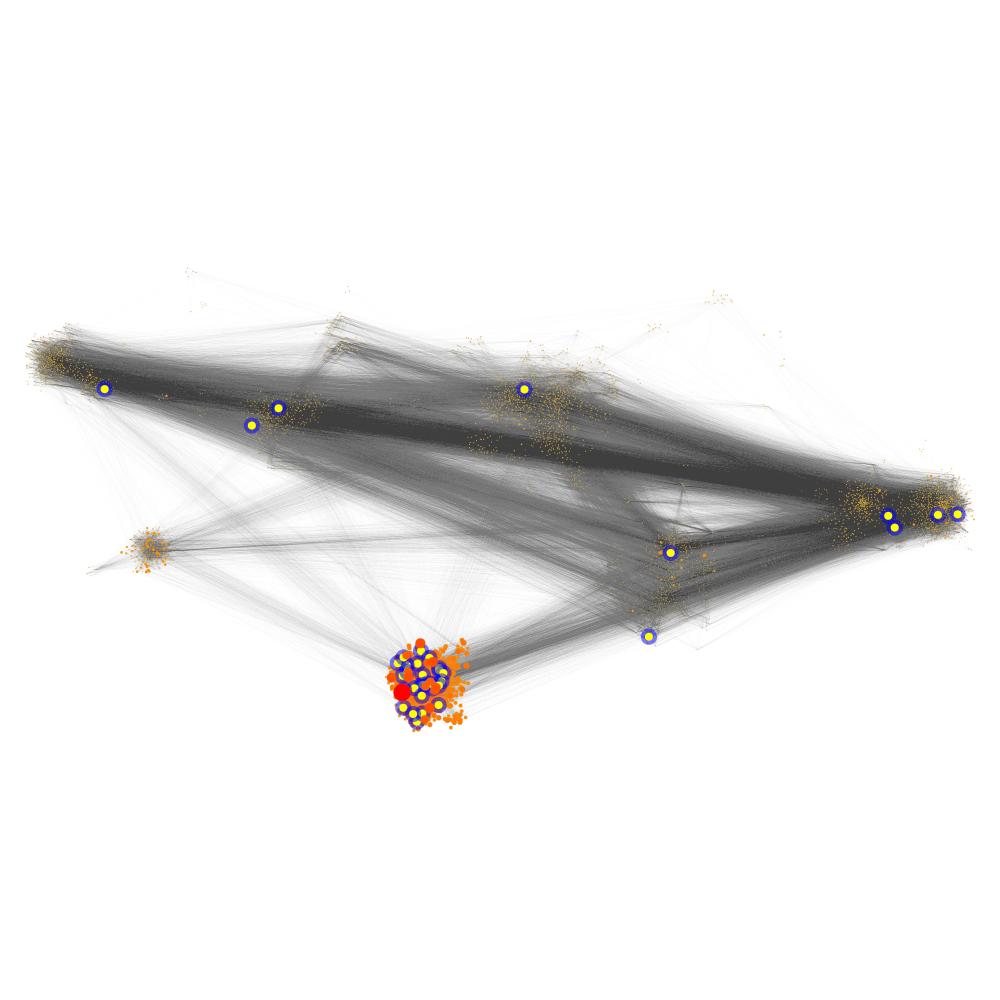}}
	\subfigure[MOV local, seed $2$]{\label{Fig_fb_spectral_4}\includegraphics[scale=0.11]{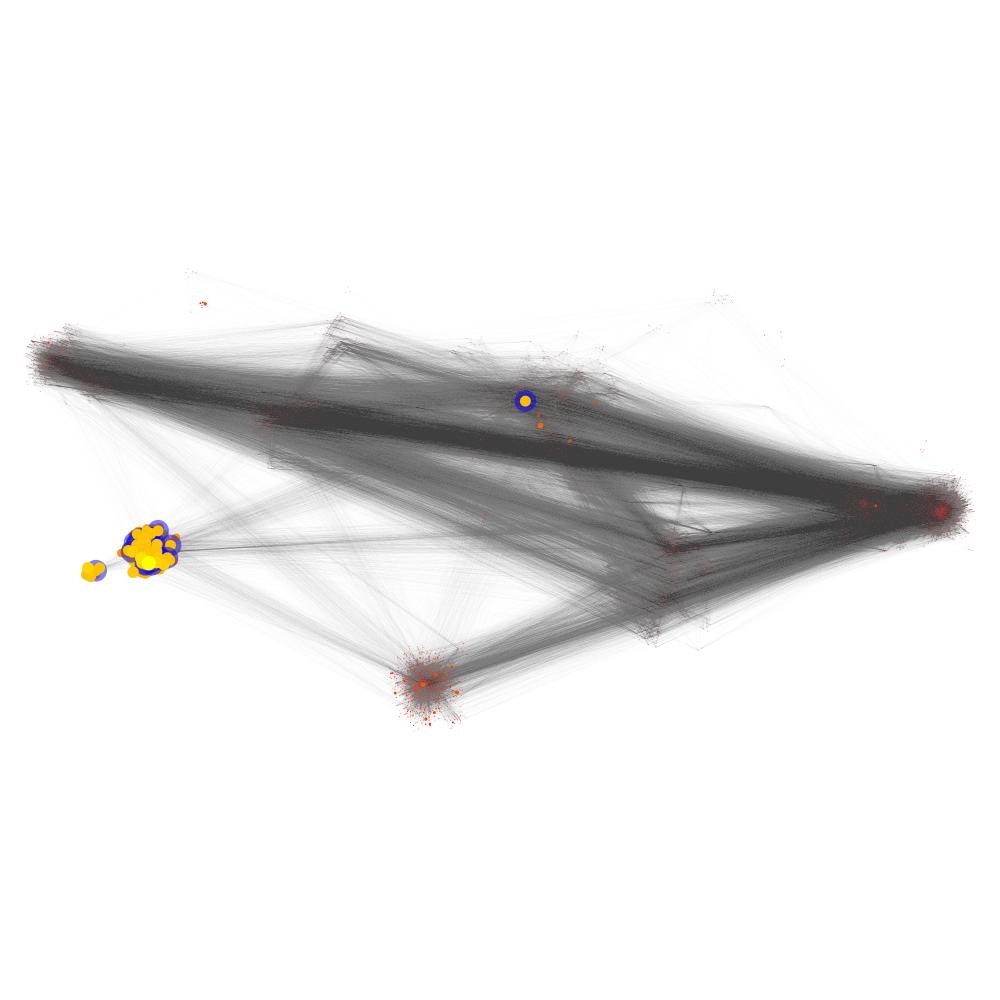}} \\
	\subfigure[$\ell_1$-regularized Page-Rank]{\label{Fig_fb_spectral_5}\includegraphics[scale=0.11]{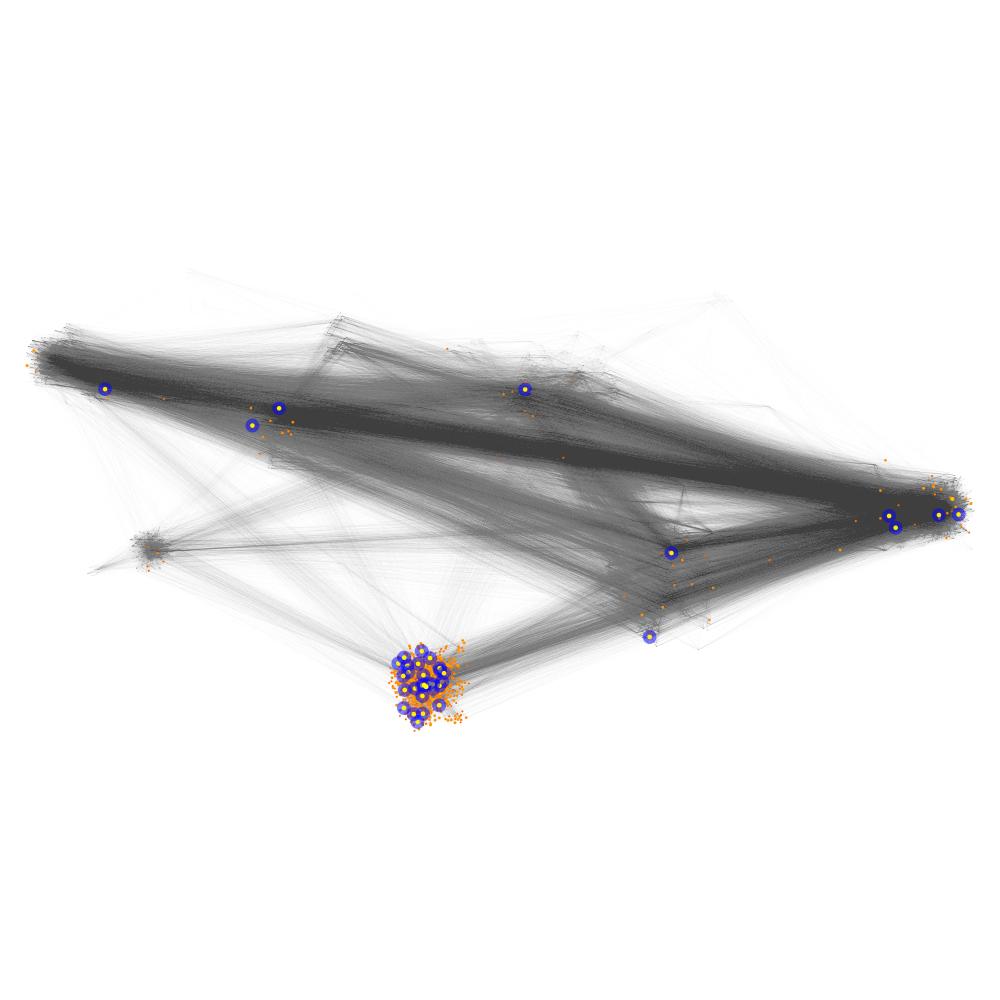}}
	\subfigure[$\ell_1$-regularized Page-Rank, seed $2$]{\label{Fig_fb_spectral_6}\includegraphics[scale=0.11]{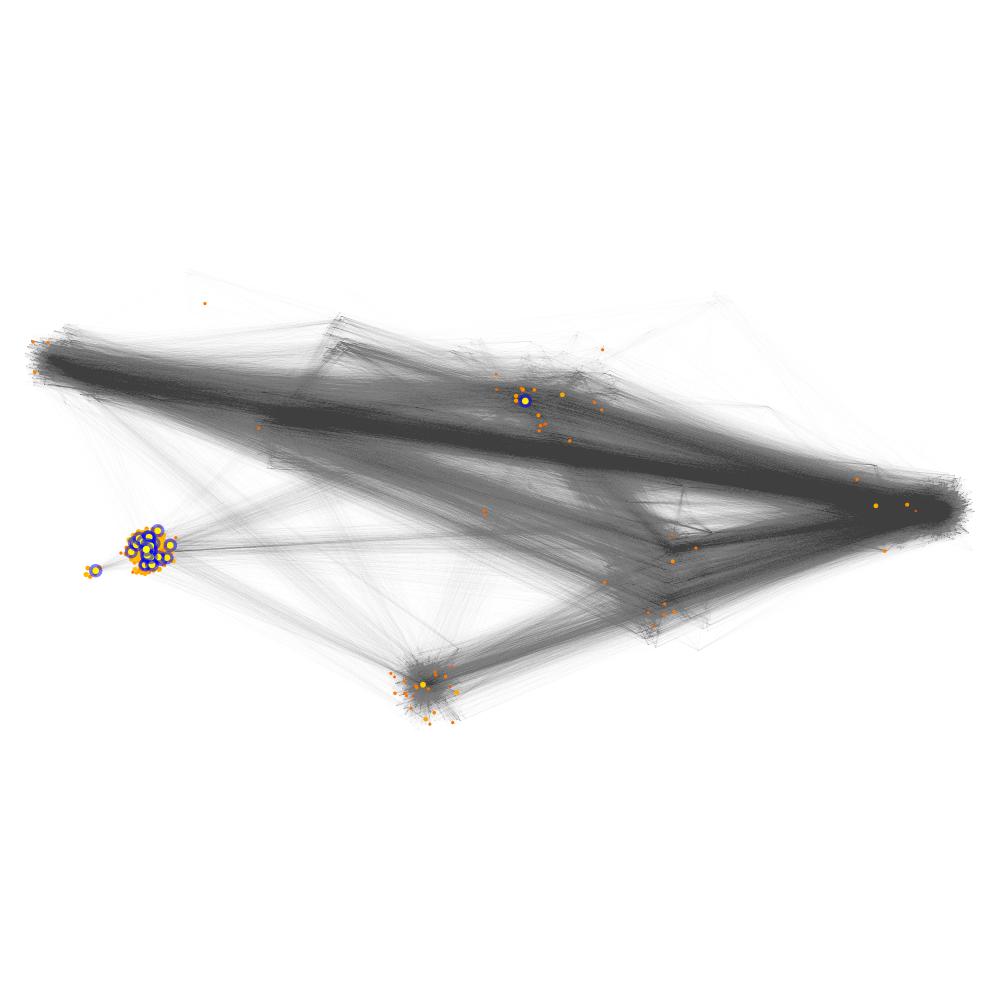}}
	\caption{FB-Johns55.\ 
	This figure shows the solutions of Spectral relaxation, MOV with global input, MOV with local input
	and $\ell_1$-regularized Page-Rank. The meaning of the colours of the nodes and its sizes is the same as in Figure \ref{Fig_senate_spectral}.}
	\label{Fig_fb_spectral}		
\end{figure}

We now use the FB-Johns55 graph which has an expander-like behaviour at all size scales, i.e., all small and large clusters have large conductance ratio.  See Figure $6$ in \cite{Jeub15} for details.
We present the results of this experiment in Figure \ref{Fig_fb_spectral}. Notice that in Figures \ref{Fig_fb_spectral_1} and \ref{Fig_fb_spectral_2} the global methods identify 
three main clusters, one small (red nodes), one medium size (orange nodes) and one large (yellow nodes). All these clusters have similar conductance ratio. 
%Depending on the application one cluster might be more important than the other. 
In Figures \ref{Fig_fb_spectral_3} and \ref{Fig_fb_spectral_4} we show that MOV can recover 
the medium or the small size clusters, respectively, by giving a localized seed set. 
In Figures \ref{Fig_fb_spectral_5} and \ref{Fig_fb_spectral_6} we illustrate that using $\ell_1$-regularized Page-Rank one can find very similar clusters while exploiting the strongly local
running time of the method.

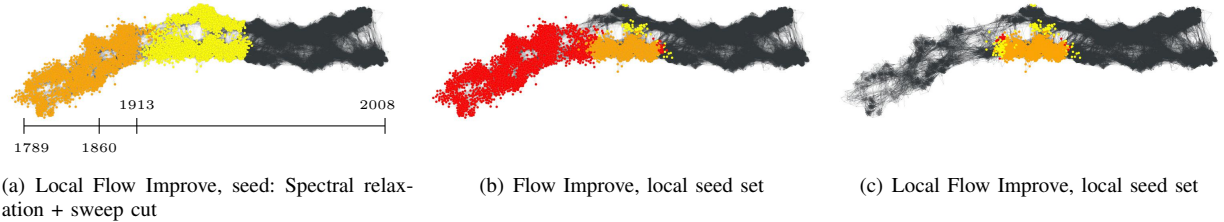
\begin{figure*}
\centering
	\subfigure[Local Flow Improve, seed: Spectral relaxation + sweep cut]{\label{Fig_senate_flow_1}\includegraphics[scale=0.15]{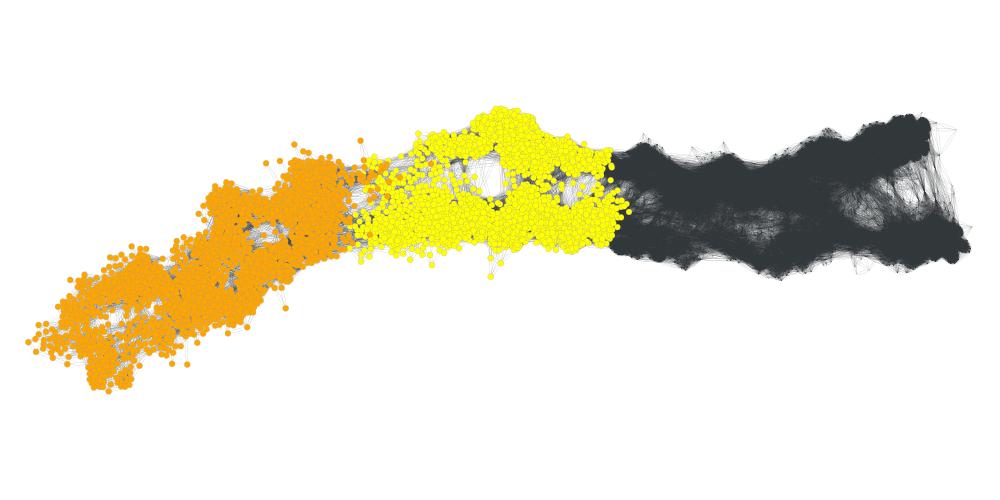}
	\hspace{-5.5cm}
	\begin{tikzpicture}
	\draw (0,0) -- (4.8,0);
	\foreach \x in {0,1,1.5,4.8}
	\draw (\x cm,3pt) -- (\x cm,-3pt);
	\draw (0.1,0) node[below=3pt] (a) {\tiny $1789$} node[above=3pt] {};
	\draw (1,0)  node[below=3pt]  {\tiny $1860$} node[above=3pt] (c) {};
	\draw (1.5,0) node[below=3pt](d) {} node[above=3pt] {\tiny $1913$};
	\draw (4.7,0) node[above=3pt] (f) {\tiny $2008$} node[below=3pt] {};
	\end{tikzpicture}
	}	
	\subfigure[Flow Improve, local seed set]{\label{Fig_senate_flow_2}\includegraphics[scale=0.15]{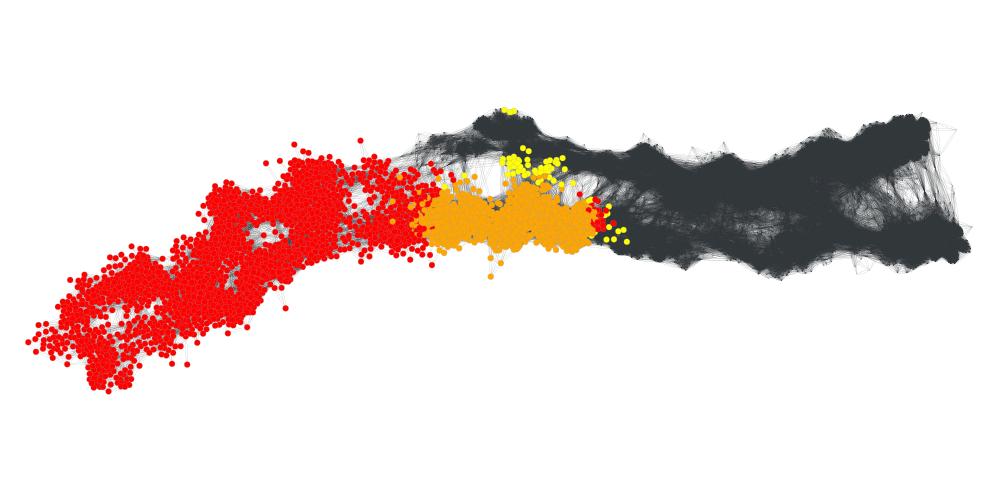}}
	\subfigure[Local Flow Improve, local seed set]{\label{Fig_senate_flow_3}\includegraphics[scale=0.15]{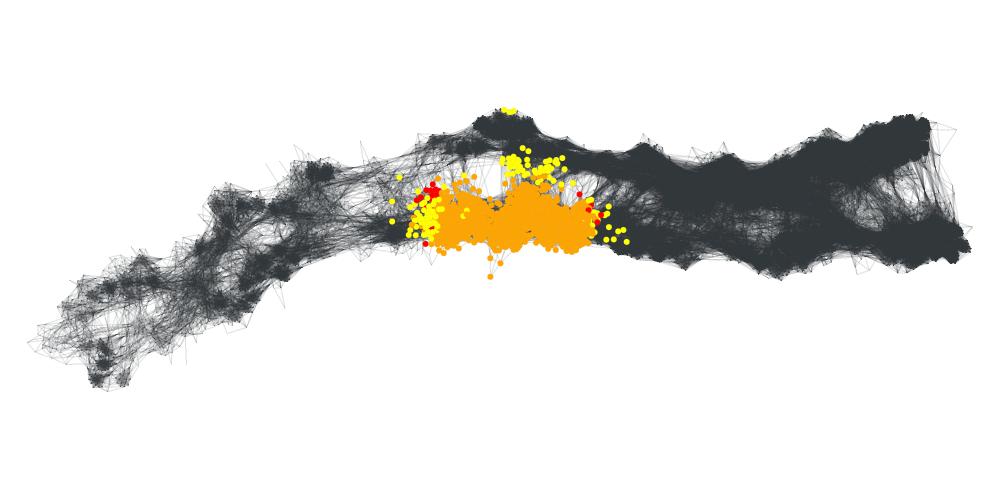}}
	\caption{US-Senate.\ This figure shows the solutions of Flow Improve and Local Flow Improve for various input seed sets. 
	Red nodes are only Flow Improve or Local Flow Improve, depending on the experiment; yellow nodes are only seed set; and orange nodes are both part of the flow algorithm and the seed set.}		
	\label{Fig_senate_flow}
\end{figure*}

Let us now present the performance of flow-based algorithms on the same graphs. We begin with US-Senate in Figure \ref{Fig_senate_flow}.
In this figure, the red nodes are part of the solution of Flow Improve or Local Flow Improve, depending on the experiment; the yellow nodes are part of the seed set only; and 
the orange nodes are in both the solution of the flow algorithm and the input seed set. In Figure \ref{Fig_senate_flow_1}, we used as an input seed set to Flow Improve the 
cluster obtained by applying sweep cut with respect to the conductance ratio on the Spectral relaxation solution. 
%Notice that Spectral relaxation and sweep cut finds a cluster with boundaries after the year $1913$, while the 
%correct separation of the graph happens at about $1860$. However, Flow Improve, which is a weakly local algorithm, 
%finds the correct boundaries at the year $1860$. 
Figures \ref{Fig_senate_flow_2} and \ref{Fig_senate_flow_3} present a clear distinction between 
Flow Improve and Local Flow Improve, weakly and strongly local algorithms, respectively. For both figures, the input seed set is located at about the middle of the graph. Flow Improve 
as a weakly local algorithm examines the whole graph and returns a cluster which includes the period before $1913$. Also, it includes big part of the input seed set in the cluster
due to the overlapping regularization term in the denominator of its objective function. See the definition of the objective function $\phi_R$ for Flow Improve in Section \ref{sec:localgraphpart}.
On the other hand, in Figure \ref{Fig_senate_flow_3} Local Flow Improve as a strongly local algorithm does not examine the whole graph and its solution is concentrated only around the input seed set.

\begin{figure}
\centering
\subfigure[Flow Improve, seed $1$]{\label{Fig_cagrqc_flow_1}
  \begin{tikzpicture}[every node/.style={anchor=center}]
    \node(a) at (8,4){\includegraphics[scale=0.095]{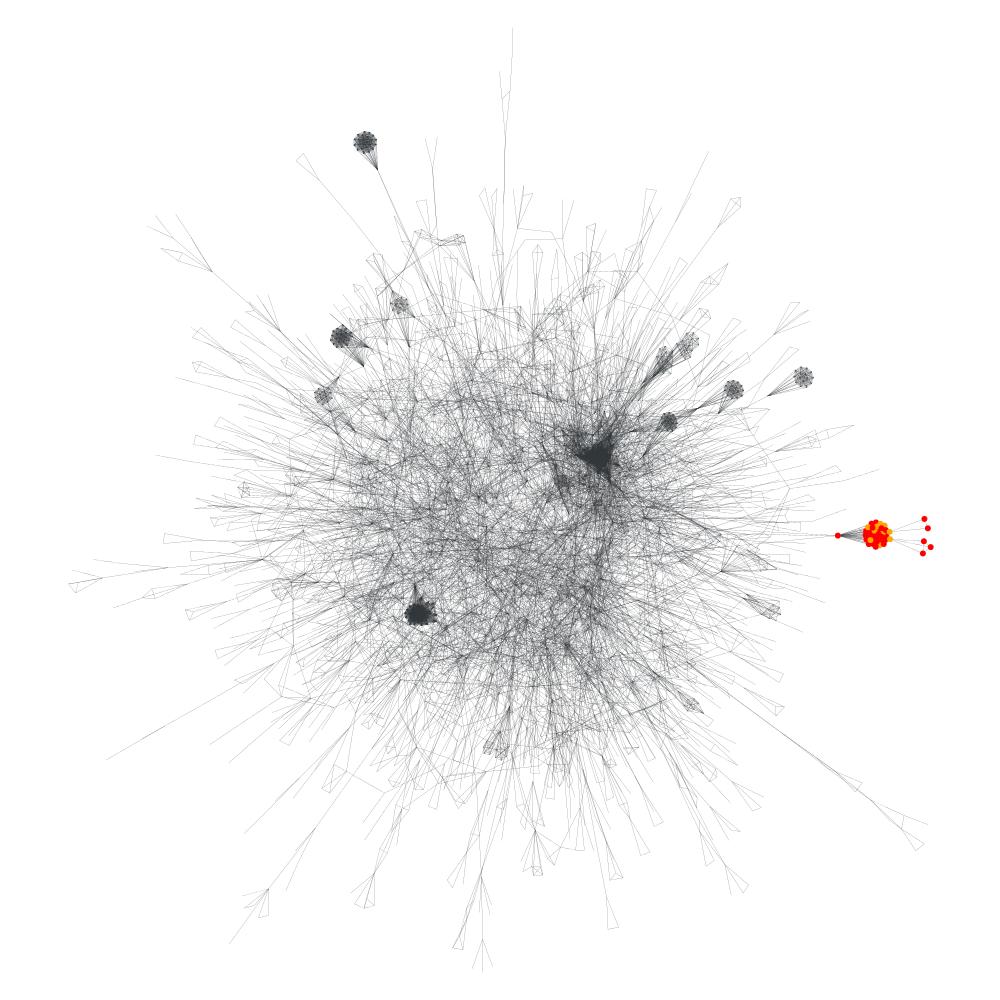}};
    \node(b) at (7.5,3.3){\frame{\colorbox{white}{\includegraphics[scale=0.04]{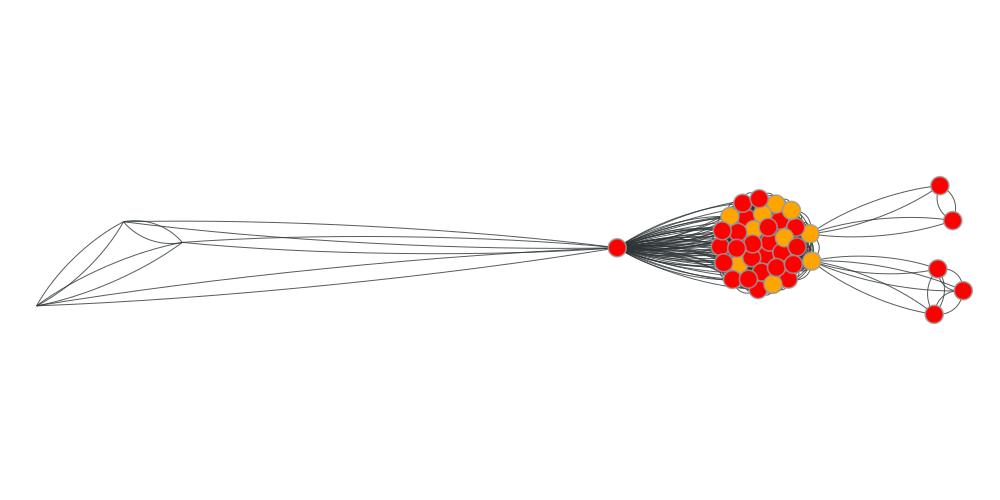}}}};
    	\draw[black] (9.0,4.05)rectangle (9.6,3.76);
    	\draw[black,->](8.33,3.3)--(9,3.75);
  \end{tikzpicture}
}
\subfigure[Local Flow Improve, seed $1$]{\label{Fig_cagrqc_flow_2}
  \begin{tikzpicture}[every node/.style={anchor=center}]
    \node(a) at (8,4){\includegraphics[scale=0.095]{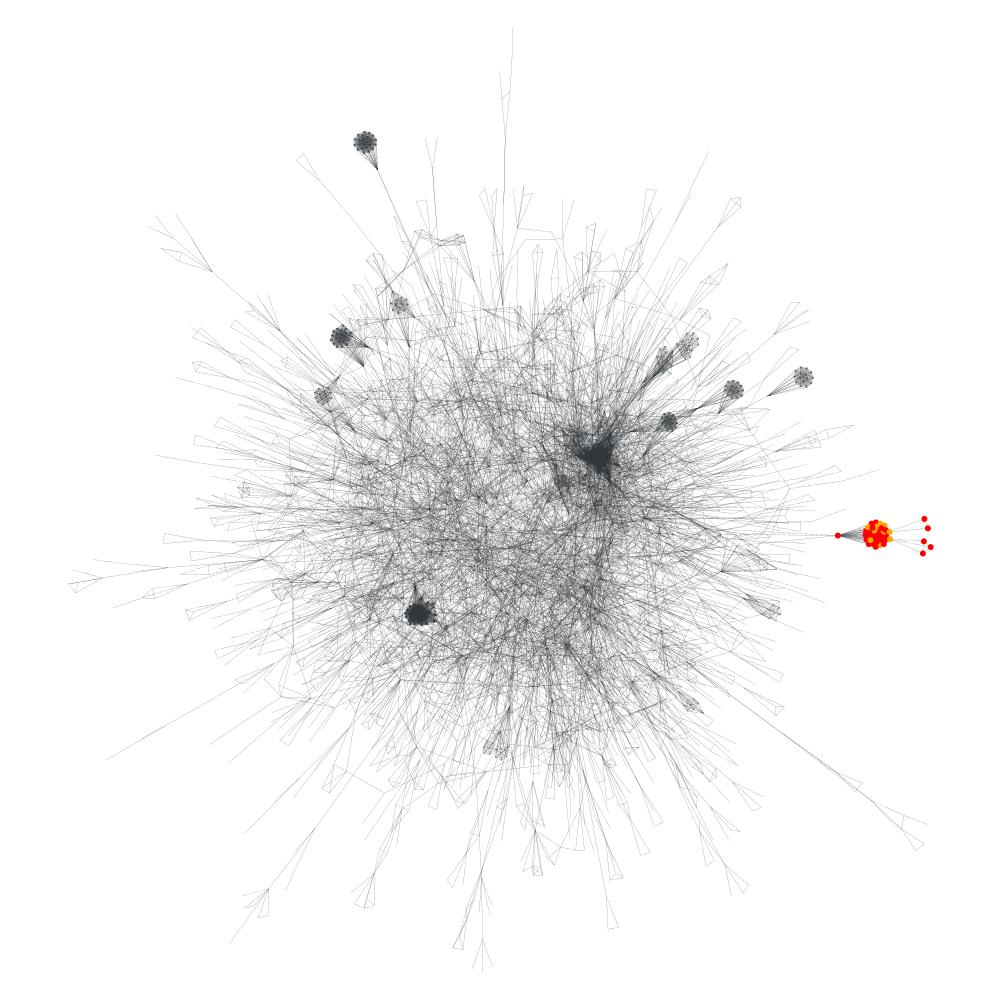}};
    \node(b) at (7.5,3.3){\frame{\colorbox{white}{\includegraphics[scale=0.04]{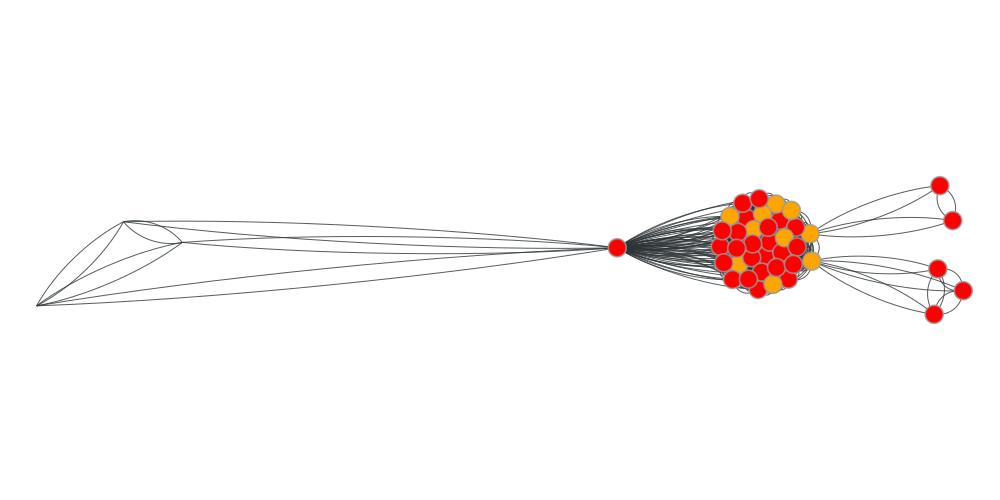}}}};
    	\draw[black] (9.0,4.05)rectangle (9.6,3.76);
    	\draw[black,->](8.33,3.3)--(9,3.75);
  \end{tikzpicture}
}\\
\subfigure[Flow Improve, local seed $2$]{\label{Fig_cagrqc_flow_3}
  \begin{tikzpicture}[every node/.style={anchor=center}]
    \node(a) at (8,4){\includegraphics[scale=0.095]{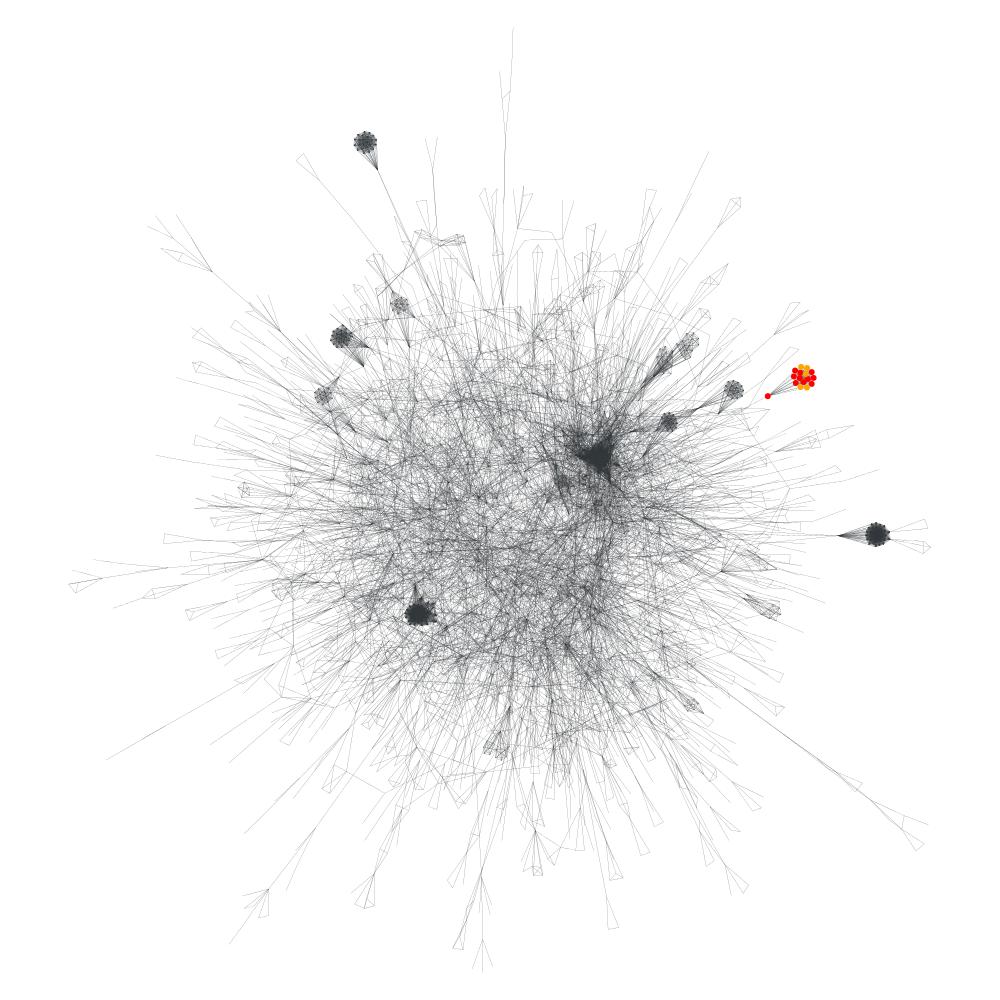}};
    \node(b) at (7.5,3.3){\frame{\colorbox{white}{\includegraphics[scale=0.04]{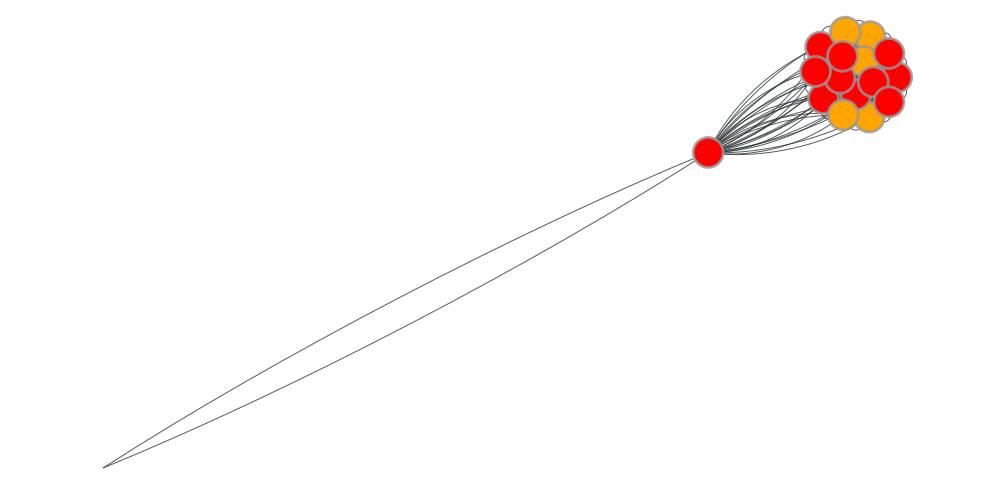}}}};
    	\draw[black] (8.8,4.53)rectangle (9.2,4.28);
    	\draw[black,->](8.33,3.3)--(8.8,4.28);
  \end{tikzpicture}
}
\subfigure[Local Flow Improve, seed $2$]{\label{Fig_cagrqc_flow_4}
  \begin{tikzpicture}[every node/.style={anchor=center}]
    \node(a) at (8,4){\includegraphics[scale=0.095]{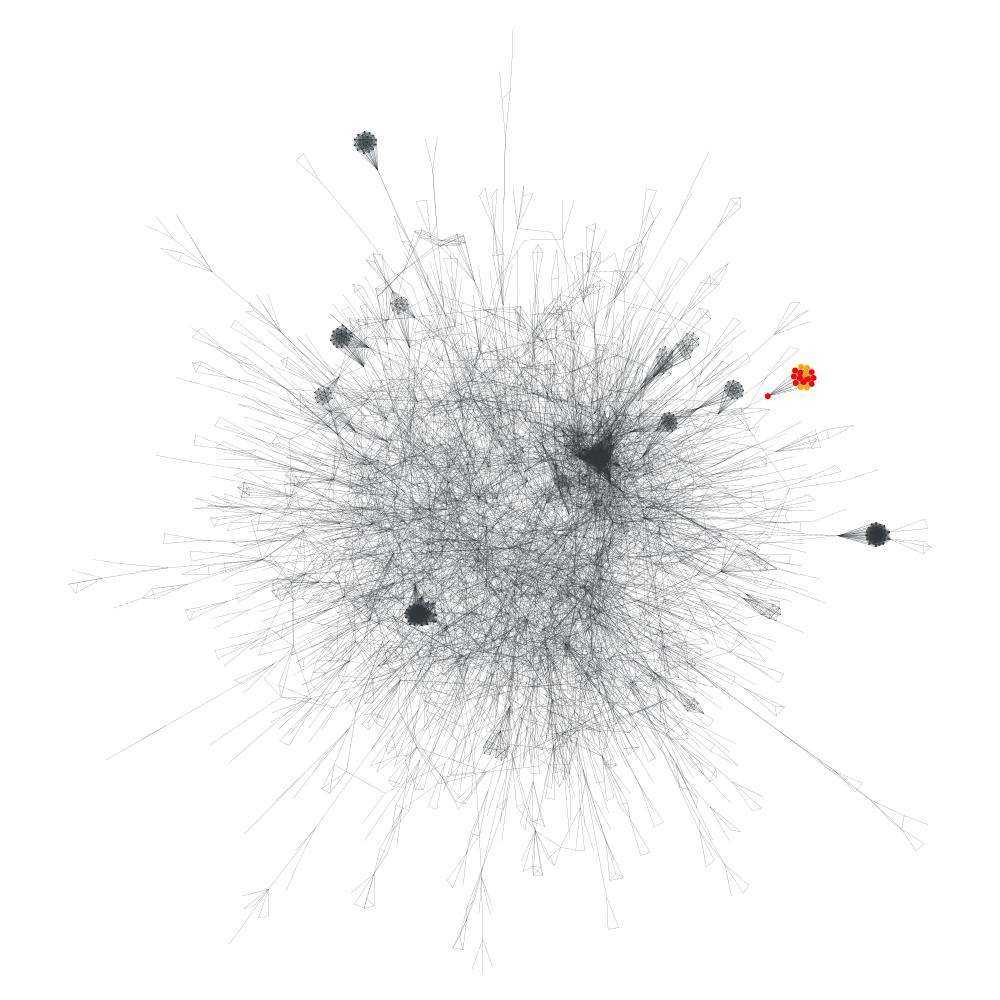}};
    \node(b) at (7.5,3.3){\frame{\colorbox{white}{\includegraphics[scale=0.04]{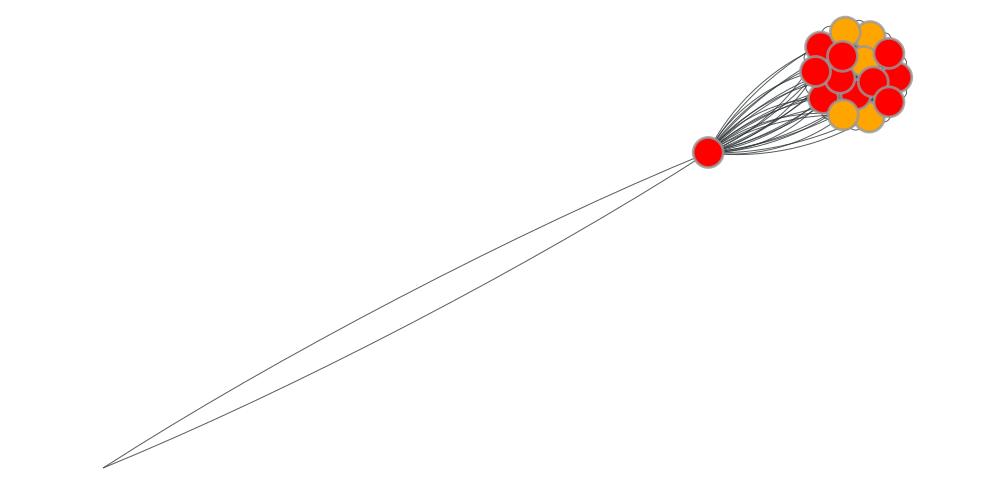}}}};
    	\draw[black] (8.8,4.53)rectangle (9.2,4.28);
    	\draw[black,->](8.33,3.3)--(8.8,4.28);
  \end{tikzpicture}
}
\caption{CA-GrQc.\ This figure shows the solutions of Flow Improve and Local Flow Improve for various input seed sets.
Red nodes are only Flow Improve or Local Flow Improve, depending on the experiment; yellow nodes are only seed set; and orange nodes are both part of the flow algorithm and the seed set.}
\label{Fig_cagrqc_flow}	
\end{figure}

\begin{figure}
\centering
	\subfigure[Flow Improve, seed $1$]{\label{Fig_fb_flow_1}\includegraphics[scale=0.11]{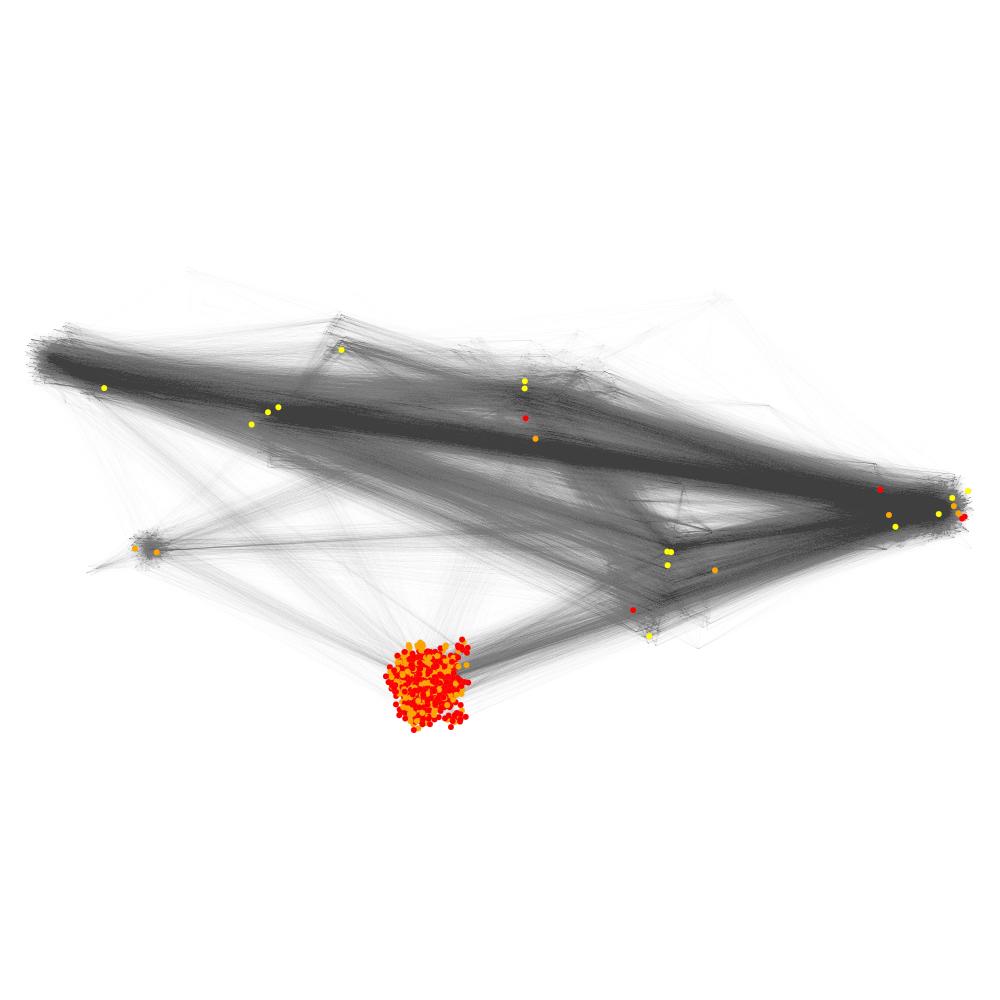}} 
	\subfigure[Local Flow Improve, seed $1$]{\label{Fig_fb_flow_2}\includegraphics[scale=0.11]{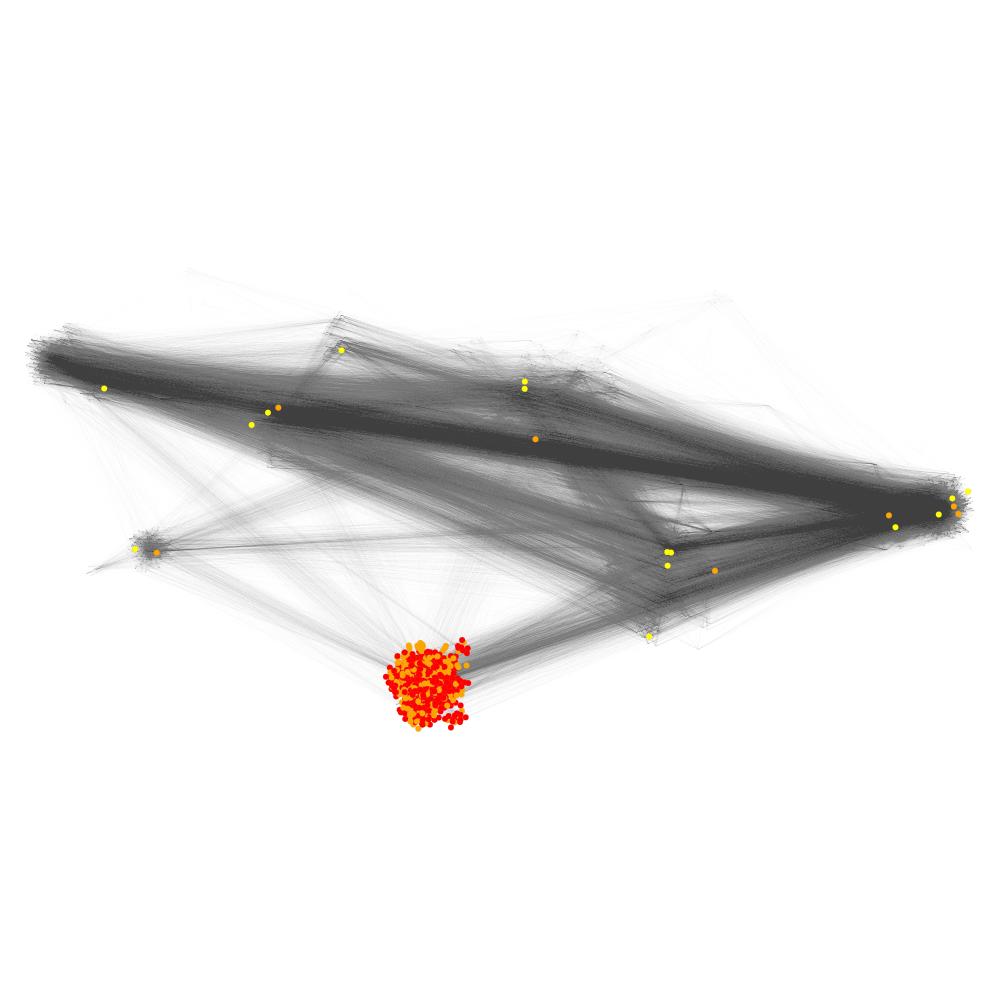}}\\
	\subfigure[Flow Improve, seed $2$]{\label{Fig_fb_flow_3}\includegraphics[scale=0.11]{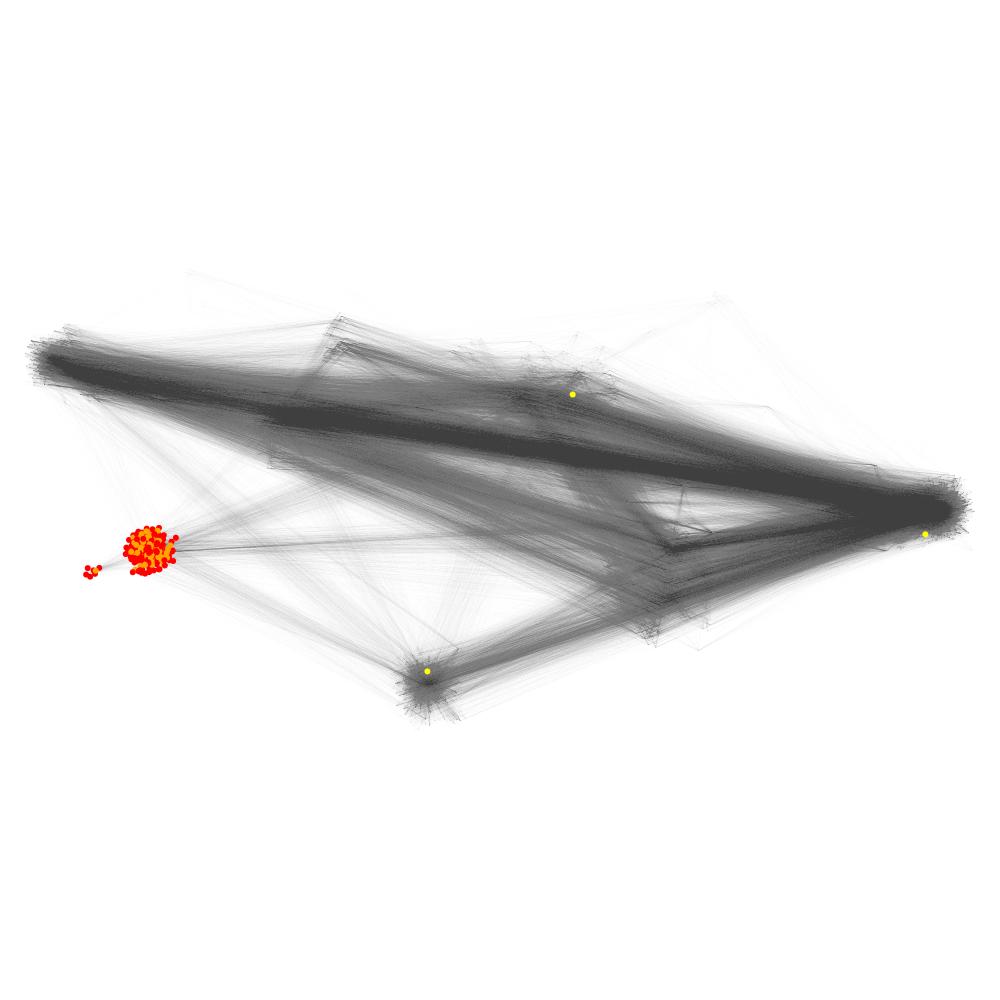}}
	\subfigure[Local Flow Improve, seed $2$]{\label{Fig_fb_flow_4}\includegraphics[scale=0.11]{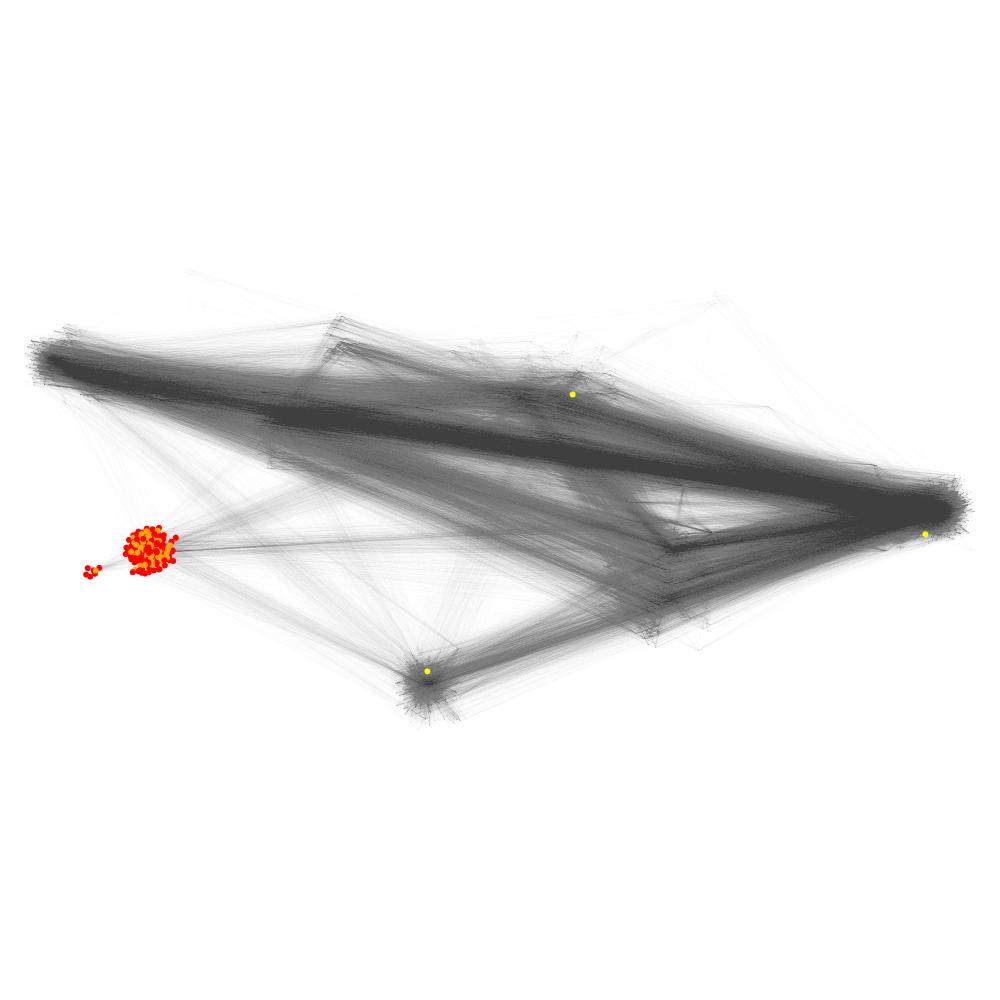}}
\caption{FB-Johns55. This figure shows the solutions of Flow Improve and Local Flow Improve for various input seed sets. 
Red nodes are only Flow Improve or Local Flow Improve, depending on the experiment; yellow nodes are only seed set; and orange nodes are both part of the flow algorithm and the seed set.}
\label{Fig_fb_flow}
\end{figure}

The distinction that we discussed in the previous paragraph between Flow Improve and Local Flow Improve is easy to visualize in the relatively well-structured US-Senate, but it is not so clear in all graphs. 
For example, in Figure \ref{Fig_cagrqc_flow} we present the performance of these two algorithms for the CA-GrQc graph. Since this graph has only small
clusters of small conductance ratio, Flow Improve and Local Flow Improve find the same clusters. This is clearly shown by comparing 
Figures \ref{Fig_cagrqc_flow_1} and \ref{Fig_cagrqc_flow_2} and Figures \ref{Fig_cagrqc_flow_3} and \ref{Fig_cagrqc_flow_4}.
A similar performance is observed for the FB-Johns55 graph in Figure \ref{Fig_fb_flow}, except that  the solution of Flow Improve
and Local Flow Improve are not exactly the same but only very similar.
%Similarly to spectral algorithms, flow-based algorithms 
%can be used to find small, medium or large scale clusters given an input seed set of nodes.

\paragraph{Flow vs. spectral, or $\ell_1$ vs. $\ell_2$}
Spectral algorithms measure distances of the nodes based on the $\ell_2$ norm. Generally this means that the nodes of the graph are embedded on the real line.
On the other hand, flow algorithms measure distances of the nodes based on the $\ell_1$ norm. The solution to flow-based algorithms that we discussed is binary,
either a node is added in the solution with weight $1$ or it is not and it has weight $0$. 
In this case, the objective function $\|Bx\|_{1,C}$ of the flow algorithms 
is a locally-biased variation on $\cut(S)$, where $S$ is constructed based on the binary $x$.\ Therefore, the flow algorithms aim to find a balance between finding good cuts and identifying the input seed set. This implies that the flow algorithms
try to minimize the absolute number of edges that cross the partition, but at the same time they try to take into account the volume regularization effect of the denominator in the objective function.
%This nice property of recovering good boundaries is lost for spectral algorithms that are based on the $\ell_2$ norm.

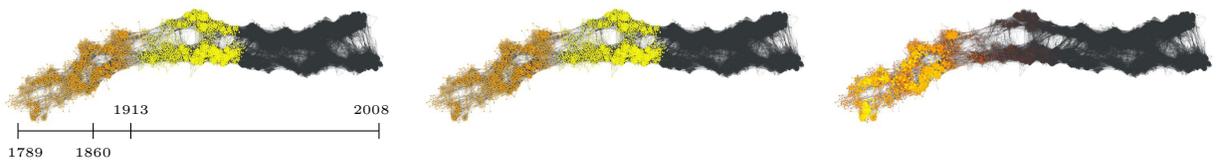
\begin{figure*}
\centering
	\subfigure[MQI, seed: Spectral relaxation + sweep cut]{\label{Fig_senate_l1vl2_1}\includegraphics[scale=0.15]{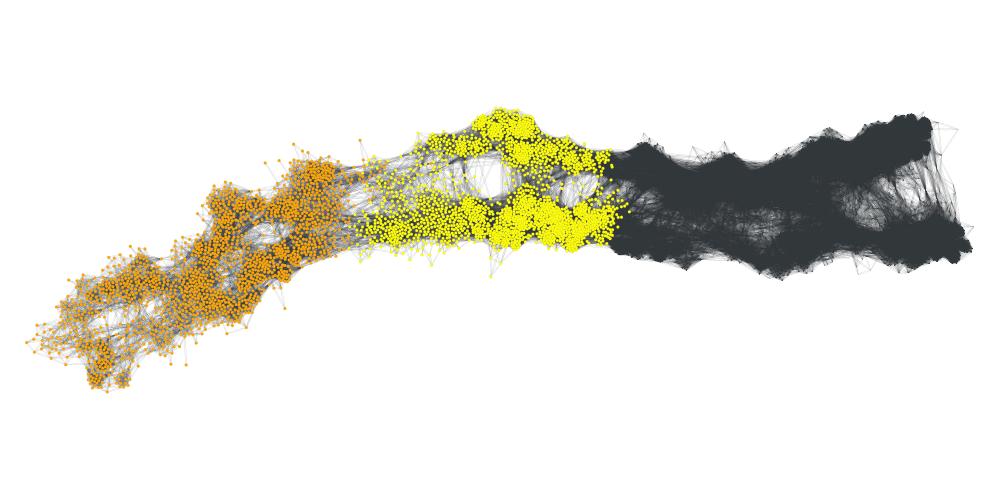}
	\hspace{-5.5cm}
	\begin{tikzpicture}
	\draw (0,0) -- (4.8,0);
	\foreach \x in {0,1,1.5,4.8}
	\draw (\x cm,3pt) -- (\x cm,-3pt);
	\draw (0.1,0) node[below=3pt] (a) {\tiny $1789$} node[above=3pt] {};
	\draw (1,0)  node[below=3pt]  {\tiny $1860$} node[above=3pt] (c) {};
	\draw (1.5,0) node[below=3pt](d) {} node[above=3pt] {\tiny $1913$};
	\draw (4.7,0) node[above=3pt] (f) {\tiny $2008$} node[below=3pt] {};
	\end{tikzpicture}
	}	
	\subfigure[Spectral MQI + sweep cut, seed: Spectral relaxation + sweep cut]{\label{Fig_senate_l1vl2_2}\includegraphics[scale=0.15]{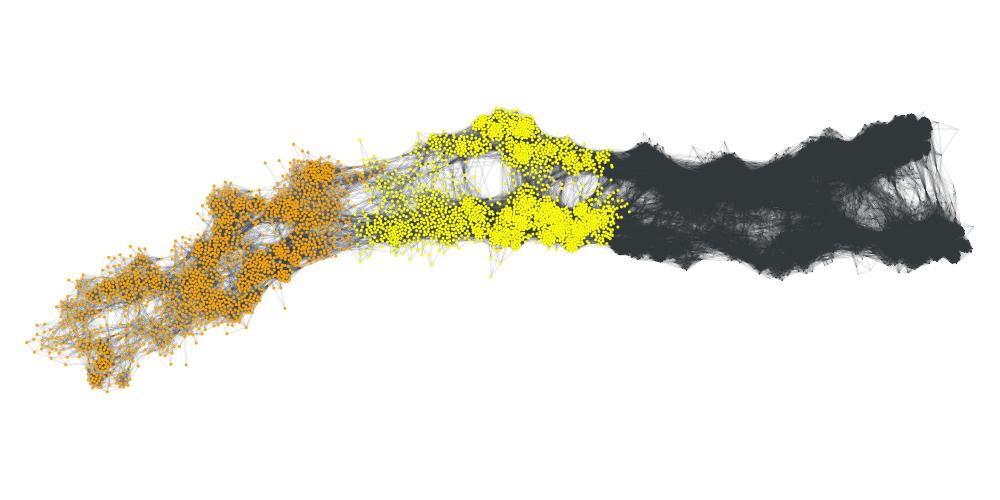}}
	\subfigure[Spectral MQI, seed: Spectral relaxation + sweep cut]{\label{Fig_senate_l1vl2_3}\includegraphics[scale=0.15]{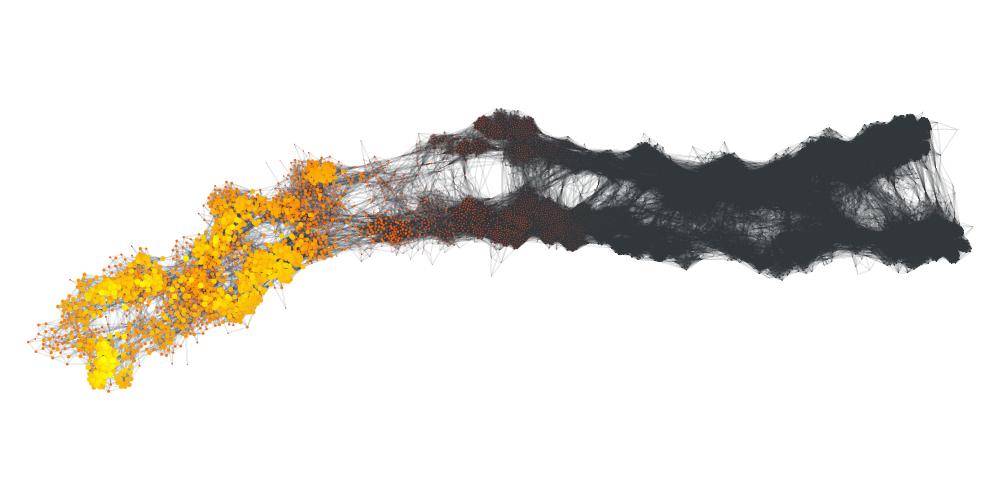}}
	\caption{US-Senate.\ This figure shows the solutions of MQI, spectral MQI, spectral MQI + sweep cut given the solution of Spectral relaxation + sweep cut as an input seed set. 
	For Figures \ref{Fig_senate_l1vl2_1} and \ref{Fig_senate_l1vl2_2} the red nodes are only MQI or spectral MQI + sweep cut depending on the experiment; yellow nodes are only seed set; and orange nodes are both part of the flow or spectral algorithm and the seed set. Figure \ref{Fig_senate_l1vl2_3} shows the solution 
	of spectral MQI without sweep cut. For Figure \ref{Fig_senate_l1vl2_3} we use a heat map to represent the weights of the nodes.
         Bright yellow means large positive and bright red means small positive. The size of the nodes shows the weights of the solution in absolute value. 
	}		
	\label{Fig_senate_l1vl2}
\end{figure*}

In this section, we will try to isolate the effect of $\ell_1$ and $\ell_2$ metrics in the output solution. We do this by employing MQI and spectral MQI, 
which are flow (i.e., $\ell_1$) and spectral (i.e., $\ell_2$) problems, respectively. The first set of results is shown in Figure \ref{Fig_senate_l1vl2}. Notice in Figure
\ref{Fig_senate_l1vl2_1} and \ref{Fig_senate_l1vl2_2} that MQI and spectral MQI + sweep cut recover the large clusters, i.e., before and after the year $1913$.
There are only minor differences between the two solutions. Moreover, observe that Spectral MQI returns a solution which is not binary. This is illustrated in Figure \ref{Fig_senate_l1vl2_3},
where the weights of the nodes are real numbers. Then sweep cut is applied on the solution of spectral MQI to obtain a binary solution with small conductance ratio, i.e., Figure \ref{Fig_senate_l1vl2_2}.

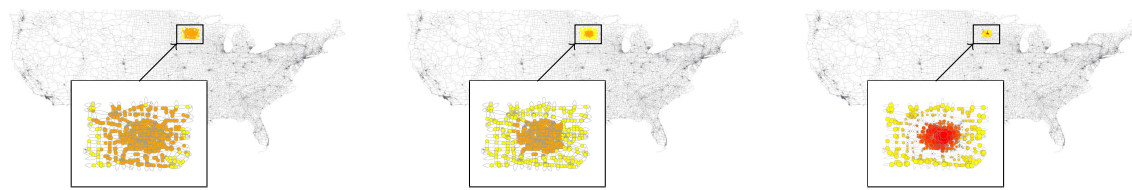
\begin{figure*}
\centering
	\subfigure[MQI, seed: Minneapolis and suburban areas]{
  	\begin{tikzpicture}[every node/.style={anchor=center}]
   	 \node(a) at (8,4){\label{Fig_usroads_l1vl2_1}\includegraphics[scale=0.133]{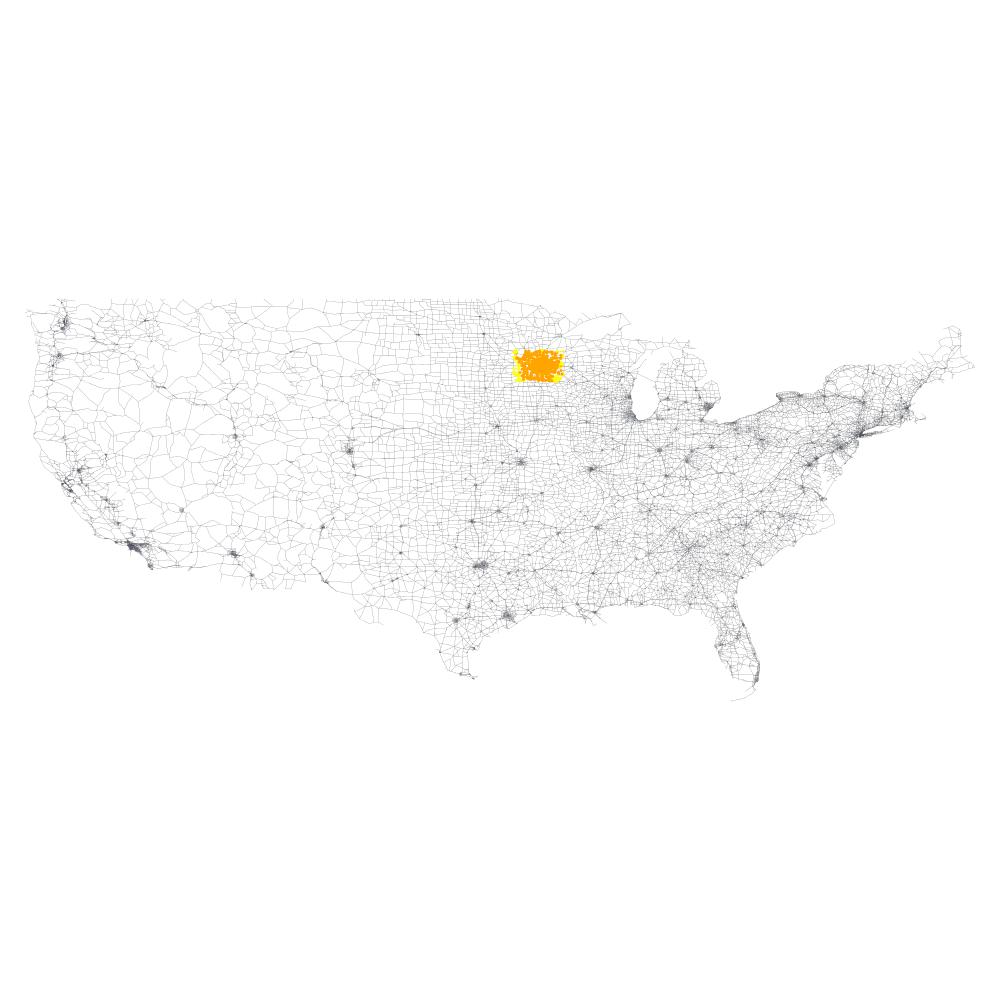}};
    	\node(b) at (7.5,3.3){\frame{\colorbox{white}{\includegraphics[scale=0.056]{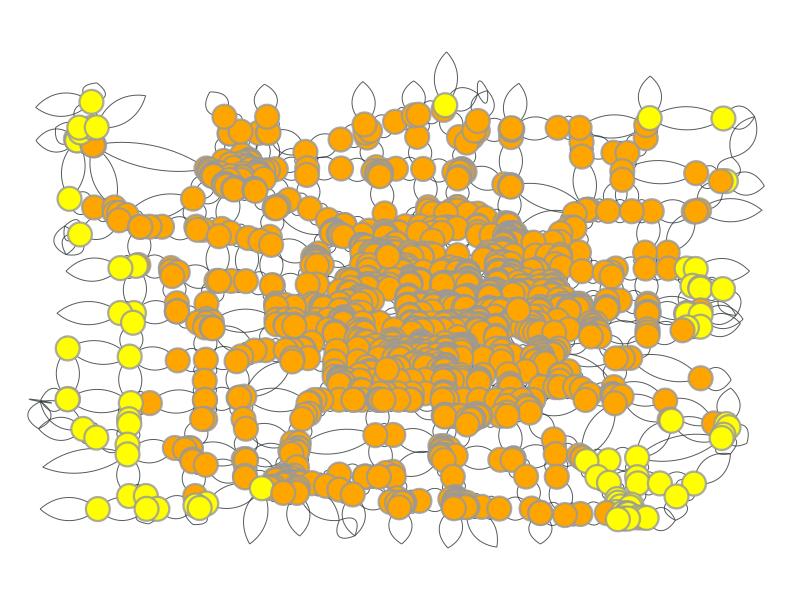}}}};
    	\draw[black] (8,4.75)rectangle (8.35,4.5);
    	\draw[black,->](7.5,4.0)--(8,4.5);
  	\end{tikzpicture}
	}
	\subfigure[Spectral MQI + sweep cut, seed: Minneapolis and suburban areas]{
  	\begin{tikzpicture}[every node/.style={anchor=center}]
   	 \node(a) at (8,4){\label{Fig_usroads_l1vl2_2}\includegraphics[scale=0.133]{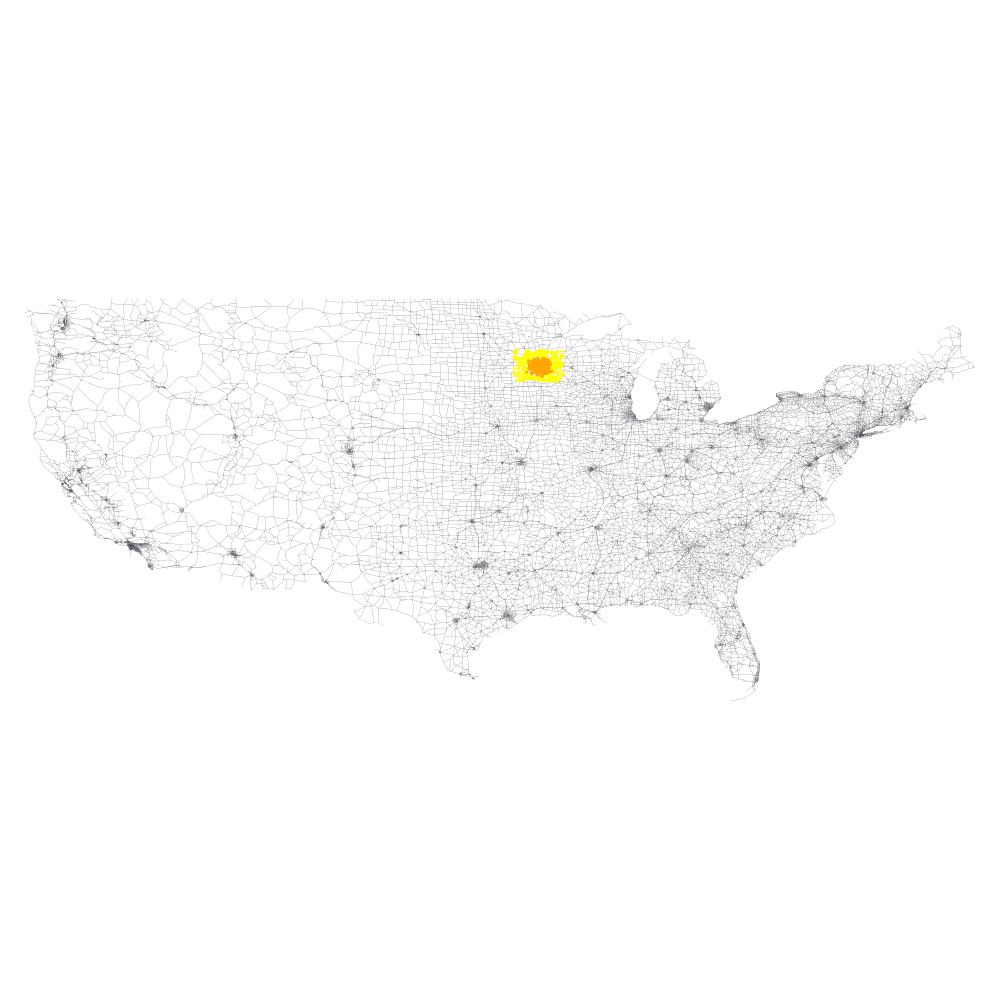}};
    	\node(b) at (7.5,3.3){\frame{\colorbox{white}{\includegraphics[scale=0.056]{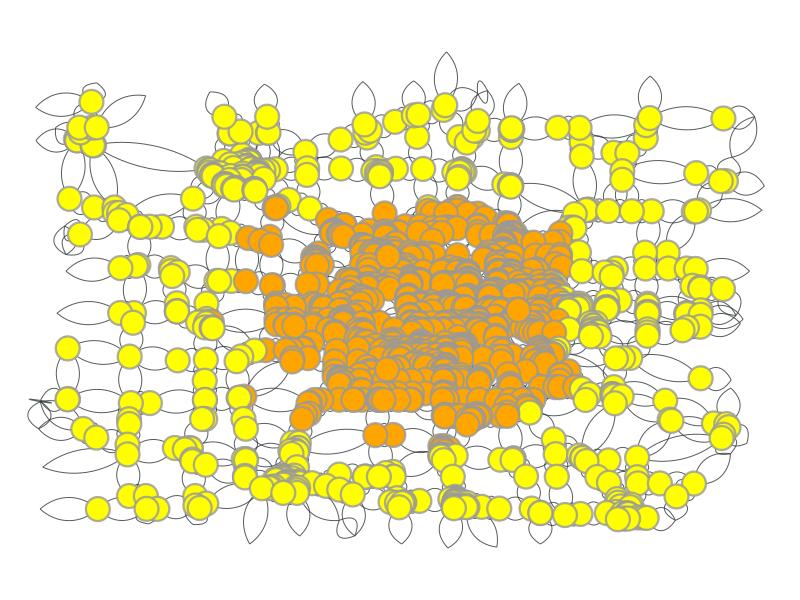}}}};
    	\draw[black] (8,4.75)rectangle (8.35,4.5);
    	\draw[black,->](7.5,4.0)--(8,4.5);
  	\end{tikzpicture}
	}
	\subfigure[Spectral MQI, seed: Minneapolis and suburban areas]{
  	\begin{tikzpicture}[every node/.style={anchor=center}]
   	 \node(a) at (8,4){\label{Fig_usroads_l1vl2_3}\includegraphics[scale=0.133]{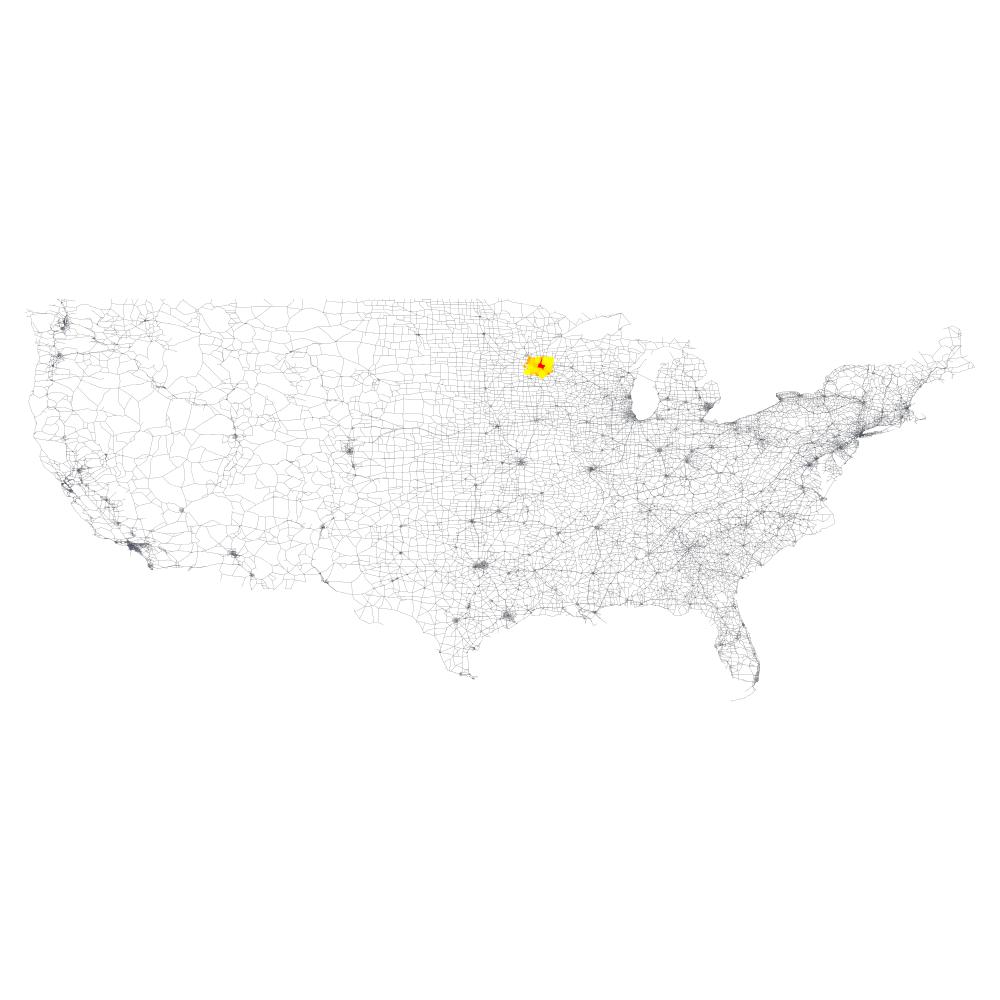}};
    	\node(b) at (7.5,3.3){\frame{\colorbox{white}{\includegraphics[scale=0.056]{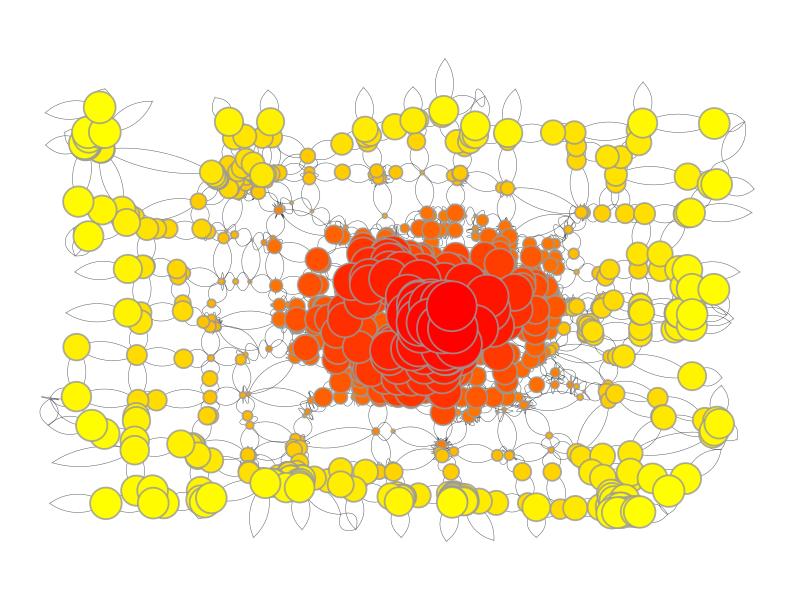}}}};
    	\draw[black] (8,4.75)rectangle (8.35,4.5);
    	\draw[black,->](7.5,4.0)--(8,4.5);
  	\end{tikzpicture}
	}
	\caption{US-Roads.\ This figure shows the solutions of MQI, spectral MQI, spectral MQI + sweep cut given Minneapolis and its suburban areas as an input seed set.
	The meaning of the colours of the nodes and its sizes is the same as in Figure \ref{Fig_senate_l1vl2}.}
	\label{Fig_usroads_l1vl2}		
\end{figure*}

The previous example did not reveal any difference between MQI and spectral MQI other than the fact that spectral MQI has to be combined with the sweep cut rounding 
procedure to obtain a binary solution. In Figure \ref{Fig_usroads_l1vl2}, we present a result showing where the solutions have substantial differences. The graph that we used for this
is the US-Roads, and the input seed set consists of nodes near Minneapolis together with some suburban areas around the city.
Notice in Figures \ref{Fig_usroads_l1vl2_1} that MQI, i.e., $\ell_1$ metric, shrinks the boundaries of the input seed set. However, 
MQI does not accurately recover Minneapolis. The reason is the volume regularization which is imposed by the denominator of the objective function of MQI. 
This regularization forces the solution to have large volume.
On the other hand, spectral MQI + sweep cut in Figure \ref{Fig_usroads_l1vl2_2} recovers Minneapolis. The reason is that 
for spectral MQI the regularization effect of the denominator is unimportant since the objective function is invariant to scalar multiplications of the solution vector.
It's the solution of spectral MQI, i.e., the eigenvector of smallest eigenvalue, which is presented in Figure \ref{Fig_usroads_l1vl2_3}, that is concentrated closely around Minneapolis.
Due to this concentration of the eigenvector around Minneapolis, the sweep is successful. Briefly, spectral MQI, which is a continuous relaxation of MQI, implicitly offers an additional
level of volume regularization, which turns out to be useful in this example.

\begin{figure*}
\centering
	\subfigure[MQI]{\label{Fig_fb_l1vl2_1}\includegraphics[scale=0.14]{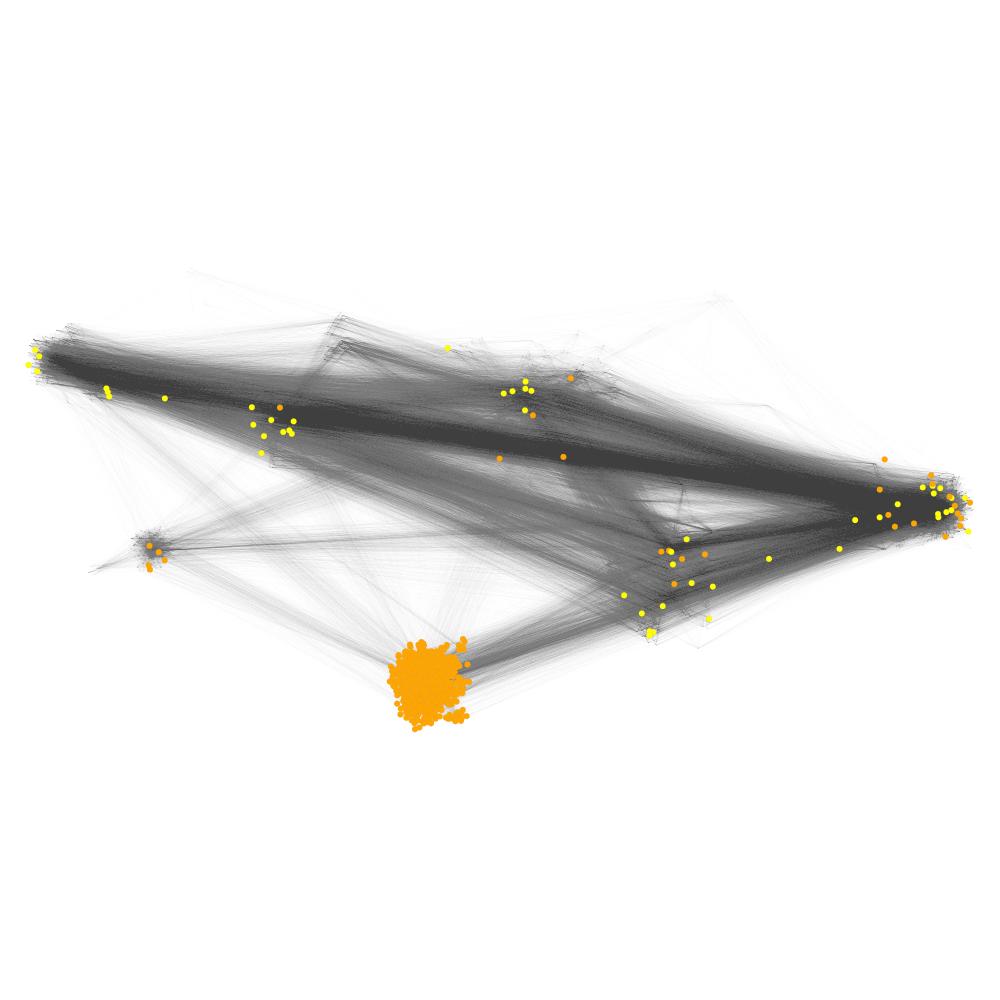}} 
	\subfigure[Spectral MQI + sweep cut]{\label{Fig_fb_l1vl2_2}\includegraphics[scale=0.14]{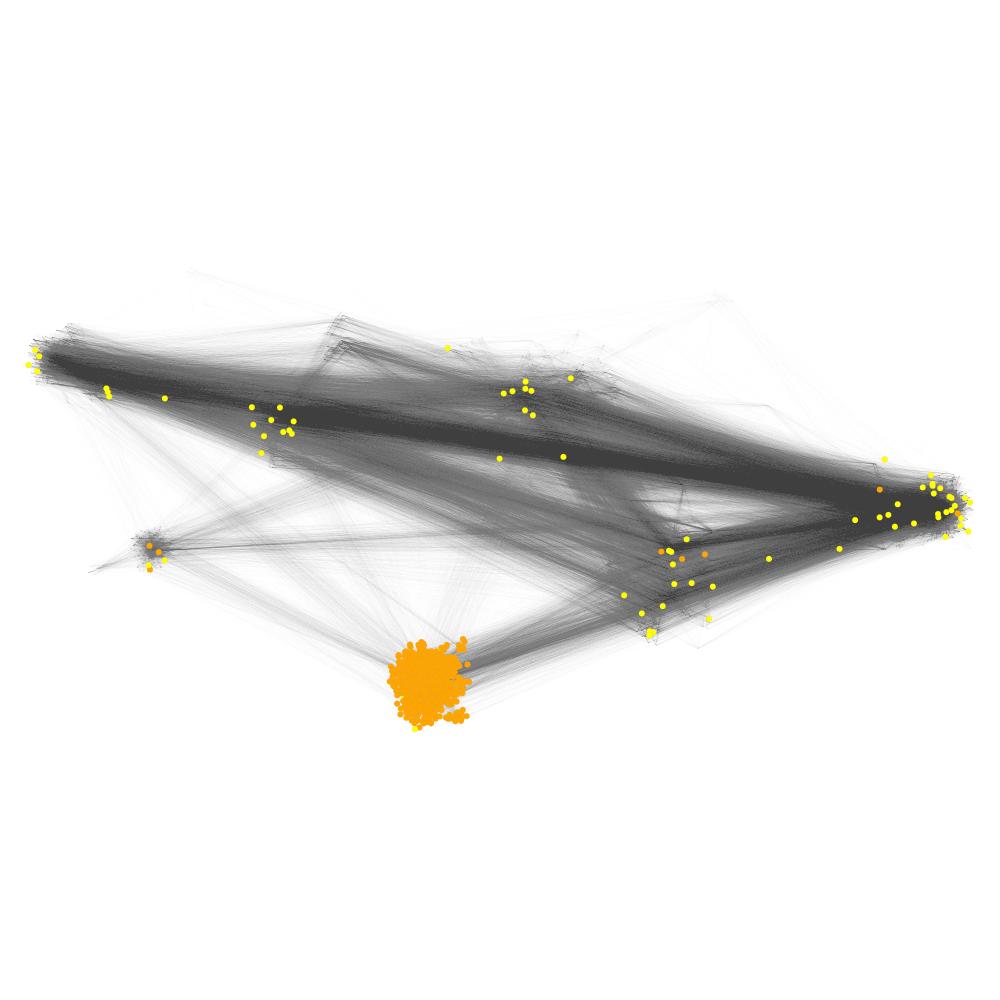}}
	\subfigure[Spectral MQI]{\label{Fig_fb_l1vl2_3}\includegraphics[scale=0.14]{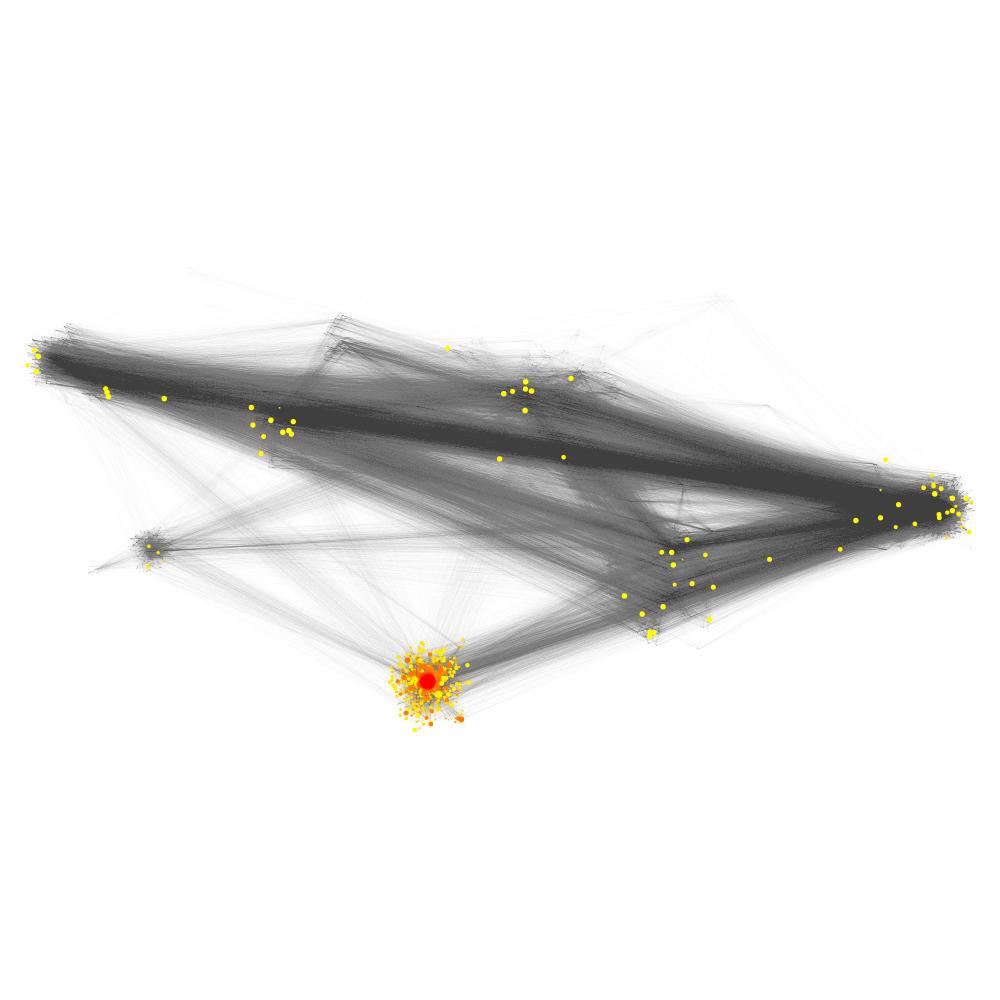}}
\caption{FB-Johns55. This figure shows the solutions of MQI, spectral MQI, spectral MQI + sweep cut for a given input seed set.
The meaning of the colours of the nodes and its sizes is the same as in Figure \ref{Fig_senate_l1vl2}.}
\label{Fig_fb_l1vl2}
\end{figure*}

Finally, we present another set of results using the FB-Johns55 graph in Figure \ref{Fig_fb_l1vl2}. As we saw before, notice that for this less well-structured graph
the solutions of MQI and spectral MQI + sweep cut are nearly the same. This happens because the regularization effect of 
the denominator of MQI and the regularization imposed by spectral MQI have nearly the same effect on this graph. This is also justified by the fact that MQI in Figure \ref{Fig_fb_l1vl2_1}
and spectral MQI without sweep cut in Figure \ref{Fig_fb_l1vl2_3} recover nearly the same cluster.

\section{Discussion and conclusion}
\label{sec:disc_conclusion}

Although the optimization approach we have adopted is designed to highlight similarities between different variants of locally-biased graph algorithms, it is also worth emphasizing that there are a number of quite different and complementary perspectives people in different research communities have adopted thus far on these methods. For example: 

%Informally, \emph{locally-biased graph algorithms} are algorithms 

%\begin{itemize}
	
%	\item 
	\textbf{Theoretical and empirical.}
	The theoretical implications of these locally-biased algorithms are often used to improve the performance of long-standing problems in theoretical computer science by improving runtimes, improving approximation constants, and handling special cases. Empirically, these algorithms are used to \emph{study} real-world data and to accelerate and improve performance on discovery and prediction tasks. Due to the strong locality, the fast runtimes for theory often manifest as extremely fast algorithms in practice. Well-implemented strongly-local algorithms often have runtimes in milliseconds even on billion-edge graphs~\cite{KG14,FCSRM16_TR}.
	
%	\item
	\textbf{Algorithmic and statistical.}
	Some of the work is motivated by having better algorithmic results, e.g., being fast and/or being a rigorous approximation algorithm, i.e., worst-case guarantees in terms of approximating the optimal solution of a combinatorial problem, while other work has provided an interpretation in terms of statistical properties, e.g., explicitly optimizing a regularized objective \cite{WSST14} or implicitly having output that are nice in some sense, i.e., well-connected output cluster \cite{ZLM13}. Often, locally-biased algorithms alone suffice as the result is an improvement to some downstream activity that will necessarily look at all the data anyway.

%	\item
	\textbf{Optimization and operational.}
	The locally-biased methods tend to result from stating an optimization problem and solving it with some sort of black box or white box. Strongly-local algorithms often arise by studying a specific procedure on a graph and showing that it satisfies some condition, e.g., that it terminates so quickly that it cannot explore the entire graph, that it leads to a solution with certain quality-of-approximation guarantees, etc. See, for instance, the spectral algorithms \cite{ACL06,AL08,AP09,C09,ST13,CS14,KG14} 
and the flow-based algorithms \cite{KKOV07_TR,OSVV08,OZ14}.

%\end{itemize}

In light of these complementary approaches as well as the ubiquity with which graphs are used to model data, we expect that locally-biased graph algorithms and our optimization perspective on locally-biased graph algorithms will find increasing usefulness in many application areas. 

%One perspective that has not yet been explored regards these methods as an information-extraction operation on the graph that reveals a signal hidden in noise, although some of the statistical perspectives approach these ideas from a different direction. This is an area we plan to investigate in the future.

%\clearpage
%\pagebreak
%\newpage
%\setlinespacing{1}
	
\bibliographystyle{IEEEtran}

\bibliography{communities,mwmbib_jrnl,mwmbib_proc,mwmbib_book,mwmbib_misc,mwmbib_drft,dgleich}

\end{document}